\documentclass{article}

\usepackage{arxiv}
\usepackage[utf8]{inputenc} % allow utf-8 input
\usepackage[T1]{fontenc}    % use 8-bit T1 fonts
\usepackage{hyperref}       % hyperlinks
\usepackage{booktabs}       % professional-quality tables
\usepackage{amsfonts}       % blackboard math symbols
\usepackage{nicefrac}       % compact symbols for 1/2, etc.
\usepackage{microtype}      % microtypograph
\microtypesetup{nopatch=footnote}
\usepackage{lipsum}		% Can be removed after putting your text content
\usepackage{graphicx}
\usepackage{natbib}
\usepackage{doi}
\usepackage{graphicx} % Required for inserting images
\usepackage{amsmath}
\usepackage{amssymb}
\usepackage{subfig}
\usepackage{float}
\usepackage{multirow}
\usepackage{caption}
\usepackage{subcaption}
\usepackage{geometry}
\usepackage{longtable}
\usepackage{setspace}
\usepackage{pdfpages}

\newcommand{\comment}[1]{}

\hypersetup{
    colorlinks=false,
    linkcolor=blue,
    filecolor=magenta,
    urlcolor=black,
    pdftitle={Källner et al. (2025): Arriving Young, Leaving Old(er)},
    pdfpagemode=FullScreen,
}
\graphicspath{{./fig},{./tmpfig}}
\title{Arriving Young, Leaving Old(er): Age-structured International Migration on Subnational Scale in Austria}

\author{\href{https://orcid.org/0009-0000-3859-0825}{\includegraphics[scale=0.06]{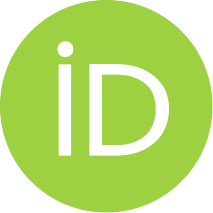}\hspace{1mm}Carsten~Källner} \\
	Complexity Science Hub\\
    Metternichgasse 8\\
	1030 Vienna, Austria \\
	\texttt{kaellner@csh.ac.at} \\
    \And
	\href{https://orcid.org/0009-0009-8099-8408}{\includegraphics[scale=0.06]{orcid.pdf}\hspace{1mm}Ola  Ali} \\
	Complexity Science Hub\\
    Metternichgasse 8\\
	1030 Vienna, Austria \\
	\texttt{ola@csh.ac.at} \\
    \And
	\href{https://orcid.org/0000-0002-3899-1175}{\includegraphics[scale=0.06]{orcid.pdf}\hspace{1mm}Andrea Vismara} \\
	Complexity Science Hub\\
    Metternichgasse 8\\
	1030 Vienna, Austria \\
	\texttt{vismara@csh.ac.at} \\
    \And
	\href{https://orcid.org/0009-0009-2261-7726}{\includegraphics[scale=0.06]{orcid.pdf}\hspace{1mm}Guillermo Prieto-Viertel} \\
	Complexity Science Hub\\
    Metternichgasse 8\\
	1030 Vienna, Austria \\
	\texttt{prieto-viertel@csh.ac.at} \\
    \And
	\href{https://orcid.org/0000-0002-0738-2633}{\includegraphics[scale=0.06]{orcid.pdf}\hspace{1mm}Rafael Prieto-Curiel} \\
	Complexity Science Hub\\
    Metternichgasse 8\\
	1030 Vienna, Austria \\
	\texttt{prieto-curiel@csh.ac.at} \\
}

\renewcommand{\headeright}{Källner et al. (2025)}
\renewcommand{\undertitle}{}
\renewcommand{\shorttitle}{\textit{Arriving Young, Leaving Old(er)}}

\hypersetup{
pdftitle={Arriving Young, Leaving Old(er): Age-structured International Migration on Subnational Scale in Austria},
pdfauthor={Carsten~Källner, Ola Megahed Ali, Andrea~Vismara, Guillermo~Prieto-Viertel, and Rafael~Prieto-Curiel},
pdfkeywords={Human Migration, Diaspora Flows, Demography, Host Societies, Austria},
}

\begin{document}
\maketitle

\begin{abstract}
%%% Modelling migration is complicated
Modelling migration is complicated, as people move for many reasons. Some leave their country for the first time, others return to places they once called home, or move on to new destinations. However, most models focus only on who arrives, missing the full picture of how migrant populations evolve.
%%% We introduce a Model of Diaspora Flows ... 
We introduce a model for diaspora flows that estimates both arrivals and exits using daily migration flow rates, disaggregated by age and nationality. Drawing on high-resolution administrative data from Austria covering over 1.8 million foreign nationals, the model allocates these movements across more than 2,000 municipalities based on the size of local diaspora communities.
%%% We find
We find that exits are not exceptions but a consistent and predictable feature across all groups. Migration rejuvenates Austria’s population, as both arriving and departing migrants are younger than the average resident. This effect has distinct age-geography patterns: younger migrants are drawn to cities, while older migrants are more evenly distributed in the country.
%%% Implications 
By capturing both arrivals and exits simultaneously, our approach provides a more comprehensive and interpretable picture of migration dynamics, how populations change over time, and how they are influenced by space, age, and national origins.
\end{abstract}

% keywords can be removed

\section{Introduction}
%%% Migration is complex  
{
Understanding how people move across borders can be challenging, as migration takes many forms. People move across borders for different reasons, at various stages of life, and often do so more than once. Some individuals leave their country of origin for the first time, while others migrate between several countries, sometimes repeatedly, or eventually return to places they have lived before, including their homeland \cite{king_return_1986, prieto_curiel_mobility_2022}. Scholars have long recognised that migration takes diverse forms, often involving movements in opposing directions. Awareness of such opposing flows, typically smaller in scale, can be traced back to Ravenstein’s influential work, \textit{The Laws of Migration}, which outlined the complexities inherent in migratory patterns \cite{ravenstein_laws_1885}. Nevertheless, migration research and policy continue to focus predominantly on immigration, while return and onward movements are less systematically understood or are treated as distinct and separate phenomena \cite{de_haas_theory_2021}. Yet, these secondary forms of migration account for approximately one-third of total global migration flows between 2010 and 2015 \cite{azose_estimation_2019}, emphasising their importance in overall human mobility. Understanding both directions of migration flows is essential for host countries, as migrants’ duration of stay and demographic composition evolve at distinct temporal and structural scales compared to those of native populations. Migrants may alter the demand for age-specific services (such as education), with direct implications for community-level integration policies. Consequently, migration cannot be understood simply in terms of who arrives or exits a country. It is equally important to focus on communities, where the everyday challenges and opportunities of migration unfold \cite{dean_migration_2024}.
}

%%% We know migration to be age selective ... many models
{
%%% MIgration is age selective
Migration is highly age-selective, with international migrants being disproportionately young adults \cite{rogers_model_1981} and shaped by the demographic structure of underlying populations \cite{welch_bringing_2024}, reflecting life-course transitions such as completing education, entering the labour market, and forming families \cite{horowitz_life_2021}. These age profiles are reflected in both the arrival and exit flows of migration \cite{raymer_modelling_2025}, providing essential insights into both individual migration trajectories and structural patterns, which determine not only the volume and direction of flows but also integration outcomes. The age of migrants influences multiple dimensions of integration, from social \cite{klok_role_2025} and political integration \cite{andersson_age_2025} to education \cite{chiswick_educational_2004}, and housing outcomes \cite{oladiran_age_2025}. Migration during early childhood is particularly associated with a higher risk of developing psychosis among minority groups \cite{kirkbride_ethnic_2017}, emphasising the importance of examining migration flows with attention to specific age cohorts. These demographic patterns matter not just for individual outcomes but also for broader population structures. Migrants help to reinforce working-age groups \cite{dorflinger_impact_2025}, contribute to trade growth, foster innovation, and support public finances through labour force participation and demographic renewal \cite{koczan_impact_2021}. Given these effects, it is essential to understand not only who arrives but also who leaves, as both inflows and outflows shape the long-term economic and social outcomes of migration.
}

%%% Migration modelling
%%% Models of migration study fall in the categories of (1) spatial, (2) statistical, and (3) agent based models models
{
Existing models of migration capture some of these dynamics. Spatial interaction models, such as the gravity model, remain most prominently used \cite{ramos_gravity_2016}. They describe patterns of human mobility by quantifying the attractiveness of spatially distinct units, origins and destinations, and the friction imposed by the distance separating them \cite{prieto-curiel_gravity_2025}. Attractiveness can be measured using GDP differences \cite{manzoor_gravity_2021}, cultural proximity \cite{grohmann_cultural_2023}, and the population size \cite{caballero_reina_gravity_2024} of both origin and destination countries. Distance is measured either in terms of physical distance \cite{lewer_gravity_2008} or through the quantification of transport, information, and psychic costs linked to migration \cite{karemera_gravity_2000}. Although radiation models represent another approach to capturing spatial interactions in migration \cite{alis_generalized_2021}, they have yet to be applied to international migration. Agent-based models have been used to represent individual migrants or entire migrant households with heterogeneous attributes (such as age, gender, income, and employment probabilities), and decision-making rules \cite{klabunde_decision-making_2016} and other elements like environmental factors (such as the risk of migration routes) \cite{gungor_simulation_2024}, the cost of migration \cite{hinsch_principles_2022}, or climate shocks \cite{entwisle_climate_2016}. As a result, they are particularly effective for modelling specific types of migration, such as climate-induced migration \cite{tierolf_coupled_2023} or forced migration in response to conflict \cite{mehrab_agent-based_2024}. Statistical models represent a third distinct category in migration flow modelling. Poisson models conceptualise migration as a stochastic process in which movements occur independently and at a constant rate. The expected migration count over a given time period (e.g., per year or per day) is given by the mean of a Poisson distribution. However, their implementation in the case of international migration is so far limited \cite{willekens_international_2016}. More frequently, time series extrapolation methods are employed, which utilise historical data to identify trends and project future migration patterns \cite{aparicio_castro_bayesian_2024}. Bayesian approaches to time series extrapolation incorporate ``expert opinion'', whereby a hypothesised distribution of migration and historical data distributions get merged in order to obtain probabilistic forecasts of migration \cite{bijak_assessing_2019}.
}

%%% Problems of modelling ... and missing data
{
%% Gravity models ...
Yet, key limitations remain in existing models. Gravity models do not capture temporal or more granular spatial dynamics \cite{beyer_gravity_2022} and their reliance on pairwise fixed effects between countries does not distinguish between demographically or socio-economically distinct sub-groups \cite{qi_modelling_2023}. Furthermore, these models generally overlook more complex mobility patterns such as return or onward migration, which fall outside their analytical scope and contribute to an incomplete understanding of migratory processes. While agent-based models are more suitable for capturing these sub-groups, they rely on complex data structures, necessitating rigorous validation and statistical calibration \cite{banks_statistical_2021}. Time series models are prone to sudden shocks and often need longer data histories to produce reliable estimates, while Bayesian approaches can be heavily influenced by the assumptions built into their prior distributions \cite{disney_evaluation_2015}. While time-series approaches, unlike spatial or individual-based models, capture the temporal dynamics of migration, they offer limited insight into the underlying mechanisms that shape these patterns.

A particular challenge in this regard is how to incorporate bidirectional movement patterns into flow models  \cite{zhang_analysing_2020}. As a result, return and onward flows from host countries have often been considered distinct forms of migration, modelled independently from other movements such as the propensity to migrate onwards or return \cite{constant_dynamics_2012}, economic determinants that influence these emigration patterns like economic and social integration \cite{van_hook_who_2011}, or human capital differentials \cite{amanzadeh_return_2024}. Therefore, as with other migration flow models, those focused on return and onward migration typically rely on a broad range of covariates. This reliance can limit their applicability across various contexts and substantially increase data demands. However, given the limited availability of flow data from both origin and destination countries, most migration flow estimates continue to depend heavily on stock data \cite{abel_estimating_2013}. Models based on stock data rely on census collections conducted at five- or ten-year intervals and most commonly use net migration estimates, which capture only the balance between immigration and emigration, without revealing the underlying flows \cite{raymer_modelling_2023}. Moreover, such temporal constraints can result in imprecise representations of migration dynamics and may introduce inconsistencies in how migration is defined across datasets \cite{sirbu_human_2021}. Although the growing availability of Big Data sources, such as social media and mobile phone records \cite{chi_measuring_2025}, offers promising opportunities for more temporally granular analyses of migration patterns, these sources raise concerns about data representativeness and validation \cite{welch_probabilistic_2022}. These limitations, both in traditional and emerging data sources, highlight the broader challenge of accurately capturing migration flows disaggregated by specific demographic characteristics, such as age \cite{ahmad-yar_anatomy_2021}. Furthermore, unless arrivals and exits are modelled jointly, underlying population dynamics remain hidden, disconnected from source population metrics, and, therefore, overall population projections become less reliable \cite{smith_practitioners_2013}.
}

%%% Adapting the Diaspora Model of human migration. Mention Austria
{
%%% Diaspora Model
To address the limitations of existing migration models and provide a more realistic account of the complex, bidirectional nature of migration flows, we propose an approach that links both immigration to and emigration from a host country with the size of its diaspora population. Shared identities and attributes, for example, nationality, ethnicity and age, play a crucial role in shaping human social relations and patterns of movement, frequently resulting in the formation of distinct localised communities \cite{mcpherson_birds_2001}. Distinct communities of migrants are referred to as diasporas, most commonly for migrant groups that share an ethnicity or country of origin. They often span from first-generation migrants to their descendants, with age and life course events shaping how these communities evolve \cite{bastia_migration_2022}. By providing social structure and information, diasporas facilitate further migration \cite{epstein_herd_2008, beine_diasporas_2011}, which makes them a valuable foundation for modelling migration flows \cite{beine_diaspora_2010}. In this study, we broaden the concept of diasporas beyond cultural or national belonging to include specific age cohorts, recognising these as distinct migrant population groups with their own age-related characteristics and migration dynamics. Our analysis builds on a framework that links migrant arrivals to the size of their diaspora in the host country and to the country’s pull and push strength, captured through total immigration volumes. This allows us to model both directions of movement within a single framework. These flows are then allocated across municipalities based on the existing spatial distribution of migrant populations, allowing us to examine where different diaspora groups tend to settle and from where they are more likely to exit within the host country. By capturing both inflows and outflows, we gain a deeper understanding of the dynamics underlying changes in migrant populations, even when overall numbers remain constant. Because it uses fewer parameters than comparable approaches and relies on routinely collected administrative data, the approach is transparent and more transferable across contexts. We account for spatial dispersion at the municipal level, reflecting the scale at which local administrations operate and where everyday processes of integration take place. Taken together, these elements provide a comprehensive view of migration dynamics on both scale and distribution.

%%% Add describtion what flow intensities are, and how the results are obtained ...
%%% The adapted Diaspora Model of Human migration allows us to 
This model for diaspora flows simplifies complex arrival and exit patterns by translating them into interpretable daily rates for each subpopulation and allocating these across subnational units. The model combines the statistical properties of a Poisson count process with spatial properties to estimate arrival and exit flows. It works in two steps. First, it estimates daily flow intensities, that is, daily migration rates, yielding the expected numbers entering and leaving each day for each diaspora group. Second, it allocates these daily flows across municipalities in proportion to the local size of each group’s diaspora.
}

%%% We find migration to be (spatially) dynamic especially when disaggregating by age-category
{
We apply our model to Austria’s residence registration data, which covers over 1.8 million non-Austrian migrants, and find that exit flows are both large and stable, a pattern consistent across nationalities and age groups. No group is characterised by either continuous build-up or drain from its population. Instead, all groups display measurable levels of arrivals and exits, indicating that migrant populations are not static but are subject to ongoing processes of self-replacement. However, the extent of these dynamics varies across different nationalities. For example, individuals with Ukrainian nationality are generally expected to stay for a shorter period, reflecting their status as temporarily displaced refugees. In contrast, people with Serbian nationality tend to exhibit longer durations of stay, a pattern consistent with their older diaspora and more established migration history in the country. Beyond nationality, these dynamics persist by age-cohorts, yet with varying degrees. This produces a net rejuvenation effect, whereby incoming migrants are younger than the resident population they join. The model reflects a clear spatial variation in migration patterns: migrants aged 18-34 tend to move into urban municipalities, where education, labour opportunities and services are concentrated \cite{buch_what_2014}, whereas older individuals are more evenly distributed or more likely to migrate to intermediate and rural areas. Taken together, these findings illustrate that international migration is both dynamic and structured, varying by age, nationality and geography.
}

\section{Results}
%%% Describtion of how results are obtained, the fit of the model is discussed in the appendix, and an overview over the results
{
Understanding how arrival and exit flows vary by nationality and age is central to capturing a country's demographic dynamics. To explore these patterns, daily arrival and exit flows by nationality and age are derived from a diaspora-flows model that captures migrant subgroups and spatial variation at high resolution. 
}

%%% Section 1: Exits are substantial and comparable to arrivals
\subsection{Emigration as measurable counterpart to immigration} \label{results:exits}
{
%%% Exits compared to arrivals
{
Between 2023 and 2024, roughly 180,000 foreign nationals left the country, while around 356,000 individuals entered (Figure \ref{fig:arrivals_exits}A), spanning all non-Austrian citizenships and age groups (excluding Austria-born children with foreign citizenship). Exits thus amounted to 50.6\% of arrivals. Although both arrivals and exits are considerable, arrivals clearly dominate: Austria welcomed nearly twice as many individuals as those who exited via onward or return migration to their country of origin. Nevertheless, exit flows remain a crucial component of migrant mobility, consistent with estimations on global return migration flows \cite{azose_estimation_2019}.
}
% FIGURE: Cumulative sum of arrivals and exits
{
\begin{figure}
    \centering
    \colorbox{white}{\includegraphics[page=1,width=\textwidth]{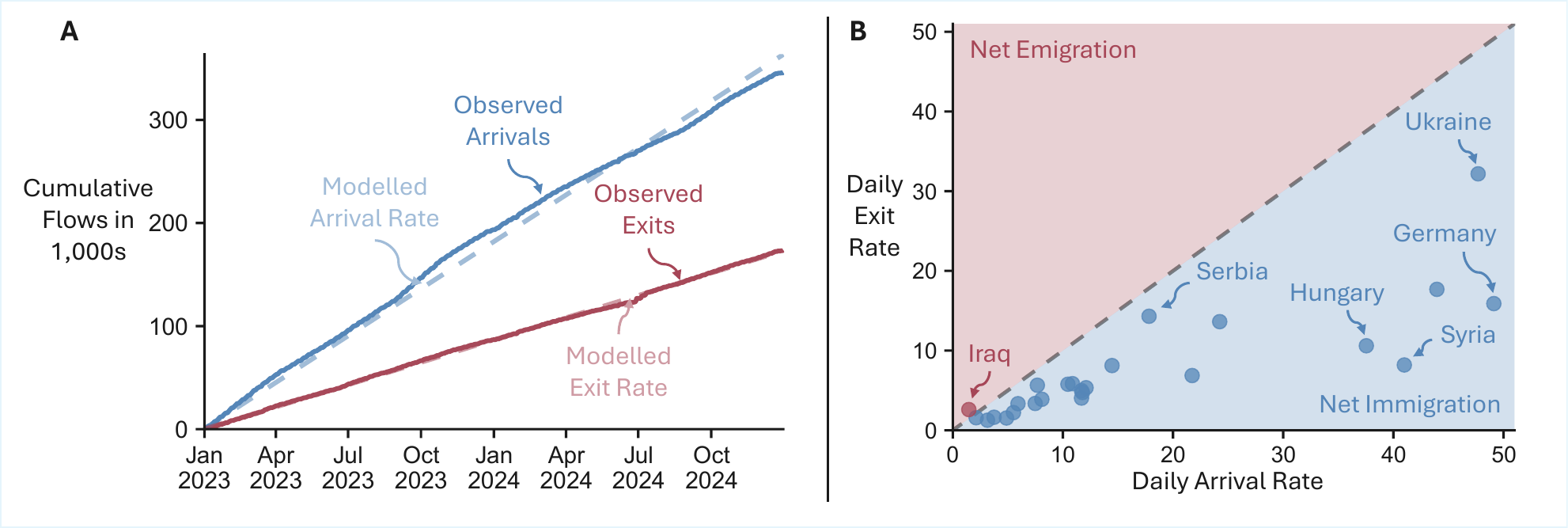}}
    \caption{\textbf{Dynamics of Arrival and Exit Flows to and from Austria.} \textbf{(A) Cumulative arrivals (blue) and exits (red)}, showing a strong match between observed data and modelled flows. Arrivals consistently exceed exits across all groups, indicating stable net immigration. \textbf{(B) Daily arrival and exit rates by citizenship}, based on the model for diaspora flows. Most groups lie below the 45-degree line, reflecting net immigration. Ukrainian and Syrian nationals show unusually high arrival rates, likely due to conflict-related migration. Iraq is the only diaspora among the 25 largest in Austria with net emigration.}
    \label{fig:arrivals_exits}
\end{figure}
}

%%% Comparable or exceeding flow stability
{
Effective migration modelling depends on the predictability of flows, allowing for robust estimates of who moves and where. To operationalise this, we translate short-term migration patterns into daily arrival and exit rates for each migrant subgroup. Arrivals follow relatively stable rates in the short term, allowing expected counts from a given country $i$ to be expressed as $\lambda_i\tau$ for the rate $\lambda_i$ of daily arriving migrants over a period $\tau$ \cite{prieto-curiel_diaspora_2024}. Building on this, we find that exits display comparable short-term stability to arrivals, allowing expected exit counts from a given country $i$ to be expressed as $\lambda_i\tau$, where $\lambda_i$ is the daily exit rate over the time period $\tau$. Across the majority of the 25 citizenship diaspora and all age cohorts in 2023 and 2024, the model for diaspora flows reproduces exits at least as accurately as it reproduces arrivals (Appendix F). When accounting for the lower overall volume of exits, the average modelling error for exits remains consistently lower than that for arrivals, indicating greater predictability. In general, only a handful of citizenships, most notably migrants from Ukraine and Syria, exhibit larger deviations in modelling accuracy for both arrivals and exits, reflecting the volatility of their recent flow. Additionally, the diasporas of Serbian and Turkish nationals exhibit fewer daily arrivals and exits than expected, likely reflecting the older nature of their diasporas in Austria, which span multiple generations. Overall, the findings confirm that exits are more stable and predictable than arrivals. Details on the derived daily rates and the model’s performance in reproducing observed migration patterns are presented in Appendix C and D.
}
}

%%% Section 1.3: Exits are experienced by every citizenship (appendix with all 25 major nationalities)
{
Building on the disaggregation by nationality, the national-level arrival and exit counts reveal that all foreign-citizen group in Austria experiences both arrivals and exits, confirming the bidirectional nature of international migration (Figure \ref{fig:diaspora_half_life}). Although the scale of these flows varies across nationalities, none lies at either extreme and, therefore, each diaspora undergoes a continuous process of self-replacement. Austria, in general, remains a country of net immigration; yet it still exhibits exit flows for each subgroup, so its population grows from persistent surpluses rather than from the absence of emigration. For every 10,000 migrants, we expect 1.76 exits per day, totalling approximately 642 per year. If this propensity to leave Austria remains constant for the initial cohort of 10,000 migrants, we expect about half of this cohort to have left Austria within about 10 years. Since some diasporas experience different rates of exits than others, these populations also differ in their proportion of exits to the initial size of the diasporas. Migrants with German nationality represent a special case due to shared language and cultural similarities; it takes 13.3 years for half of their initial diaspora of 120,000 to have left Austria. For Serbians (who have a long history of migration to and from Austria, resulting in the largest diaspora in the country of about 145,000), they experience a smaller daily exit rate, so it takes 18.7 years for half of the diaspora to exit. In contrast, Ukrainian migrants have a shorter history of migration, and following the Russian invasion, migration has become more dynamic. For an initial population of 78,000 Ukrainians, it takes only 4.1 years for half of the initial population to leave Austria (Appendix E). These variations illustrate that arrival and exit flows occur within diverse migration contexts and across populations with distinct histories and mobility patterns.
}

% FIGURE: Diaspora Hlf Life
{
\begin{figure}
    \centering
    \colorbox{white}{\includegraphics[page=1,width=\textwidth]{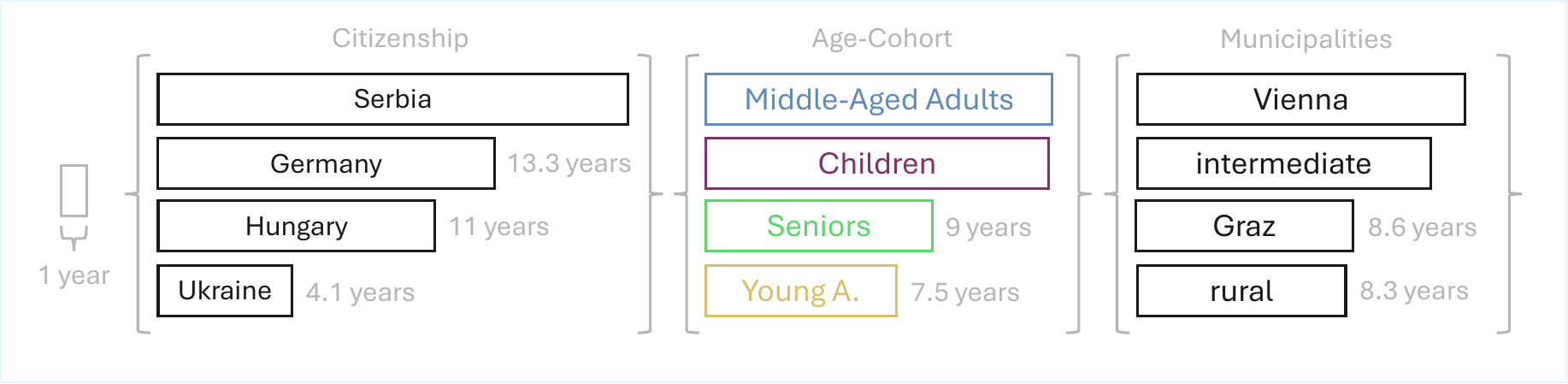}}
    \caption{ \textbf{Diaspora Half-Lifes}, or the time it takes for half of the pre-existing population size to leave Austria without accounting for any further entry, varies by citizenship, age, and location. Ukrainians leave fastest, while Germans and Serbians stay longer. Younger migrants and those in rural areas move on sooner, whereas older groups and city residents, especially in Vienna, remain longer.}
    \label{fig:diaspora_half_life}
\end{figure}
}

%%% Section 2: age-dynamics of arrivals & exits (compare to Austria)
\subsection{Age-dynamics of arrival and exit flows}  \label{results:flows_by_age}
{
%%% Age as alternative subpopulation and similar findings
{
Each age group can be regarded as a distinct migrant population, growing through the arrival of individuals at that age and shrinking through exits \cite{rogers_model_1981}. We group ages into five age-cohorts that represent different stages of life: children (2-17), young adults (18-34), middle-aged adults (35-54), older adults (55-69), and seniors (70+). This perspective allows tracing how people move at different points in their lives, revealing when migration is most common and how it changes as individuals age (Figure \ref{fig:age_dynamics}). Similar to nationality-based flows, we can assess the predictability of arrivals and exits for age-based diasporas. These flows are generally stable across age groups, with exits showing slightly higher predictability than arrivals. The only notable exception is the senior group, where predictability is roughly equal for both directions of movement (Appendix F). This outcome is likely due to administrative irregularities in exit recording rather than reflecting meaningful behavioural patterns.
}

%%% ... underlying importance of life-course perspective on in- and outmigration
{
International migration in Austria is most pronounced during the early stages of the life course, with children, young adults, and middle-aged individuals accounting for the majority of both arrivals and exits. All age groups experience both directions of movement, underscoring the demographic dynamism of each cohort. As migrants age, exit flows gradually converge with arrivals, leading to net emigration among older cohorts. This life-course pattern is consistent with age-scheduled migration flows \cite{raymer_modelling_2025}, in which young adults are the primary actors in both immigration and emigration, and middle-aged adults and children also contribute significantly to net immigration. A distinct peak in emigration is observed at age 62, highlighting the tendency of international migrants to return to their countries of origin upon retiring \cite{cobb-clark_return_2013}.
}
}

% FIGURE: Migration dynamics by age
{
\begin{figure}[H]
    \centering
    \colorbox{white}{\includegraphics[page=1,width=0.7\textwidth]{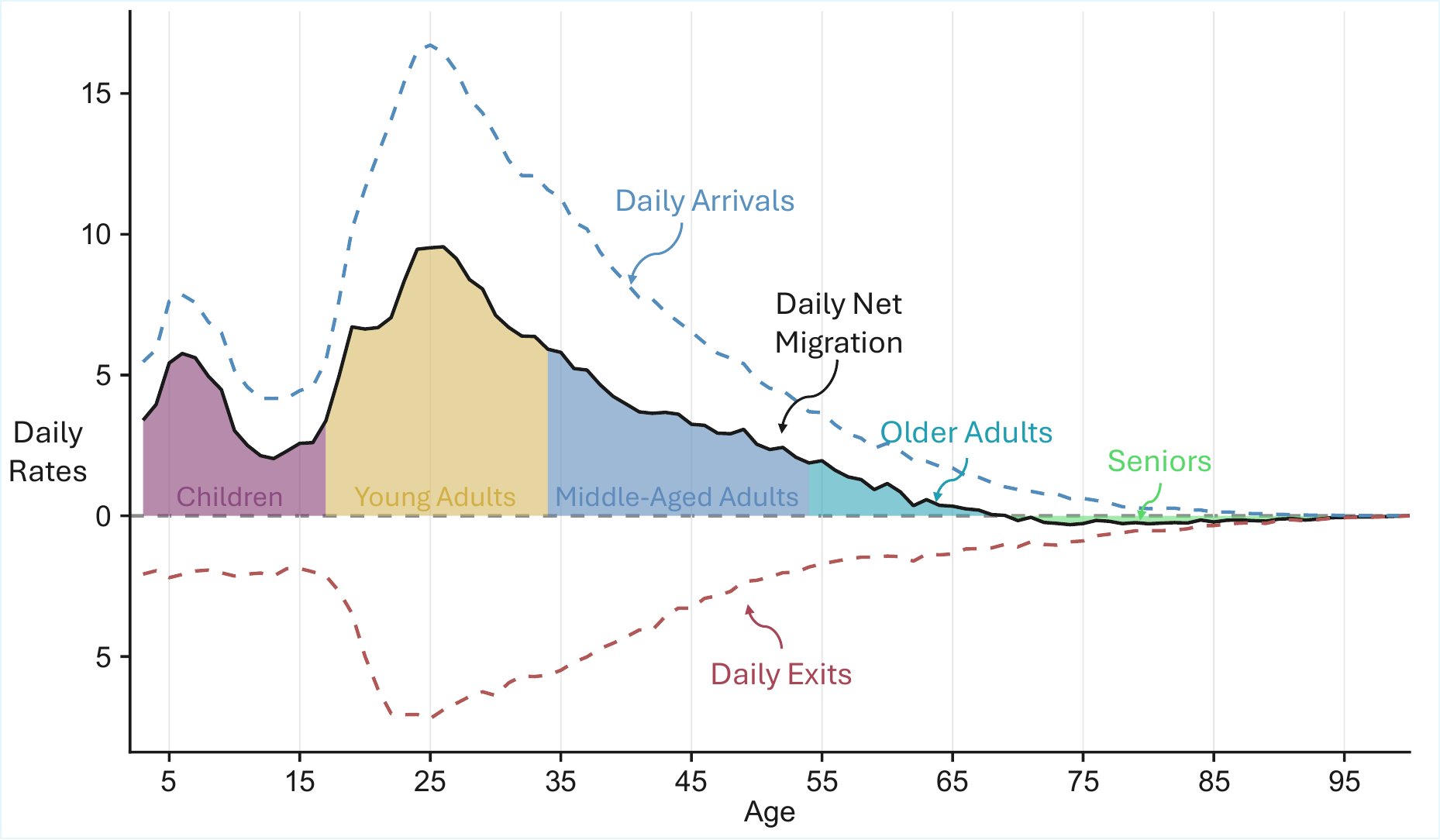}}
    \caption{\textbf{Age-specific arrival and exit dynamics:} Daily migration flows are shown for each age from 2 to 100, with arrivals (blue dashed) and exits (red dashed) indicating movement rates across the life course. The black line represents net migration, categorised into five age groups: children (2–17), young adults (18–34), middle-aged adults (35–54), older adults (55–69), and seniors (70+), marked by shaded segments. Migration is highest among young adults, who drive the majority of the net population growth. In contrast, net migration declines steadily with age, approaching zero or becoming negative among older groups. The pattern mirrors the age schedule of migration \cite{rogers_model_1981}, while extending the perspective to include exit flows alongside arrivals.}
    \label{fig:age_dynamics}
\end{figure}
}

%%% Comparison to age structure of Austrian natives
{
Based on our analysis of Austrian migration data, the average age of arriving migrants is 30 years, considerably younger than the national mean age of 43.6 years. In contrast, those leaving the country have an average age of 34.7 years. This 4.7-year age differential between arrivals and exits reflects a net inflow of younger individuals. Migration dynamics, therefore, have two distinct impacts on the host society. First, both incoming and outgoing migrants are substantially younger than the native population, which helps alleviate demographic pressures associated with population ageing. Second, the nearly five-year age gap between arrivals and exits suggests that many migrants spend a portion of their most economically productive years in the host country before moving on. This pattern highlights the critical role of international migrants in the labour force \cite{international_labour_organization_ilo_2024}, particularly during a formative stage of life when career trajectories and educational pathways are established.
}

%%% Age-diaspora exits
{
When exits are considered relative to the size of each age-based diaspora, our analysis reveals that flows occur in distinct contexts (similar to the patterns observed among nationality-based diasporas). Considering the relatively short period considered here (of less than three years) we model that people remain in the same age group over the projection period. Results show that young adults show the highest propensity to exit, a pattern mirrored among seniors (Figure \ref{fig:arrivals_exits}C). In contrast, middle-aged adults and children exhibit similar exit propensities, likely reflecting their tendency to migrate together, such as in family contexts (Appendix F).
}

%%% Section 3: different contributions of migration spatially 
\subsection{Spatial disaggregation of age-related migration dynamics} \label{results:flows_by_urbanisation}
{
%%% Introducnig spatial grouping shows ... by age
{
Given that international migration exhibits distinct characteristics across the degree of urbanisation \cite{chen_migration_2025}, we examine arrival and exit dynamics at the municipal level. To spatially locate migration movements to and from Austria, we define the arrival municipality as the first municipality of registration, and the exit municipality as the last municipality of registration. As a result, internal movements within Austria are not captured, so a migrant may arrive in one municipality but eventually leave another. Municipalities are classified into rural, intermediate, or urban areas based on spatial data from the national statistics office \cite{statistik_austria_urbanisation_2025}. By recognising the differences between urban, intermediate, and rural areas, this categorisation reflects how variations in local government structures, service availability, and overall connectivity shape migrants' lived experiences \cite{ensle-reinhardt_immigrant_2022}, especially refugees \cite{wong_non-metropolitan_2023}.
}

%%% Relation of age-diasporas and migration flows
{
We find that both arrival and exit flows at the municipal level increase with both the citizenship-specific and age-specific size of the local diaspora, confirming that out-migration is tied to diaspora presence just as strongly as in-migration (Appendix D). Overall, the national-level relationship in which exit rates amount to approximately 50\% of arrival rates remains consistent across all levels of urbanisation. Urban areas attract 258 daily arrivals and 124 daily exits, twice as many arrivals and exits as intermediate areas, and around two and a half times more than rural areas. These national-level patterns are reflected in local contexts, where individual cities and municipalities display similar dynamics, scaled to their diaspora size. Graz, Austria’s most populous single district and second largest city, records 28.6 arrivals per day and 15.2 exits. This means daily exits are about 53\% of daily arrivals. In contrast, intermediate and rural municipalities tend to exhibit lower levels of migration activity, yet follow the same national-level pattern. Feldkirch, a small town in the far west, records 2.2 arrivals and one exit per day, again mirroring the national pattern, with exits about 45\% of arrivals. The majority of rural municipalities, however, show little to no migration activity, a pattern linked to lower accessibility for international migrants \cite{mclean_rural_2022} and to smaller local diasporas. Of Austria’s 2,112 municipalities, 41\% each expect fewer than 10 migrants a year, and 5\% fewer than one migrant a year. 
}

%%% add "nice" figure with countryside
{
\begin{figure}
    \centering
    \colorbox{white}{\includegraphics[page=1, width=\linewidth]{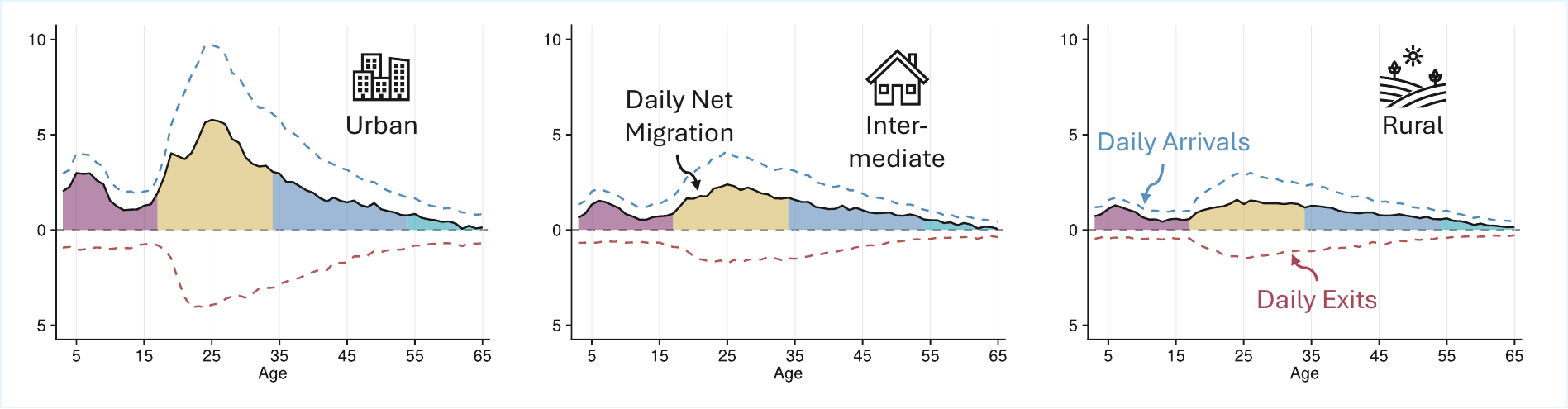}}
    \caption{\textbf{Age-specific migration flows by degree of urbanisation:} Daily arrivals (blue dashed), exits (red dashed), and net migration (black line) are shown by age for urban, intermediate, and rural areas. Urban centres see the highest inflows, especially among young adults. In contrast, net migration for middle-aged adults, older adults, and seniors is more evenly distributed across intermediate and rural areas, suggesting a greater likelihood of settling outside cities at older ages.}
    \label{fig:age_degree_urbanisation}
\end{figure}
}

%%% Dynamism of urban areas driven by younger age groups
{
The age composition of international migrants differs notably between municipalities, depending on their respective levels of urbanisation (Figure \ref{fig:age_degree_urbanisation}). Although international migration is commonly characterised as concentrated in urban areas \cite{buch_what_2014}, our analysis demonstrates that migration flows occur across all levels of urbanisation and age groups. Children and young adults are most concentrated in the urban areas. Urban areas welcome nearly twice as many children and more than twice as many young adults as intermediate areas, and almost three times as many as rural regions. The strong pull of urban areas begins to shift among middle-aged adults: urban areas attract twice as many as rural areas, while intermediate areas fall in between. For older adults, the differences narrow but remain: urban areas still attract more than intermediate and rural areas, though by an increasingly smaller margin. Among seniors, the trend reverses. Rural areas show a slight net gain, while cities see more people exiting than arriving. These patterns highlight how the relationship between age and regions shapes migration flows: urban areas serve as key entry points, especially for younger migrants, while intermediate and rural areas play a more prominent role later in life.

To better understand how migration affects population structure, we compare the average age of migrants with that of the resident population across different levels of urbanisation. Comparing the average age of migrants in the flows with the average age of residents at each level of urbanisation shows that migration rejuvenates populations in urban, intermediate and rural areas alike. In urban areas, residents have an average age of 41.6 years, while arrivals average 29.2 years and exits average 34.7 years. In intermediate areas, the resident average age is 43.7 years; arrivals average 30.3 and exits 33.3. In rural areas, residents average 44.4 years; arrivals 31.4 and exits 33.8 \cite{statistik_austria_meanage_2023}. Taken together, the migrants who arrive and those who exit are consistently younger than residents, thereby refreshing the age structure across urbanisation levels.
}
}

%%% Section 3.1: Spatial heterogeneitites by age and urbanisation
{
%%% Disaggregation by Austrian state introduced 
{
By integrating Austrian states into our analysis, we can reflect more spatially distributed migration patterns for each age cohort and analyse the intersection of age cohort, urbanisation degree, and regional context (Figure \ref{fig:age_cohort_state_urbanisation}A). Vienna, as the capital and largest city in the country, consistently attracts a large number of migrants, particularly young adults, with 82 arrivals expected each day. This pattern is mirrored at lower intensities among children and middle-aged adults. Outside the capital, Lower Austria and Upper Austria show more balanced but positive migration dynamics among younger and middle-aged migrants. In contrast, peripheral or alpine states (such as Tyrol, Vorarlberg, Carinthia, and Burgenland) appear less demographically dynamic or even declining, with migration being less active in these regions, indicating lower migratory attractiveness. Among older adults and seniors, the dynamics shift again: across most regions, the ratio of arrival and exit flows converges to parity, reflecting a general decline in migration intention with age. Vienna is experiencing considerable exit by its' senior migrant population with an expected net migration of 5 migrant seniors exiting Vienna each day. 

Spatial differences across age cohorts reveal a clear concentration of young migrants arriving in and around urban areas (Figure \ref{fig:age_cohort_state_urbanisation}B), whereas older adults exhibit more evenly distributed migration patterns, with arrivals and exits occurring at comparable levels across regions (Figure \ref{fig:age_cohort_state_urbanisation}C). Altogether, these findings illustrate that international migration is not evenly distributed but instead appears in distinct spatial regions with differing patterns. Some states and urban areas constitute fluid regions characterised by high migration volumes, while others exhibit features of settlement zones, where mobility declines. Although Vienna is highly dynamic, with the highest levels of both arrivals and exits, this fluidity is not fully reflected in the estimated diaspora half-life. Assuming a steady likelihood of exit, it would take approximately 13 years for half of Vienna’s current diaspora population to exit from Austria. In intermediate regions, this process is somewhat faster, with a half-life of approximately 11.6 years, reflecting a more moderate level of population turnover (Appendix E). In contrast, smaller cities such as Graz and rural regions would reach the same point more quickly, in about 8.6 and 8.3 years, respectively (Figure \ref{fig:arrivals_exits}C). The resulting geography of migration is not static but is shaped by demographic flow patterns, in which age, degree of urbanisation, and regional opportunities influence when, where, and whether migrants move, and where they eventually settle.
}

% FIGURE: Arrival and exit intensities by urbanisation degree, state, and age-cohort
{
\begin{figure}[H]
    \centering
    \colorbox{white}{\includegraphics[page=1, width=\linewidth]{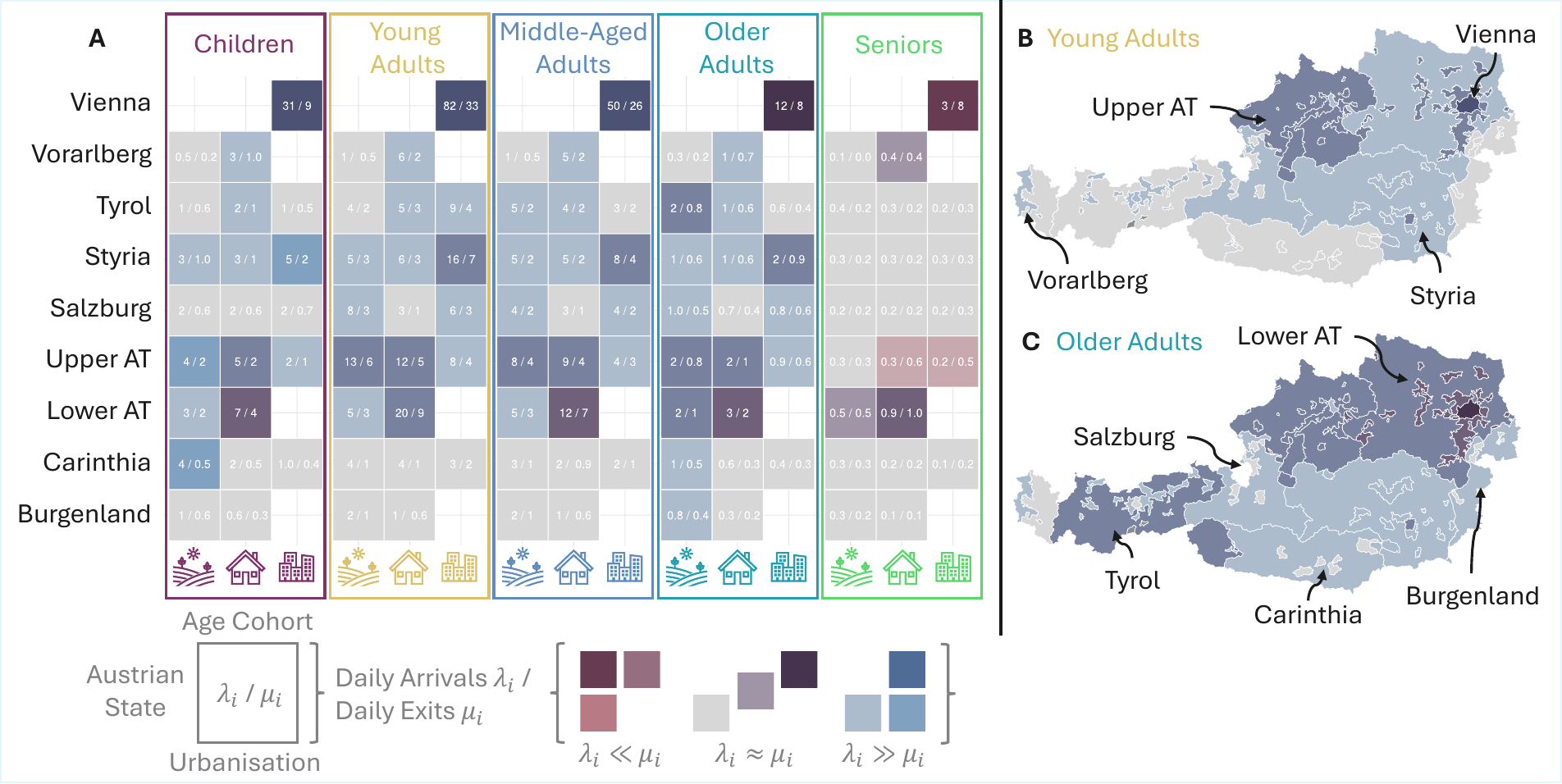}}
    \caption{\textbf{Arrival and exit rates by age-cohort, Austrian state, and degree of urbanisation:} \textbf{(A)} Daily arrival and exit rates by Austrian state and age group, disaggregated by urbanisation level. Cells show flow intensities with $\lambda_i$ for arrivals and $\mu_i$ for exits. Each cell is shaded by the balance of arrivals and exits (blue = net immigration, red = net emigration). Young adults account for the highest migration volumes, especially in urban centres like Vienna and Upper Austria. Older cohorts show lower flows, which are more balanced or negative, and are spatially more evenly distributed. \textbf{(B)} Daily arrival rates for young adults, concentrated in urban regions such as Vienna, Upper Austria, and Styria. \textbf{(C)} Arrival rates for older adults are more evenly spread, with balanced flows across urban, intermediate, and rural areas, particularly in Lower Austria, Upper Austria, and Tyrol.}
    \label{fig:age_cohort_state_urbanisation}
\end{figure}
}
}
%%% now we can cater services more to actual age dynamics
\section{Conclusion}
%%% Migration is complex, we show that exits are predictable (spatially), exits are substantial for different nationalities and ages
{
Migration processes go beyond simple, unidirectional flows that begin in sending countries and conclude in receiving ones. People arrive, settle, and in some cases move on or return to their country of origin. These patterns highlight both the complexity of migration and its place within an individual's life course. For host countries, tracking these patterns is essential to understanding migrant population dynamics, particularly at the municipal scale, where administrative and integration services are provided. From the perspective of the host country, Austria, we propose a diaspora-based model that utilises diaspora size to estimate daily arrival rates and municipal-level diaspora counts, thereby identifying where these movements occur. We then add exit flows and their spatial distribution, so the same framework captures both directions of movement. Furthermore, to better reflect the role of life stage in international migration, the definition of diaspora is extended to include not only a migrant’s nationality but also their age cohort. This adjustment recognises the age-specific characteristics and patterns that shape migration dynamics. Our approach addresses key limitations of existing migration flow models by using gross rather than net flows to more accurately capture population dynamics, using fewer covariates, and producing daily estimates of arrivals and exits at a subnational scale. Crucially, it also incorporates the analysis of migrant subgroups, enabling a deeper understanding of diaspora and migration flow dynamics.
}

%%% We look at arrivals and exits for ages and find that ... 
{
Consistent with national-level estimates of subsequent migration flows \cite{azose_estimation_2019}, our analysis indicates that exit flows represent a substantial component of overall mobility. While not matching arrival flows in absolute terms, exit flows occur across all citizenship groups, age cohorts and exhibit distinct spatial patterns. Moreover, disaggregation by age group reveals a rejuvenation effect associated with migration. Younger cohorts contribute most significantly to both arrival and exit flows and tend to represent net-immigration subpopulations. In contrast, older age groups are more likely to emigrate, indicating a relative outflow among this demographic. Overall, both arriving and exiting migrants are significantly younger than the overall Austrian population underlying two effects of migration dynamics: First, migrants help to rejuvenate Austria’s population; Second, with an average age at arrival of 30 and at exit of 34, those who eventually leave still spend several years in Austria during highly productive stages of the life course \cite{koczan_impact_2021}. Looking more closely at where migrants arrive and where they eventually leave within the host country, arrivals and exits occur at every level of urbanisation and across all age groups. Urban areas, particularly Vienna, register the highest overall migration volumes. This concentration of movement is primarily shaped by younger migrants aged 18 to 40. Intermediate areas attract a broader mix of age groups, though with more moderate and balanced flows. In contrast, rural areas exhibit considerably lower migration levels, yet their movements still contribute to local demographic shifts. Across all levels of urbanisation, migrants are consistently younger than the resident population, indicating that migration acts as a source of demographic rejuvenation. 
}

{
We calculated how quickly diaspora groups renew, offering insight into the typical length of residence in these communities. This understanding supports the development of integration policies that reflect migrants' settlement journey and highlight which phases are most prevalent in each community. These spatial and age-related differences highlight the importance of accounting for age-cohort dynamics when analysing the geography of migration and its longer-term implications.
}

%%% Impact, limitations, and further research ... 
{
Understanding arrivals and exits for distinct subpopulations provides essential insights into the dynamics of migration flows and enables more targeted policy responses, particularly by tailoring services to specific age groups or citizenships. Equally important is the combined consideration of spatial and subpopulation dynamics, since processes of social, economic, and demographic integration unfold primarily at the municipal level, where pressures on services and opportunities for inclusion are most directly experienced. The available register data often underestimates exit flows \cite{danko_assessing_2024}. Moreover, the model does not capture internal movements of international migrants. As a result, individuals may enter one municipality and exit another without these movements being reflected in our analysis. While internal movements by distinct migrant populations within host countries remain an underexplored field \cite{robiglio_multiscale_2025}, it has been shown that non-refugee migrant populations have a low likelihood of moving \cite{ali_quantifying_2024}. Finally, the formation and evolution of a diaspora, e.g., due to shocks, changing migration drivers, or declining migration costs, remain outside the scope of this study but represent an important avenue for future research.
}

%%% THIS IS THE END: challenges of migration process (juvinating effect of migraiton in Austria)
{
Overall, our findings demonstrate that both arrivals and exits of international migrants contribute to sustaining a dynamic population structure in Austria, shaped by age composition and spatial distribution. Age constitutes a key dimension in our analysis, with younger cohorts contributing disproportionately to urban arrivals and exits. In comparison, older cohorts are comparatively more likely to migrate to and from rural areas, though at relatively lower rates. The movement of migrants into and out of Austria significantly contributes to demographic renewal. Many engage in the labour market during their most productive years, making critical economic contributions before eventually returning to their country of origin or moving elsewhere. This dynamic helps sustain Austria’s workforce and offsets the challenges associated with an ageing population. 
}

\section{Methods}
\subsection{Data Source}
%%% BMI data on non-Austrian residents excluding births and deaths
{
We use high-resolution administrative data from Austria’s Ministry of the Interior (BMI), covering the years 2023 and 2024. This dataset includes all legally registered non-Austrian residents and tracks changes in residential status through the national registration system (''Meldezettel''), which requires individuals to declare their municipality of residence. The dataset includes around 1.8 million individuals and contains key variables such as citizenship, age, gender, and municipality of residence. It is updated daily, offering an unusually detailed view of short-term dynamics in international migration. To accurately capture migration flows, only individuals with complete registration records are included. Children born in Austria to non-citizen parents and recorded deaths were excluded to focus on international migration patterns. A full description of the dataset and the applied data filtering methods is provided in Appendix A.
}
\subsection{Modelling Migration Intensity}
%%% We define arrivals and exits as ... our model
{
We define an arrival as the first recorded residence for an individual within the observation window and an exit as the last record associated with a departure. Our model estimates the daily arrival and exit rates for each migrant group using a Poisson process, treating migration as a count-based event that occurs over time. The number of arrivals $A_i(t)$ and exits $E_i(t)$ for a given subpopulation $i$ up to time $t$ is modelled as:
\[ A_i(t) \sim \text{Pois}(\lambda_it), \quad E_i(t) \sim \text{Pois}(\mu_it).\]
Here, $\lambda_i$ and $\mu_i$ represent the average daily arrival and exit rates for subpopulation $i$, respectively. To reflect individual-level behaviour, we assume that exits follow a binomial process, where each migrant from diaspora $D_i$ independently exits with probability $r_i$ over the observed time period (Appendix E). Since a binomial process conditional on a Poisson also follows a Poisson distribution, daily exits can be modelled as $E_i(t) \sim \text{Pois}(\mu_it)$. The net migration flow $N_i(t)$ for subpopulation $i$, therefore, follows a Skellam distribution, representing the difference between two independent Poisson variables:
\[N_i(t)=A_i(t)-E_i(t) \sim \text{Skellam}(\lambda_it, \mu_it).\]
This allows us to estimate not only whether a group is growing or shrinking, but also the scale of movements behind those changes.
}
\subsection{Allocating Flows Across Municipalities}
%%% 
{
The next step involves assigning arrivals and exits to Austria’s 2,112 municipalities. We use a multinomial distribution to allocate flows in proportion to each municipality’s share of the diaspora:
\[\pi_{ij} = \tfrac{R_{ij}}{R_i}, \]
where $R_{ij}$ is the number of migrants from group $i$ in municipality $j$ as of the end of 2022, $R_i$ is the number of migrants from group $i$, and $\pi_{ij}$ is the probability that a migrant from group $i$ either arrives in or exits from that municipality. Given total flows $A_i(t)$ and $E_i(t)$, spatially disaggregated flows $A_{ij}(t)$ and $E_{ij}(t)$ are modelled as:
\[A_{ij}(t) \sim \text{Pois}(\pi_{ij}\lambda_it), \quad \text{and} \quad E_{ij}(t) \sim \text{Pois}(\pi_{ij}\mu_it).\]
Thus, reflecting the idea that larger local diasporas both attract new arrivals and generate more exits.
}
\subsection{Model Capabilities and Limitations}
%%% Strengths
{
A key strength of our approach lies in its ability to estimate both arrival and exit flows of migrant subpopulations at the local level, disaggregated by characteristics such as age and nationality. This enables a more nuanced understanding of migration dynamics than is typically possible with national-level or net-migration models, particularly in capturing the scale and direction of flows across municipalities. Another advantage of the model is that its predictability improves with larger populations: although the total magnitudes of arrival and exit flows become more variable as more people are involved, this variation grows more slowly than the population itself. As a result, the model performs especially well for larger diasporas, offering more stable and reliable estimates where migration is most concentrated.
}

%%% Weaknesses
{
Our model captures only international movements and is based on administrative registration data, which excludes unregistered or undocumented migration. It does not account for internal relocation within Austria, meaning individuals may arrive in one municipality and later exit another without those internal moves being reflected. In addition, the model does not simulate the formation or evolution of diaspora over time; instead, it uses a fixed snapshot of the existing resident migrant population. Diasporas often form over extended periods and may eventually include second- or third-generation migrants, by which point the drivers of initial migration may have already weakened. In contrast, other migration movements are shaped by more recent developments and are therefore observed either in an early formation phase or during a temporary surge that may gradually decrease. Despite these constraints, the framework offers a transparent, transferable tool for analysing short-term migration flows in a disaggregated, interpretable way. A detailed description of the model structure, estimation procedures, and underlying assumptions can be found in Appendix B.
}

\bibliographystyle{unsrt}

\appendix
\section{Appendix} \label{appendix:data} 
\subsection{Data Source}
%%% Daily provided by the Austrian Ministry of Interior BMI
{
Our analysis is confined to persons of foreign citizenship who have been formally registered as residents in Austria under the statutory registration requirement ("\textit{Meldepflicht}"). Migrants are therefore operationally defined as foreign citizens. As Austria grants citizenship primarily by descent (\textit{ius sanguinis}), the zero-year age cohort is excluded to avoid including infants born in Austria to two non-citizen parents, who are not considered international migrants. Age 1 is also omitted, as only the year of birth is recorded in the data. This can lead to misleading results, especially for children born late in the year and registered at the start of the next year, who may appear to be age one despite being only a few days or weeks old. These cases reflect delayed registration rather than true migration.
}
\subsection{Definition of migrants}
%%% Both voluntary and forced migration
{
The dataset incorporates both voluntary and forced forms of migration. Accordingly, the dataset includes individuals classified as migrants, referring to those who relocate for economic, educational or personal reasons, as well as individuals with refugee or asylum-seeking status, who are involuntarily displaced due to war, conflict or environmental disasters.
}
\subsection{Data Structure}
%%% Overall Number & description
{
For the selected period of 2023 and 2024, the dataset comprises 2,370,607 entries representing 1,787,848 unique individuals (see Figure \ref{tab:variable_construction}). To build realistic estimates of diaspora sizes, data from 2022 was also incorporated. Further details on the construction of the diaspora data can be found in Section \ref{diaspora_overview}. The dataset includes various personal characteristics of each migrant, such as:
}
%%% TABLE: Variables
{
\begin{table}[H]
\centering
\renewcommand{\arraystretch}{1.25} 
\begin{tabular}{| c | p{0.75\textwidth} |} 
 \hline
 \textbf{Variable} & \textbf{Description} \\ \hline\hline
 \textit{person\_id} & Unique identifier assigned to each migrant; allows tracing across entries. \\ \hline
 \textit{range} & Date range for the respective data entry. The range may be open-ended if the entry is ongoing (i.e., the end date is missing). \\ \hline
 \textit{is\_current} & Boolean variable indicating whether the entry is considered current (\texttt{TRUE} or \texttt{FALSE}). An entry can still be marked as \texttt{TRUE} even if it has a closed date range, for instance, if the individual has died or left the country. \\ \hline
 \textit{citizenship} & ISO2c country code of citizenship. Can take the value \texttt{STATELESS}, for unknown citizenship, or contain two codes in cases of dual citizenship. \\ \hline
 \textit{year\_of\_birth} & Year of birth of the individual, recorded as an integer. \\ \hline
 \textit{year\_of\_death} & Year of death of the individual, recorded as an integer. If the individual is not deceased, the value is set to \texttt{-1}. \\ \hline
 \textit{municipality\_code} & Austrian municipality code of residence, represented as an integer. \\ \hline
\end{tabular}
\vspace{5pt}
\caption{Overview of individual-level variables in the dataset.}
\label{tab:data_variables}
\end{table}
}
\subsection{Variable Construction Procedures and Data Filtering}
%%% Construction of age (defined as year of observation - year of birth) and singular citizenship (defined as the first entered citizenship; multiple citizenships are possible and ordered alphabetically)
{
As part of the data preparation procedure, additional variables were introduced, namely \textit{age} and \textit{singular\_citizenship}, and entries were filtered where the data did not meet reasonable criteria (see Table \ref{tab:variable_construction}). In particular, individuals recorded as age 0 or 1 were excluded, as only the year of birth is available. This means that those born late in one year and registered in the next may be incorrectly categorised as age $1$ instead of $0$, due to the absence of exact birth dates.
}
%%% TABLE: Data filtering
{
% TABLE: Number of data entries -> double citizenship -> STATELESS -> TOTAL NUMBER OF INDIVIDUALS
\begin{table}[H]
\centering
\renewcommand{\arraystretch}{1.25} 

\begin{tabular}{| c | p{0.5\textwidth} | p{0.2\textwidth} |}
 \hline
 \textbf{Variable} & \textbf{Description} & \textbf{Number} \\ \hline
 \multicolumn{2}{|c|}{Total number of individuals} & 1,865,211 \\ \hline \hline
 \textit{age} & \multicolumn{2}{| p{0.7\textwidth} |}{Constructed as the difference between the year of the \textit{update\_first\_date} variable and the individual's recorded \textit{year\_of\_birth}.} \\ \hline
 \multicolumn{2}{|c|}{Removed due to \textit{age} equal to $0$} & 41,206 \\ \hline
 \multicolumn{2}{|c|}{Removed due to \textit{age} equal to $1$} & 25,586 \\
 \hline \hline
 \textit{singular\_citizenship} & \multicolumn{2}{| p{0.7\textwidth} |}{Derived from the \textit{citizenship} variable. In cases of multiple citizenships, the first citizenship in alphabetical order is selected.} \\ \hline
 \multicolumn{2}{|c|}{Individuals with simplified double citizenship} & 72 \\ \hline
 \multicolumn{2}{|c|}{Removed due to \textit{singular\_citizenship} being \texttt{STATELESS}} & 10,571 \\ \hline \hline
 \multicolumn{2}{|c|}{\textbf{Total number of individuals after data filtering}} & \textbf{1,787,848} \\ \hline
\multicolumn{2}{|c|}{Total number of data entries after filtering} & 2,370,607 \\
 \hline
\end{tabular}
\vspace{5pt}
\caption{Overview of construction of \textit{age} and \textit{singular\_citizenship} and data filtering.}
\label{tab:variable_construction}
\end{table}
}
%%%  Arrivals and exits of individuals are designed as follows:
{
To determine individual migrant arrivals and departures on specific days, the following definitions were applied:
\begin{table}[H]
\centering
\renewcommand{\arraystretch}{1.25} 

\begin{tabular}{| c | p{0.7\textwidth} |}
\hline
 & \textbf{Description} \\ \hline \hline
 \textbf{Arrivals} & \textit{update\_first\_date} of the initial recorded entry of a specific \textit{person\_id} within the period from 1 January 2023 to 31 December 2024. Instances where a migrant exits and subsequently re-enters the country, provided both events occur within the observed time frame, are not treated as distinct occurrences. Only the first arrival is considered. \\ \hline
 \textbf{Exits} & \textit{update\_last\_date} of the most recent recorded entry of a specific \textit{person\_id} within the period from 1 January 2023 to 31 December 2024. Additionally, \textit{is\_current} must be set to \texttt{TRUE}, indicating that the migrant's departure has been officially recorded within the administrative data. \\ \hline
\end{tabular}
\vspace{5pt}
\caption{Overview of the construction of arrival and/or departure dates for individual migrants.}
\label{tab:definition_arrivals_exits}

\end{table}
}
\subsubsection{Size of initial diaspora} \label{diaspora_overview}
%%% Size of Diaspora
{
To estimate the size of the diaspora at the beginning of 2023, we utilise the first day of observation (22 November 2022) to identify all migrants registered as current at that time. Newly arrived migrants between 23 November and 31 December 2022 are added, while those who exited Austria or died during the same period are subtracted.
}
%%% TABLE: Diaspora construction
{
% TABLE: as of 22.11.2022 -> Arrivals -> Exits -> Diaspora 
\begin{table}[H]
\centering
\renewcommand{\arraystretch}{1.25} 
\begin{tabular}{| p{0.7\textwidth} | c |}
\hline
\textbf{Registered migrants on the 26 November 2022} & \textbf{1,421,999} \\ \hline \hline
Arrivals between the 27 November 2022 and 31 December 2022 & 19,774 \\ \hline
Exits \& Deaths between the 27 November 2022 and 31 December 2022 & 7,943 \\ \hline \hline
\textbf{Diaspora size on 1 January 2023} & \textbf{1,433,830} \\ \hline
\end{tabular}
\vspace{5pt}
\caption{Construction of the diaspora size.}
\label{tab:construction_diaspora}
\end{table}
}

\subsection{Data Limitations}
{
A key limitation of the dataset concerns the undercounting of exits, which has been shown to be an issue in Austrian administrative data \cite{danko_assessing_2024}. Exit events are approximated using the dropout rate from the registration records, which depend on individuals formally notifying the authorities when they leave the country. However, not all migrants follow this procedure, particularly citizens of European Union member states who face minimal administrative barriers to movement and may not consider deregistration necessary. As a result, the data may overestimate the duration of stay and stability of specific migrant populations. At the same time, the observed stability of exit flows could also reflect active administrative practices, such as regular follow-up procedures or systematic data cleaning, which help identify and remove individuals who are no longer present. These factors create uncertainty in estimating actual outflows and the long-term presence of diaspora groups.
}

\section{Model Description} \label{appendix:model}
\subsection{The adapted Diaspora Model of Migration}
%%% Following Prieto et al. (2024) ... 
{
%%% Following the framework by Prieto et al. (2024) the model consists of two distinct parts examined in the following
We analyse two distinct components. The migration intensity is the expected number of daily arrivals \cite{prieto-curiel_diaspora_2024}. It is quantified using a Poisson fitting procedure, enabling the modelling of count-based migration events over time.
%%% Second, assessing assortativity of arriving and exiting migrants through a multinominal distribution.
Second, the assortative patterns of arrival and exit among migrants are evaluated using a multinomial distribution, enabling analysis of group-specific migration dynamics.
%%% Novelty of the combination of arrivals and exits into a singular framework
A key innovation of this approach is the integration of arrivals and exits into a unified analytical framework, enabling a more comprehensive understanding of migratory flows.
}
\subsubsection{Modelling migration flows}
%%% We describe migration as...
{
Let \(\lambda_i\) represent the daily arrival rate of migrants of the subpopulation \(i\) to a receiving country, and let \(\mu_i\)  represent the daily emigration rate. We model both the number of daily arrivals and daily exits using a Poisson distribution such that the cumulative arrivals \(A_i(t)\) and the cumulative exits \(E_i(t)\) until \(t\) days are independent random variables, characterised by the distributions
\[A_i(t) \sim \text{Pois}(\lambda_i t), \quad \text{and} \quad E_i(t) \sim \text{Pois}(\mu_i t).\]
Exits are assumed to follow as a binomial process, where each migrant from diaspora $D_i$ has a small independent probability $r_i$ to exit from the host country (Appendix E). A binomial process conditional on a Poisson-distributed count also follows a Poisson distribution, so the number of daily exits can be expressed as$E_i(t) \sim \text{Pois}(\mu_i t)$.
}
%%% Combine Poisson models to Skellam ...
%%% RPC please ask me about this definition: Therefore, given that the daily rate arrivals exits, we can describe net migration as Skellam distribution with net migration is given by ...
{
Under these assumptions, the net cumulative migration up to day $t$, expressed as $N_i(t) := A_i(t) - E_i(t)$, follows a Skellam distribution with
\[\kappa_i \stackrel{\text{def}}{=} \lambda_i - \mu_i.\]
}
%%% Therefore, cumulative migration described as ...
{
Then, the cumulative net migration \(N_i(t)\) of individuals of the subpopulation \(i\) is following a Skellam distribution, such that:
\[M_i \sim \text{Skellam}(\lambda_it,\mu_it) \quad \text{with $t \in \mathbb{R}_0^+$}.\]
}
%%% ... with expected migration for days t
{
The expected value of the Skellam distribution is given by the difference of the means of the two underlying distributions. Hence, the expected value for cumulative net migration
\[\mathbb{E}[M_i(t)] = \lambda_it - \mu_it = (\lambda_i - \mu_i)t,\]
represents the average net change in the population of migrants of the subpopulation\(i\) over the period \(t\). If \(\lambda_i > \mu_i\), we expect the population of migrants \(i\) in the receiving country to grow, and, conversely, if \(\lambda_i < \mu_i\), we expect this population to decline.
}
%%% variance of skellam: The variance of a Skellam distribution is XX, meaning that (it has more variance for larger flows but proportionally gets better)
{
The variance of a Skellam distribution is the sum of the variances of its two underlying Poisson distributions:
\[Var[M_i(t)] = \lambda_it  + \mu_it .\]
Thus, the variance of net migration increases with the total number of people moving, $\lambda_it  + \mu_it$. Even when arrivals and exits are expected to be equal ($\lambda_i  = \mu_i$), leading to an expected migration of $\mathbb{E}[M_i(t)]=0$, the variance of net migration remains substantial with $Var[M_i(t)]=2\lambda_it>0$. This indicates that despite no change in the size of the migrant population, the underlying arrival and exit flows can be large. Such high turnover leads to considerable variance, posing challenges purely from the magnitude of movements.
}
%%% Standard deviation of Skellam: proportional to the sqrt of size, therefore, decreasing proportionally to the mean
{
While the variance increases in proportion to the sum of people moving, the standard deviation of the Skellam distribution is:
\[\text{SD}(M_i(t))=\sqrt{\lambda_it  + \mu_it}.\]
Therefore, the standard deviation grows with the square root of the overall migration size, $\sqrt{\lambda_it  + \mu_it}$. As a result, larger migration flows, although exhibiting greater variability in absolute terms, are relatively more predictable when measured against their overall scale. In contrast, smaller flows display higher relative uncertainty.
}
%%% Obtaining parameters by ... 
{
Our method analyses arrivals and exits separately \cite{prieto-curiel_diaspora_2024}. The rate for both distributions, \(A_i(t)\) and \(E_i(t)\) is obtained by defining the error term for day \(j\) as \(e_j(t)=A_i(t)-\lambda_it\). Thus, the sum of squared errors over all observed days \(n\) is given by \(f(\lambda_i) = \sum_{t=0}^{n}e_j(t)^2\). By setting \(f'(\lambda_i) = 0\) we obtain that
\[\lambda_i^* = \tfrac{6\sum_{t=0}^{n}tA_i(t)}{n(n+1)(2n+1)}.\]
Given that \(f(\lambda)\) is a continuous function and the second derivative \(f''(\lambda_i) = n(n+1)(2n+1)/3 >0\), the \(\lambda_i\) minimises the error. This holds for both arrivals \(\lambda_i\) and exits \(\mu_i\), where instead of arrivals \(A_i(t)\) exits \(E_i(t)\) are used for calculations.
}
%%% In order to estimate arrvial and exit rates by diaspora size, we consider that
{
To estimate arrival and exit rates as a function of diaspora size, we assume that the flow of individuals is proportionally related to the size of the respective diaspora group, so that
\[A_i(t) = \rho R_it, \quad \text{and} \quad E_i(t) = \psi R_i t ,\]
where \(R_i\) denotes the size of diaspora of migrants of the subpopulation \(i\), \(\rho > 0\) is a fixed pull rate, and \(\psi > 0\) is a fixed push rate for all migrant subpopulations.
}
%%% Thus, daily arrivals and exits by
{
Therefore, for a given diaspora size from country of origin \(i\), denoted \(R_i\), we assume a fixed pull rate \(\rho\) and a fixed push rate \(\psi\), which respectively characterize the rates of arrivals to and exits from the destination country of international migrants holding citizenship from country \(i\).
}
%%% Values for rho and psi are obtained by minimising the errors such that the error e = Ai(t) - rho Ri t and e = Ei(t) - psi Ri t.
{
To determine both migration rates, we employ the same approach as outlined above, minimising the errors so that \(e_{A,i} = A_i -\rho R_it\) and \(e_{E,i} = E_i -\psi R_it\).
}
%%% Thus, we obtain that ...
{
Therefore, the sum of squared errors over the observed days is given by \(g(\rho) = \sum_i e_{A,i}^2\) and \(g(\psi) = \sum_i e_{E,i}^2\). By setting both \(g'=0\), we obtain that \[\rho^*=\tfrac{\sum_{i=1}^\mu A_iR_i}{t\sum_{i=1}^{\mu}R_i^2}, \quad \text{and} \quad \psi^*=\tfrac{\sum_{i=1}^\mu E_iR_i}{t\sum_{i=1}^{\mu}R_i^2}.\]
}
%%% Second derivative > 0, therefore, minimising the error
{
For both \(g\) the second derivates is given by \(2\sum_{i}R_i^2t^2>0\), so the squared error minimises the value for both \(\rho^*\) and \(\psi^*\).
}
%%% For Austria we obtain the pull rate 3.56 * 10^-4 and the push rate 1.76 * 10^-4
{
In the case of Austria, we obtain for the pull-rate \(\rho_{Aus}=3.56 \times 10^{-4}\) and for the push-rate \(\psi_{Aus}=1.76 \times 10^{-4}\) for all subpopulations. In practical terms, this means that for a diaspora of 10,000 individuals, approximately 3.56 people are expected to migrate to Austria each day, while about 1.76 leave. When we express the estimated push and pull rates for subpopulation \(i\) by diaspora size we obtain \[\lambda_i = \rho R_i, \quad \text{and} \quad \gamma_i = \psi R_i.\]
}
%%% Over- and underestimation by arrivals and exits and examples
%%% Ola: I remember an equation we had on change in a diaspora size in time R' = R (1+ lamda) or something like that... now this should change if you add exits to it, making the increase in diaspora size caped at times, dimensions or increasing .. I know you showed in figure 2 that it should only increases but might be worth writing it down here. 
{
}
\subsubsection{Modelling spatial dispersion}
%%% Assortativity defined as probability of choosing destination
{
We define the spatial dispersion of migration flows at the sub-national level as the distribution of destinations within the country for incoming international migrants, as well as the locations of last registration prior to departure. In the first step, we measure the volume of arrivals and exits, denoted as \(A_i(t) = a\) and \(E_i(t)=e\), respectively.  The spatial allocation of these flows is assumed to depend on the size of the diaspora of subpopulation \(i\) in area \(i\), with the probability of settlement or last residence modelled as:
\[\pi_{ij}=\tfrac{R_{ij}}{R_i},\]
where \(R_{ij}\) is the diaspora size of subpopulation \(i\) in area \(j\). Consequently, larger diaspora concentrations tend to attract more incoming migrants and generate higher volumes of migrant exits from the country.
}
%%% Thus, destination and departure point alike as a multinominal draw
{
The process of subnational arrival and exit flows is then modelled as a multinomial distribution following:
\[A_{ij} | A_i(t) \sim \text{Mult}(a, \bar{\pi}_i), \quad \text{and} \quad E_{ij} | E_i(t) \sim \text{Mult}(e, \bar{\pi}_i),\]
where \(A_{ij}\) are the arrivals and \(E_{ij}\) the exits of subpopulation \(i\) in area \(j\) and \(\bar{\pi}_i = (\pi_{i1}, \pi_{i2}, ..., \pi_{i\nu})\) is a vector of all the probabilities of migration flows occurring in area \(j\) with \(\sum_k\pi_{ik}=1\).
}
%%% Since a multinominal distribution conditional on a Poisson distribution...
{
Since the conditional allocations of arrivals \(A_{ij}|A_i(t)\) and exits \(E_{ij}|E_i(t)\) are multinomial and the totals \(A_i\) and \(E_i\) are Poisson distributions, it can be shown that the multinomial distributions also follow Poisson distributions such that
\[A_{ij}(t) \sim \text{Pois}(\pi_{ij}\lambda t), \quad \text{and} \quad E_{ij}(t) \sim \text{Pois}(\pi_{ij}\gamma t).\]
Through the estimation of arrival and exit flows conditional on diaspora sizes, \(\lambda_i = \rho R_i\) and \(\gamma_i = \psi R_{ij}\), and the description of flow dispersion using \(\pi_{ij}=R_{ij}/R_i\), we get that
\[A_{ij}(t) \sim \text{Pois}(R_{ij}\lambda t), \quad \text{and} \quad E_{ij}(t) \sim \text{Pois}(R_{ij}\gamma t).\]
Therefore, the estimate for arrivals and exits for \(t\) time periods is given by the diaspora size \(R_{ij}\) of subpopulation \(i\) in area \(j\),
}
\section{Poisson Parameters} \label{appendix:parameter}
Based on the Poisson-Skellam framework outlined in Appendix \ref{appendix:model}, we estimate daily arrival $\lambda_i$ and exit $\mu_i$ rates for each migrant subgroup. This section presents the derived parameters in two parts: first, disaggregated by citizenship, and second, by age cohort. For each group, we include a table listing the estimated rates and accompanying plots comparing cumulative observed arrivals and exits with the modelled trajectories. Shaded bands in the plots indicate the 99\% confidence intervals derived from the model, allowing visual assessment of uncertainty and model fit over time.
\subsection{By Citizenships} \label{appendix:parameter_citizenship}
%%% Analysis of the 25 top diaspora
{
The analysis focuses on the 25 largest citizenship groups to ensure statistical robustness, minimise noise from small sample sizes, and support meaningful evaluation of migration flows. These groups represent 89\% of the total diaspora population, 84\% of arrivals, and 79\% of exits, underscoring their central role in recent migration patterns (see Table \ref{tab:top_25_diaspora}).
}
%%% TABLE Top 25 Diasporas
{
\begin{table}[H]
\centering
\renewcommand{\arraystretch}{1.25} 
\begin{tabular}{| p{0.4\textwidth} | c | c | c |}
\hline
\textbf{Citizenship} & \textbf{Diaspora Size} & \textbf{Arrivals} & \textbf{Exits} \\ \hline
\textbf{All Citizenships} & \textbf{1,433,830} & \textbf{346,075} & \textbf{174,077} \\ \hline \hline
Serbia & 143,391 & 12,469 & 10,396 \\ \hline
Turkey & 133,985 & 16,388 & 10,311 \\ \hline 
Germany & 114,634 & 35,249 & 12,294 \\ \hline 
Bosnia \& Herzegovina & 107,605 & 9,916 & 6,526 \\ \hline 
Romania & 99,688 & 29,904 & 13,606 \\ \hline 
Croatia & 88,252 & 14,862 & 5,327 \\ \hline 
Syria & 81,404 & 27,495 & 5,992 \\ \hline 
Ukraine & 76,174 & 34,158 & 21,955 \\ \hline 
Hungary & 63,358 & 26,442 & 8,308 \\ \hline 
Afghanistan & 51,424 & 6,908 & 4,203 \\ \hline 
Poland & 41,777 & 8,253 & 4,090 \\ \hline 
Russia & 40,407 & 6,984 & 4,191 \\ \hline 
Slovakia & 31,417 & 8,194 & 3,702 \\ \hline 
Bulgaria & 27,330 & 8,011 & 3,845 \\ \hline 
Kosovo & 26,027 & 4,143 & 2,503 \\ \hline 
North Macedonia & 25,742 & 3,640 & 1,687 \\ \hline 
Italy & 21,091 & 8,408 & 3,070 \\ \hline 
Iran & 17,723 & 5,838 & 3,120 \\ \hline 
Iraq & 14,595 & 933 & 1,868 \\ \hline 
China & 14,425 & 5,948 & 2,523 \\ \hline 
Slovenia & 13,891 & 3,401 & 1,158 \\ \hline 
India & 13,057 & 5,375 & 3,908 \\ \hline 
Somalia & 10,236 & 1,932 & 985 \\ \hline 
Czechia & 9,721 & 2,643 & 1,256 \\ \hline 
United Kingdom & 8,701 & 1,515 & 1,182 \\ \hline
\hline
\textbf{Top 25 Diasporas} & \textbf{1,276,055} & \textbf{289,009} & \textbf{138,006} \\ \hline
Share & 89.25\% & 83.51\% & 79,29\% \\ \hline
\end{tabular}
\vspace{5pt}
\caption{The sizes of the top 25 diasporas as of 1 January 2023, along with their corresponding arrival and exit counts for 2023 and 2024. Additionally, their combined proportional shares of total diaspora and migration flows.}
\label{tab:top_25_diaspora}
\end{table}
}
\subsubsection{Daily Arrival and Exit Rates by Citizenship}
%%% TABLE Arrival & Exit Rates
{
\begin{table}[H]
\centering
\renewcommand{\arraystretch}{1.25} 
\begin{tabular}{| p{0.35\textwidth} | c | c | c |}
\hline
\textbf{Citizenship} & \textbf{Daily Arrivals} \boldmath$\lambda_i$ & \textbf{Daily Exits} \boldmath$\mu_i$ & \textbf{Net Migration} \\ \hline \hline
Serbia & 17.82 & 14.34 & 3.47 \\ \hline
Turkey & 24.24 & 13.65 & 10.58 \\ \hline 
Germany & 49.12 & 15.91 & 33.21 \\ \hline 
Bosnia \& Herzegovina & 14.46 & 8.14 & 6.32 \\ \hline 
Romania & 43.95 & 17.75 & 26.20 \\ \hline 
Croatia & 21.74 & 6.88 & 14.86 \\ \hline 
Syria & 41.01 & 8.21 & 32.80 \\ \hline 
Ukraine & 47.69 & 32.23 & 15.46 \\ \hline 
Hungary & 37.55 & 10.64 & 26.91 \\ \hline 
Afghanistan & 10.87 & 5.86 & 5.00 \\ \hline 
Poland & 12.10 & 5.35 & 6.75 \\ \hline 
Russia & 10.46 & 5.79 & 4.66 \\ \hline 
Slovakia & 11.77 & 4.77 & 7.00 \\ \hline 
Bulgaria & 11.69 & 5.01 & 6.67 \\ \hline 
Kosovo & 5.93 & 3.35 & 2.58 \\ \hline 
North Macedonia & 5.51 & 2.23 & 3.28 \\ \hline 
Italy & 11.69 & 4.06 & 7.62 \\ \hline 
Iran & 8.09 & 3.90 & 4.19 \\ \hline 
Iraq & 1.46 & 2.61 & -1.15 \\ \hline 
China & 7.49 & 3.38 & 4.12 \\ \hline 
Slovenia & 4.88 & 1.55 & 3.33 \\ \hline 
India & 7.68 & 5.66 & 2.02 \\ \hline 
Somalia & 3.16 & 1.28 & 1.87 \\ \hline 
Czechia & 3.75 & 1.65 & 2.11 \\ \hline 
United Kingdom & 2.13 & 1.58 & 0.55 \\ \hline
\hline
\textbf{Sum of 25} & \textbf{416.24} & \textbf{185.81} & \textbf{230.44} \\ \hline
\end{tabular}
\vspace{5pt}
\caption{Estimated daily arrivals $\lambda_i$, daily exits $\mu_i$, and daily net migration for the top 25 diasporas as of 1 January 2023, based on migration model estimates.}
\label{tab:daily_flows_top25}
\end{table}
}
%%% PLOTS
{
% First page of figures (15 images)
\renewcommand{\arraystretch}{1.0}
\setlength{\tabcolsep}{4pt}
\begin{longtable}{rrr}
\caption{\textbf{Flow Intensities by Citizenship:} Cumulative number of arrivals (blue solid line) and exits (red solid line). Overlaid dashed lines represent the modelled daily rates, derived from a Poisson process, for both arrivals and exits. The shaded areas surrounding these dashed lines indicate the 99\% confidence intervals of the modelled rates. Ordered by diaspora size.} \label{fig:appendix_citizenship_flows} \\
\endfirsthead

\multicolumn{3}{r}%
{{\bfseries \tablename\ \thetable{} -- continued from previous page}} \\
\\
\endhead

\includegraphics[width=0.31\textwidth]{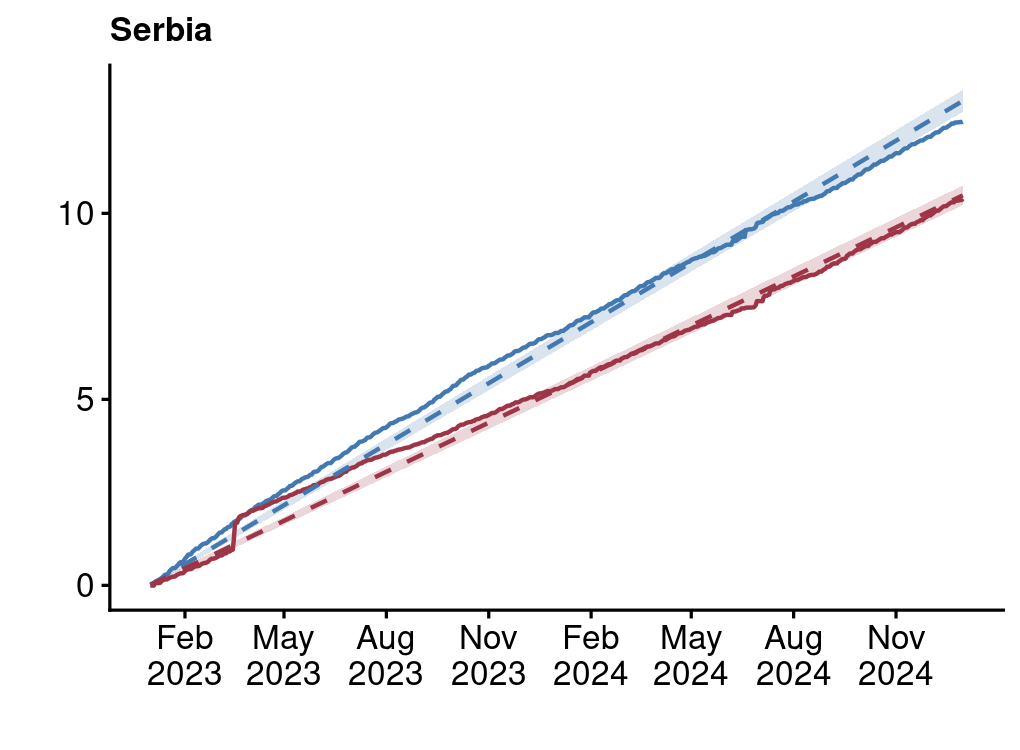} &
\includegraphics[width=0.31\textwidth]{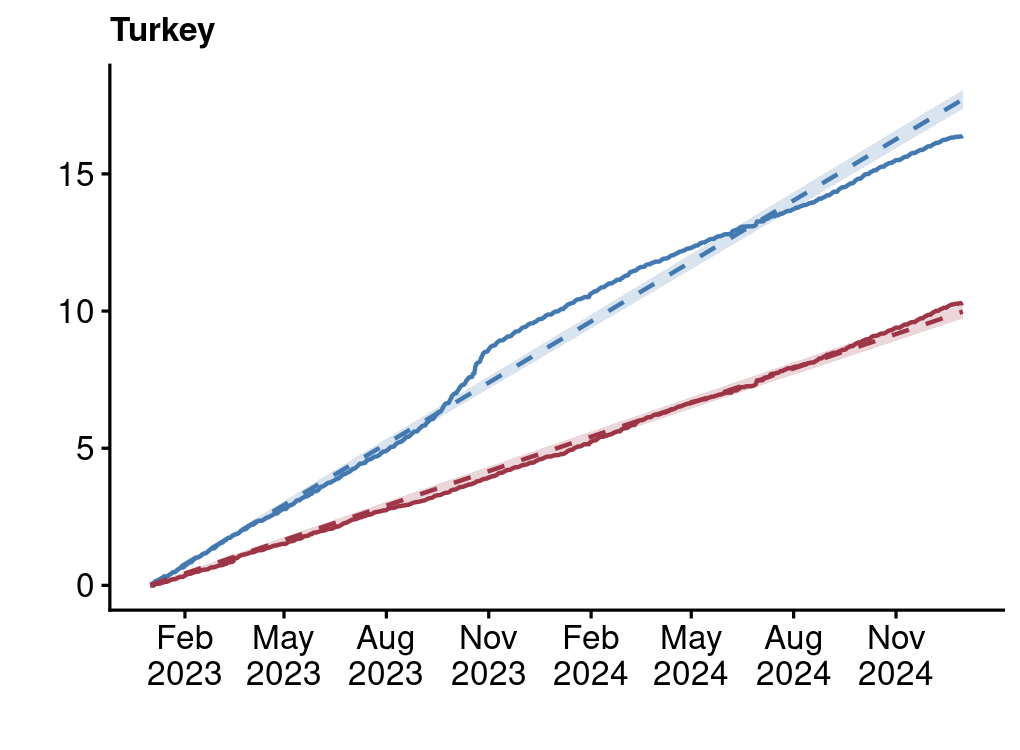} &
\includegraphics[width=0.31\textwidth]{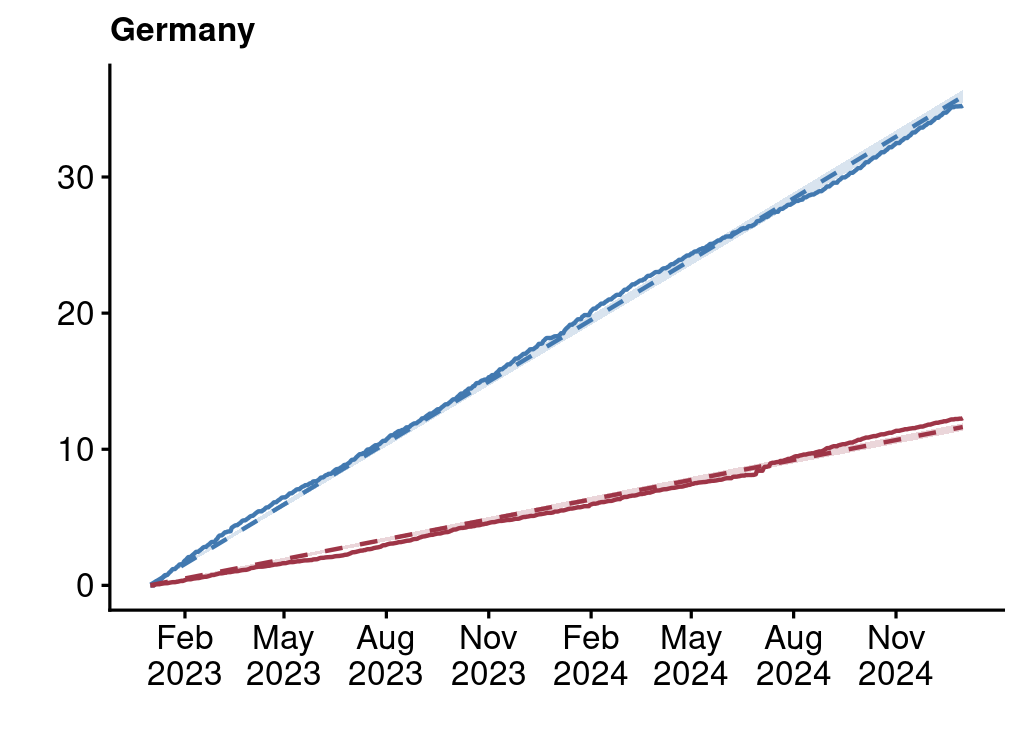} \\

\includegraphics[width=0.31\textwidth]{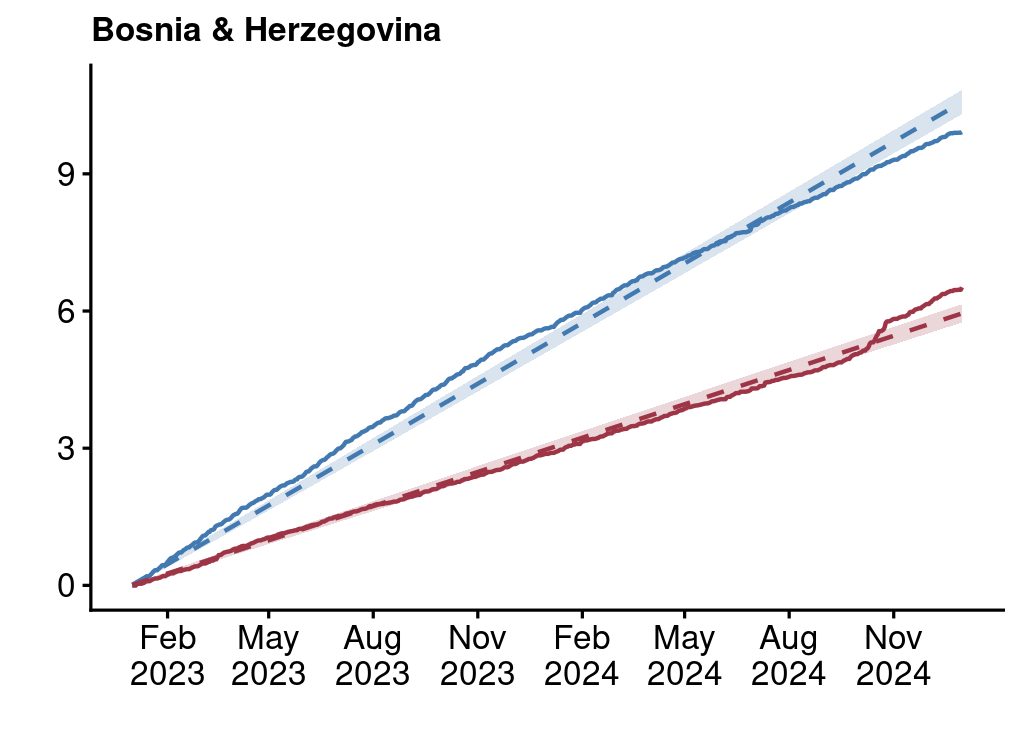} &
\includegraphics[width=0.31\textwidth]{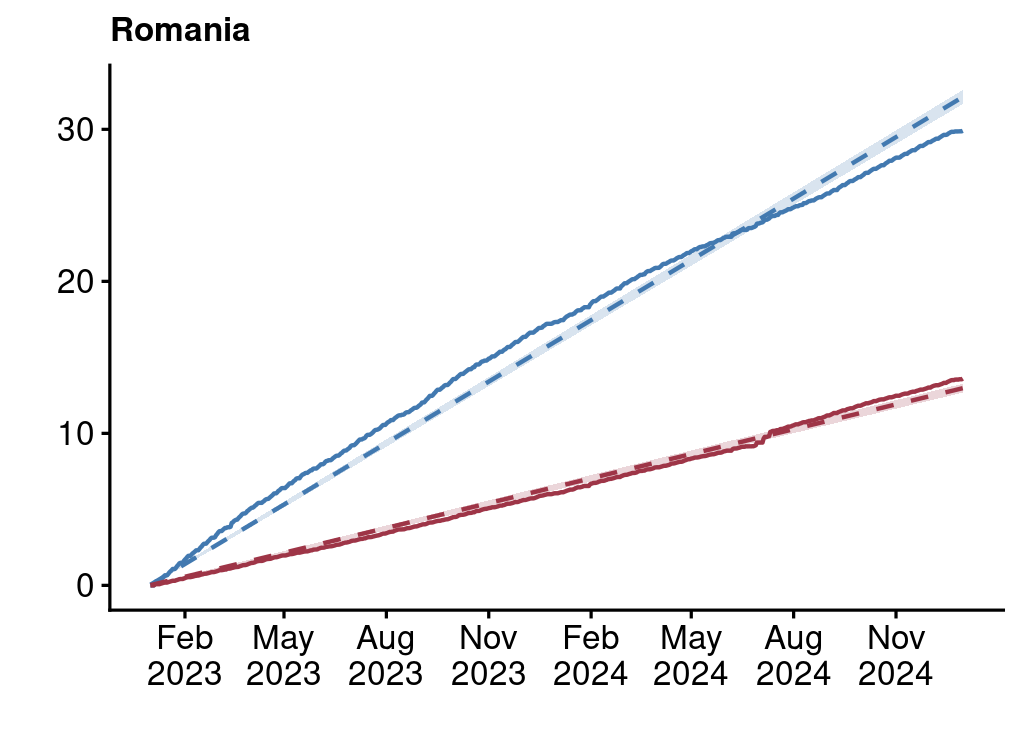} &
\includegraphics[width=0.31\textwidth]{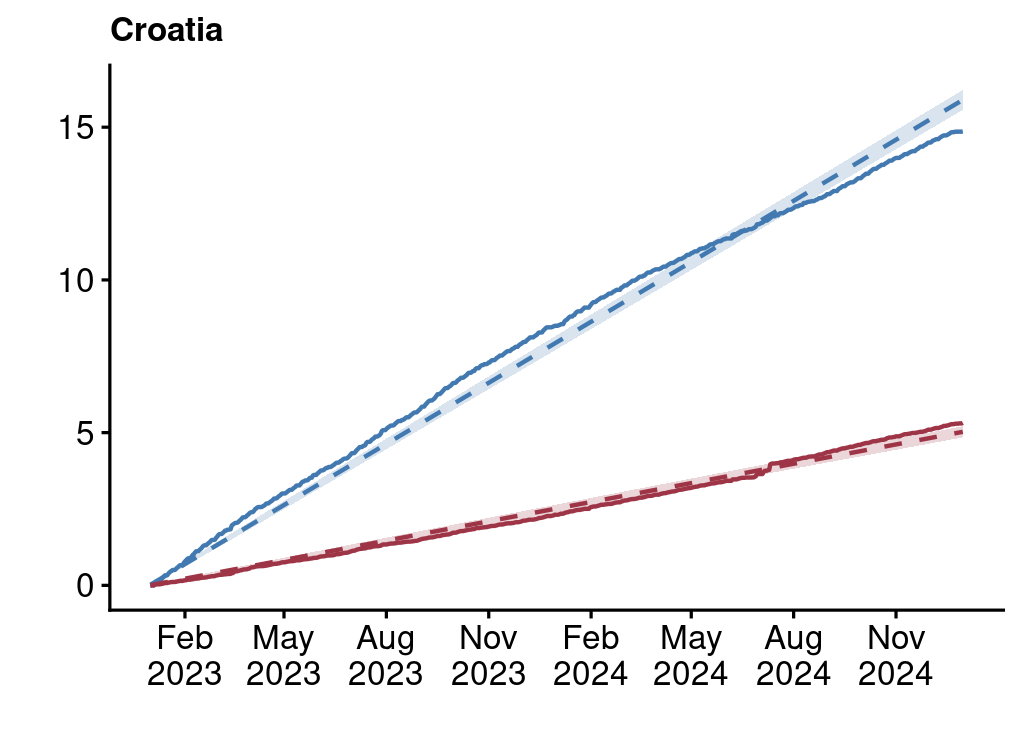} \\

\includegraphics[width=0.31\textwidth]{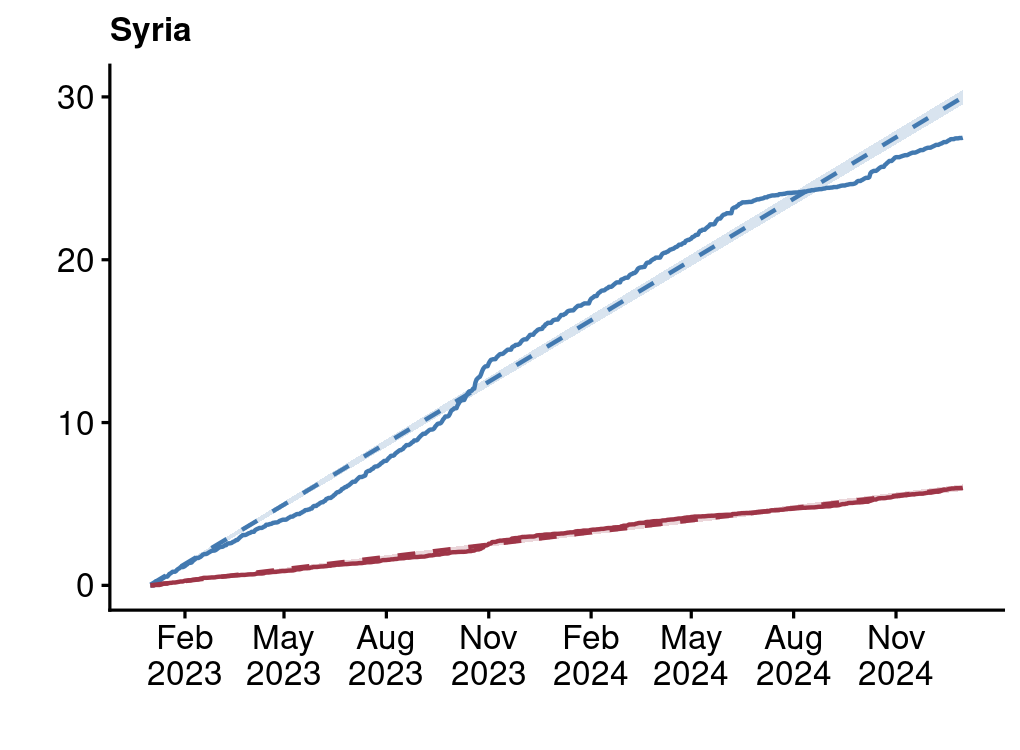} &
\includegraphics[width=0.31\textwidth]{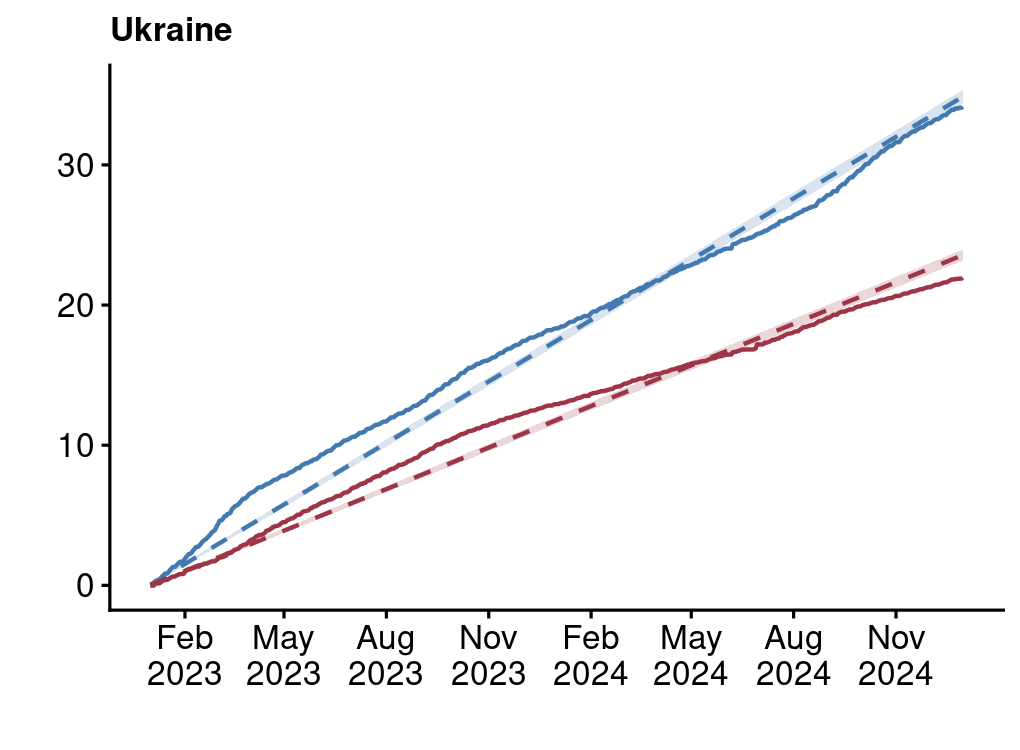} &
\includegraphics[width=0.31\textwidth]{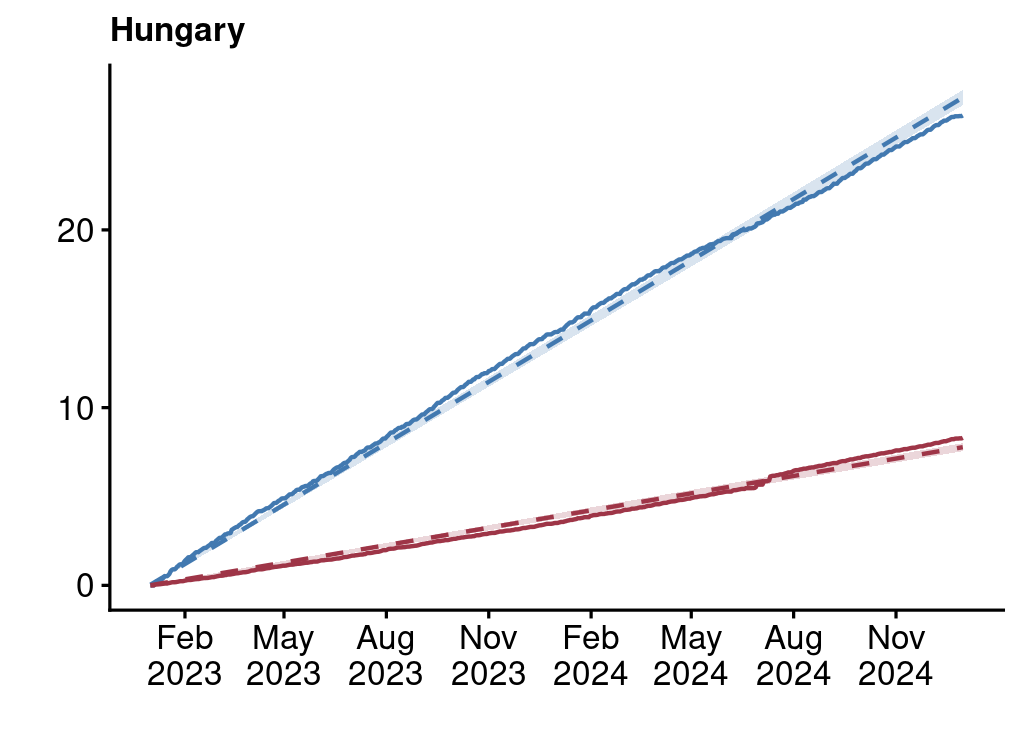} \\

\includegraphics[width=0.31\textwidth]{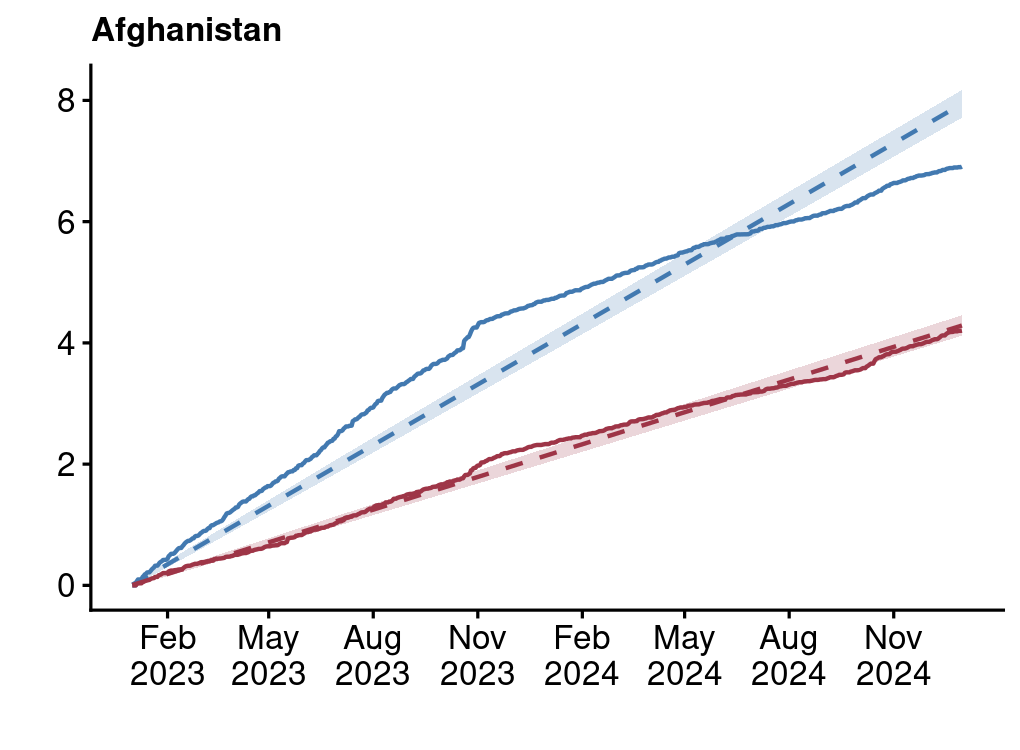} &
\includegraphics[width=0.31\textwidth]{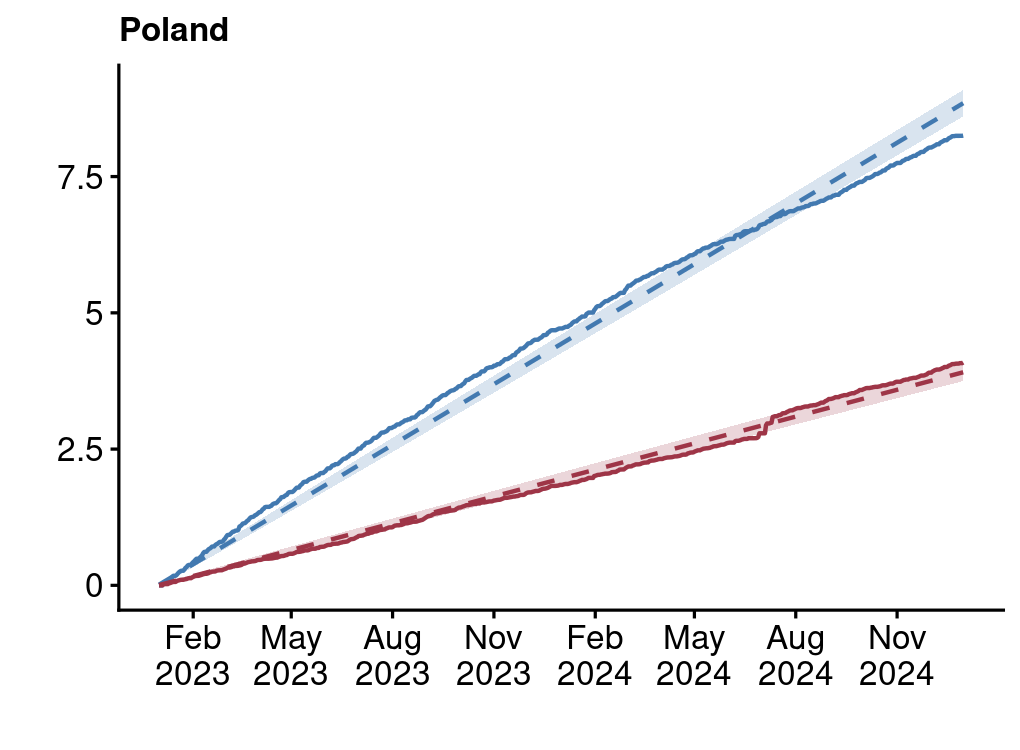} &
\includegraphics[width=0.31\textwidth]{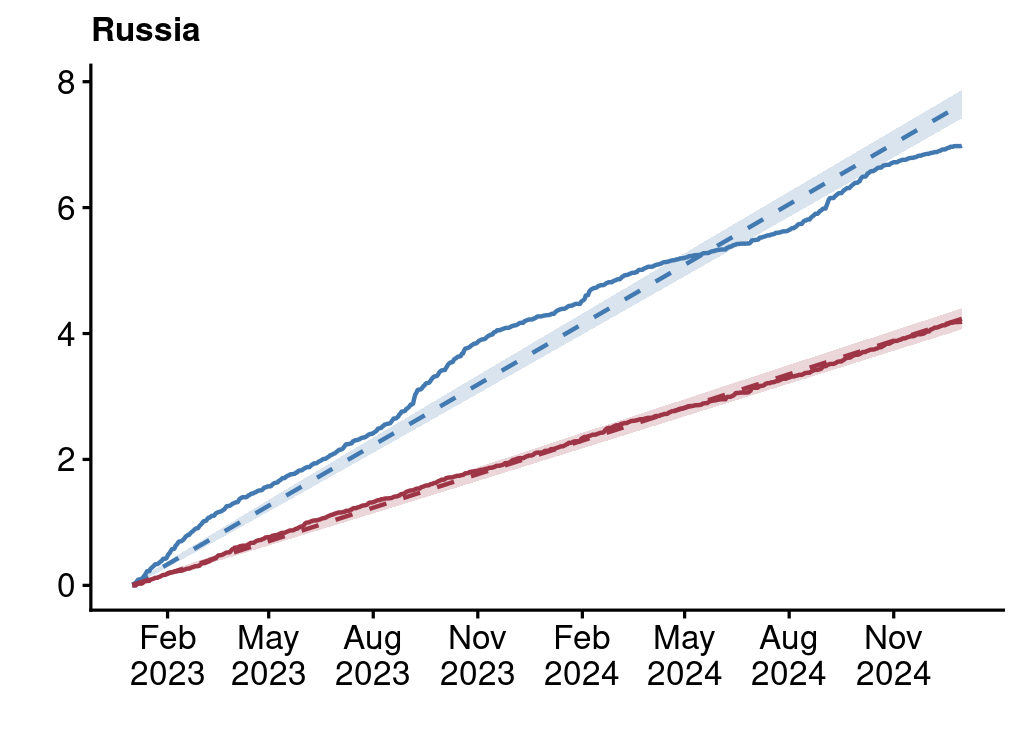} \\

\includegraphics[width=0.31\textwidth]{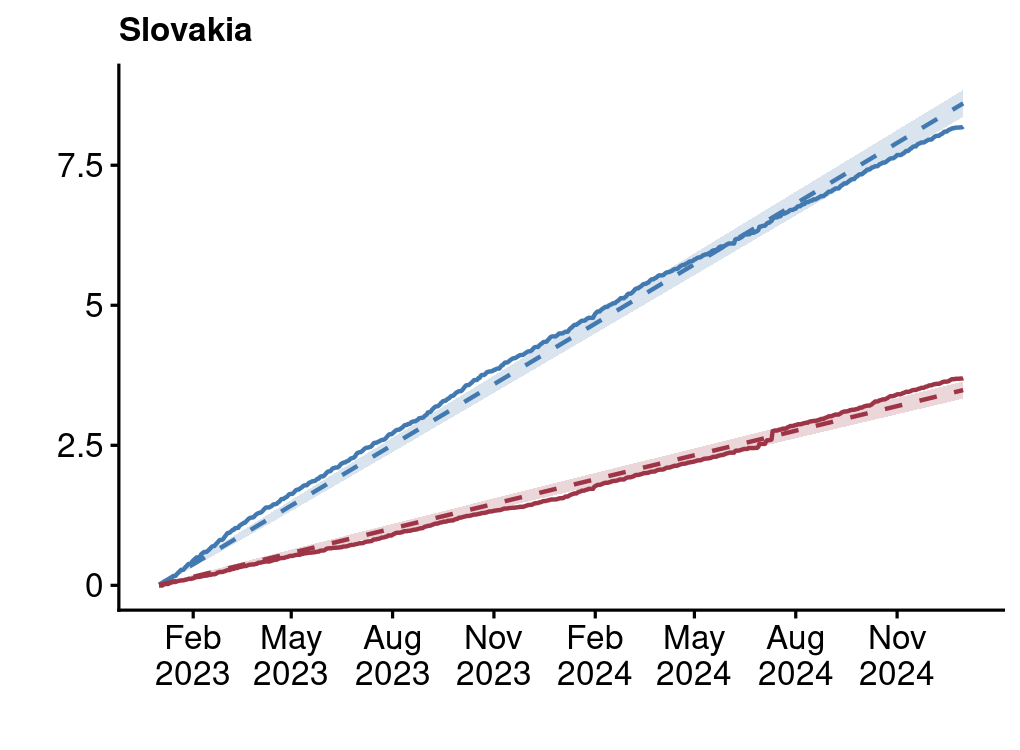} &
\includegraphics[width=0.31\textwidth]{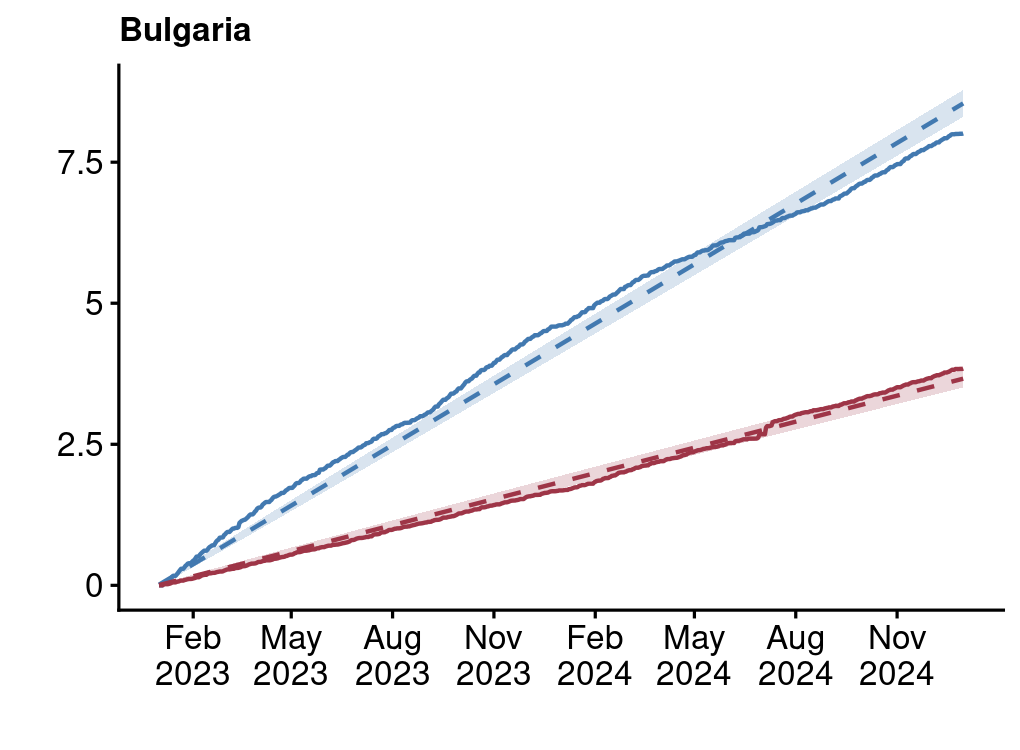} &
\includegraphics[width=0.31\textwidth]{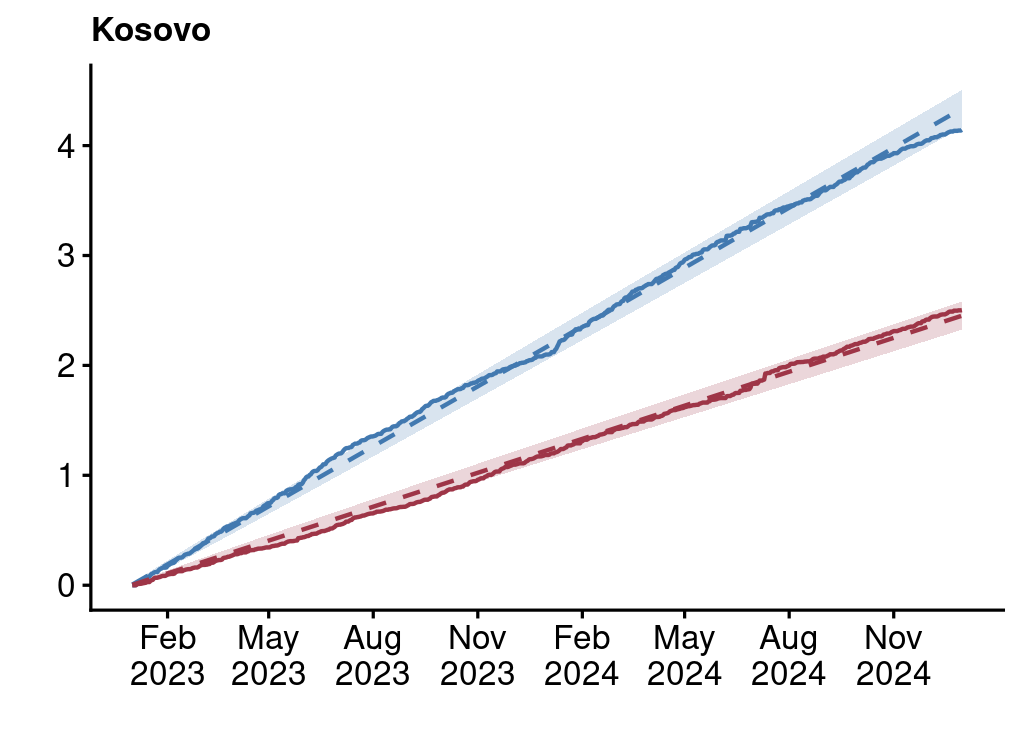} \\

\includegraphics[width=0.31\textwidth]{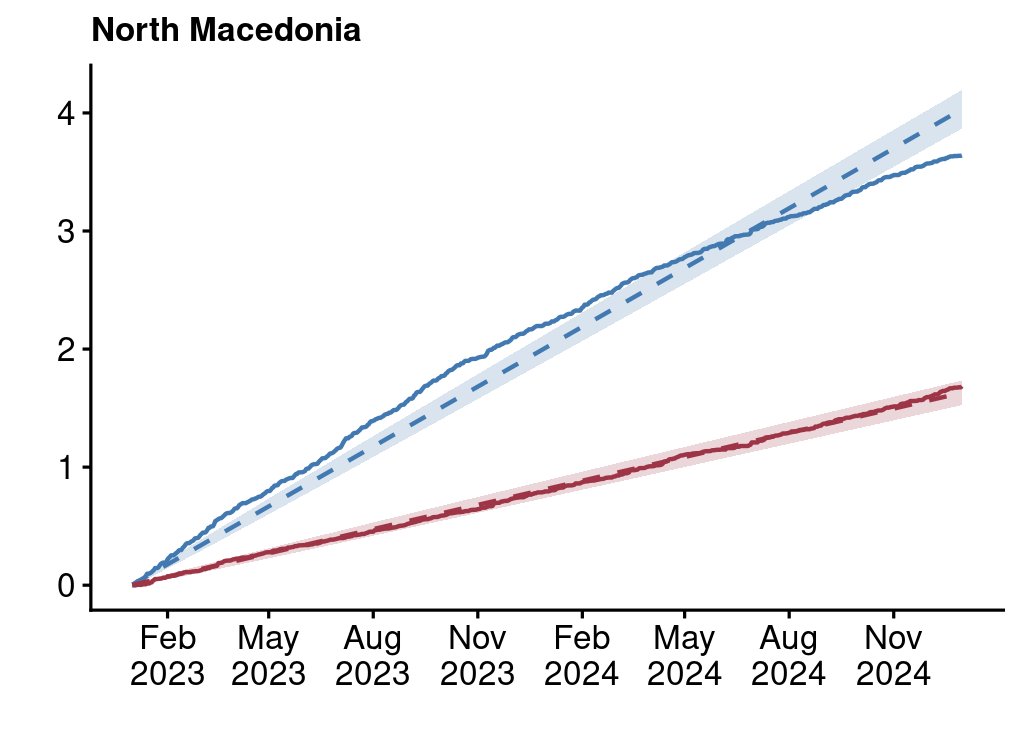} &
\includegraphics[width=0.31\textwidth]{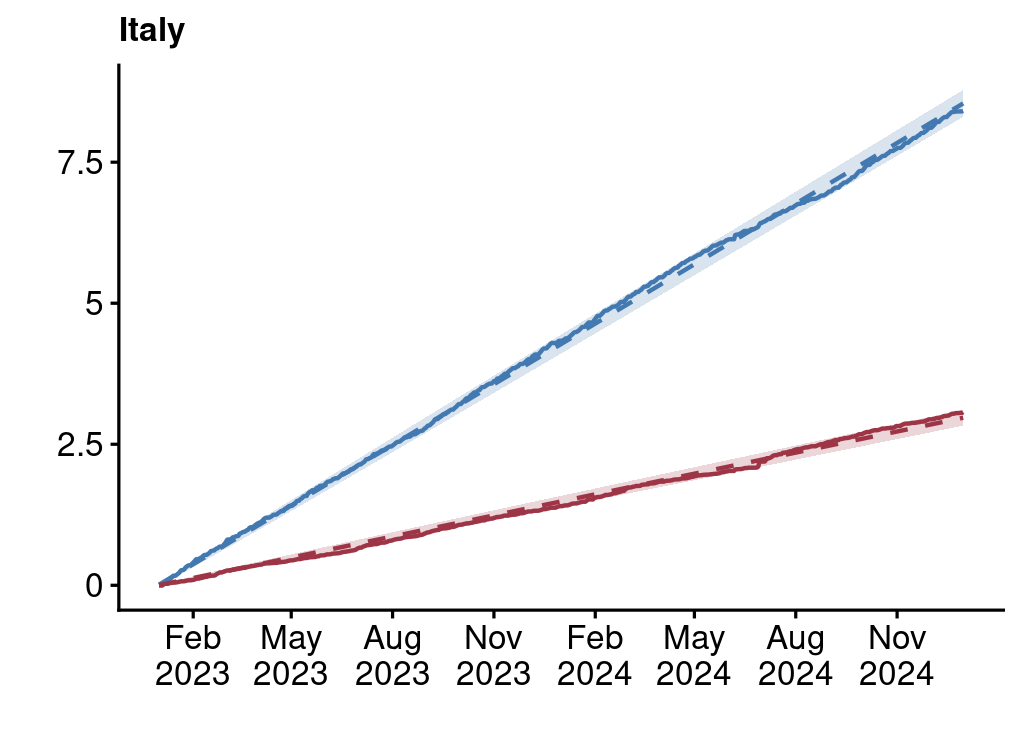} &
\includegraphics[width=0.31\textwidth]{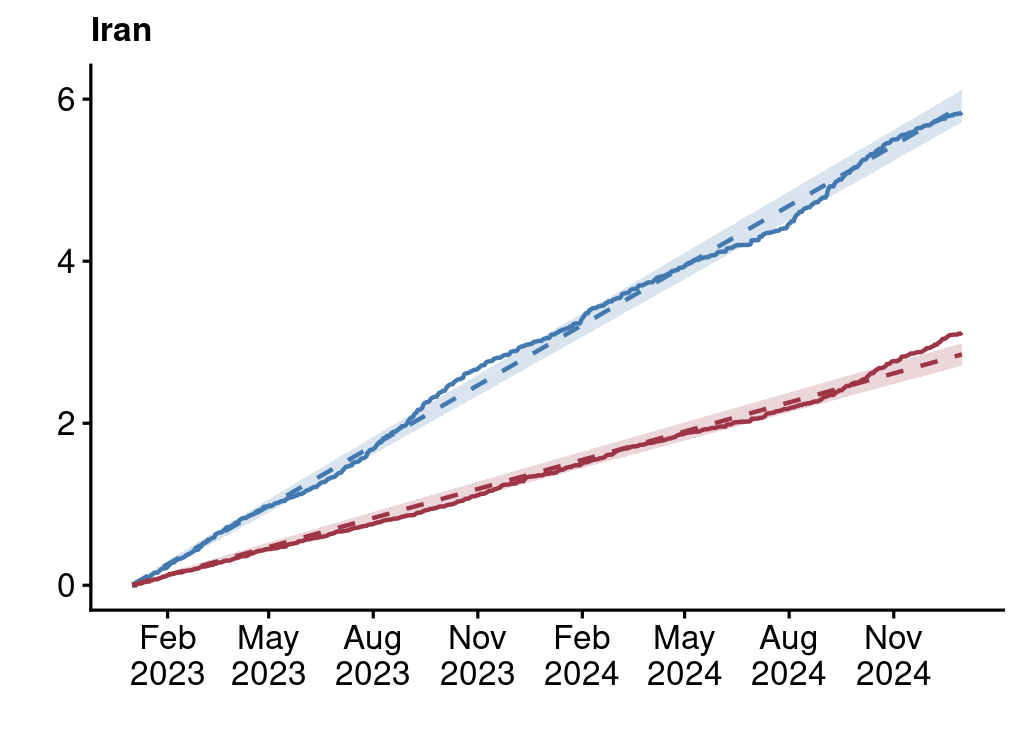} \\

\includegraphics[width=0.31\textwidth]{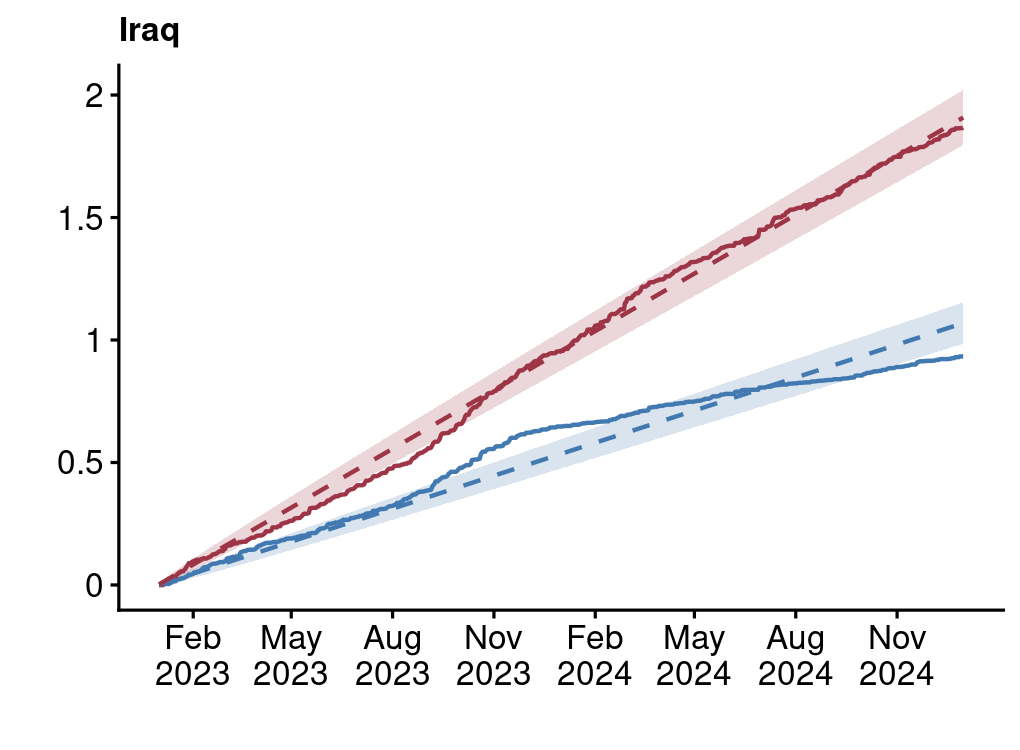} &
\includegraphics[width=0.31\textwidth]{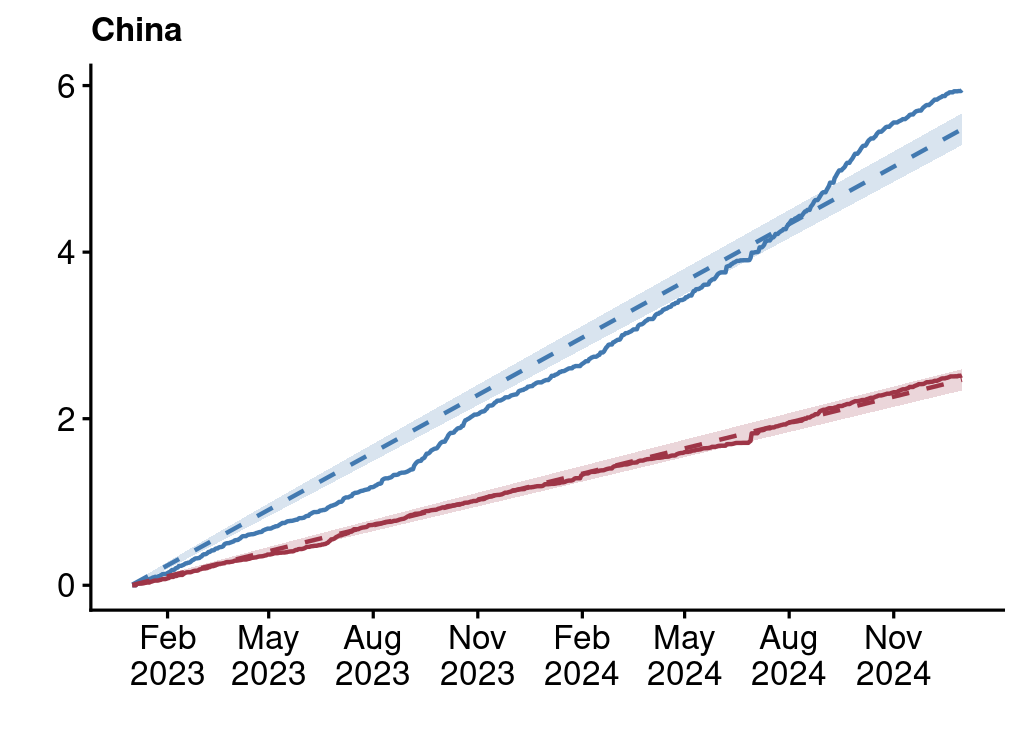} &
\includegraphics[width=0.31\textwidth]{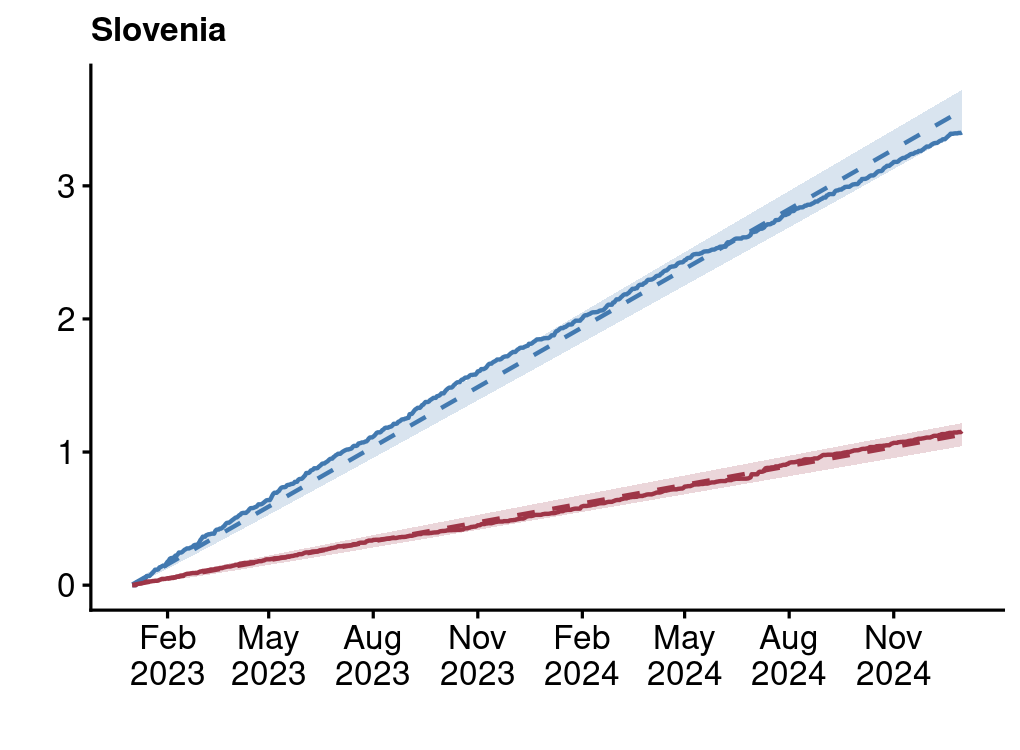} \\

\includegraphics[width=0.31\textwidth]{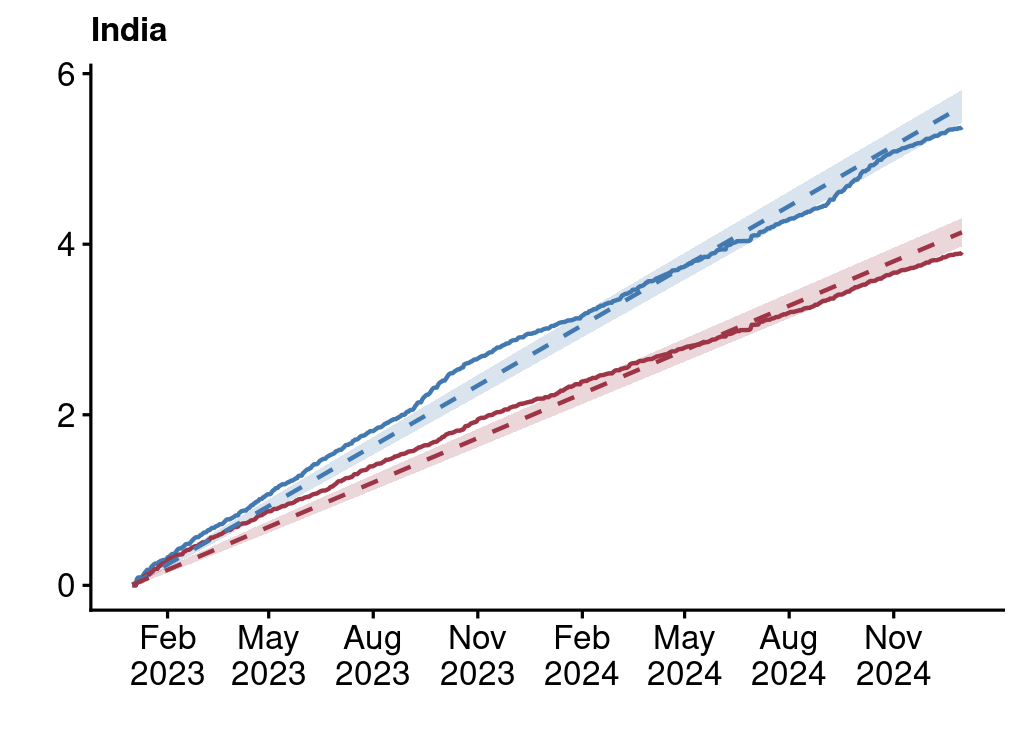} &
\includegraphics[width=0.31\textwidth]{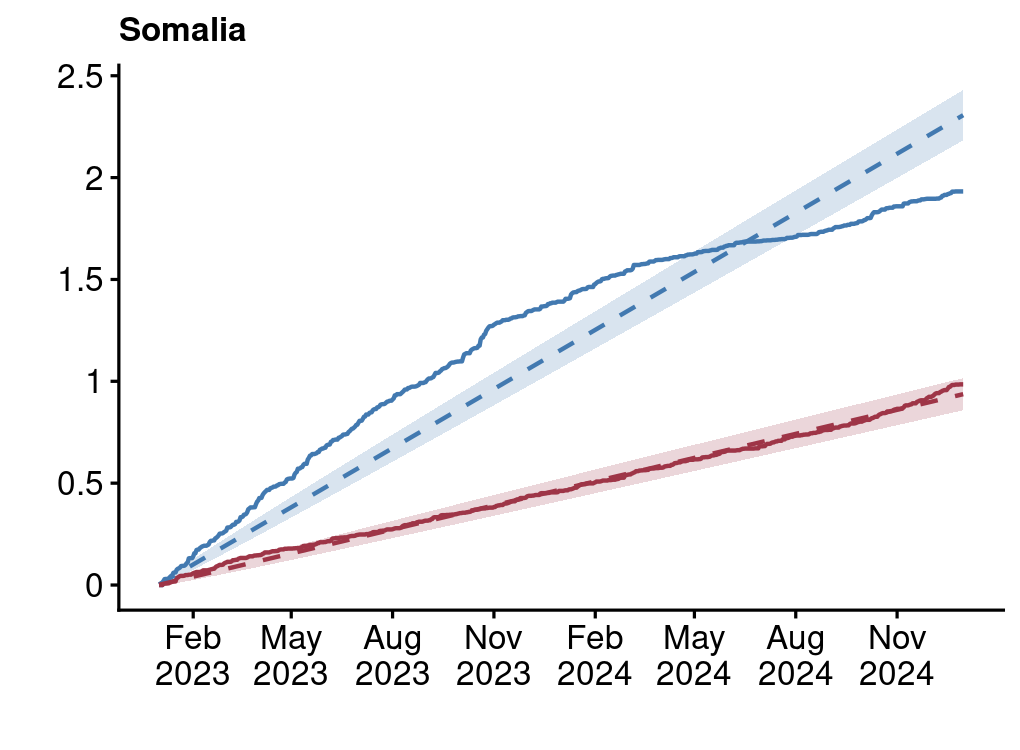} &
\includegraphics[width=0.31\textwidth]{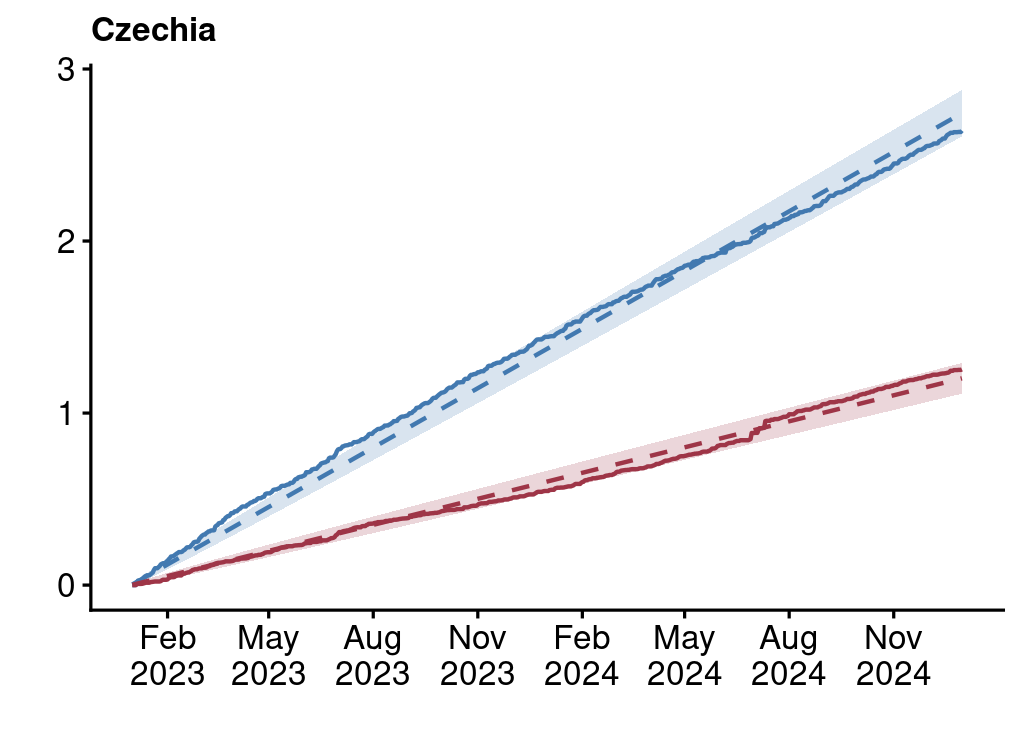} \\

\includegraphics[width=0.31\textwidth]{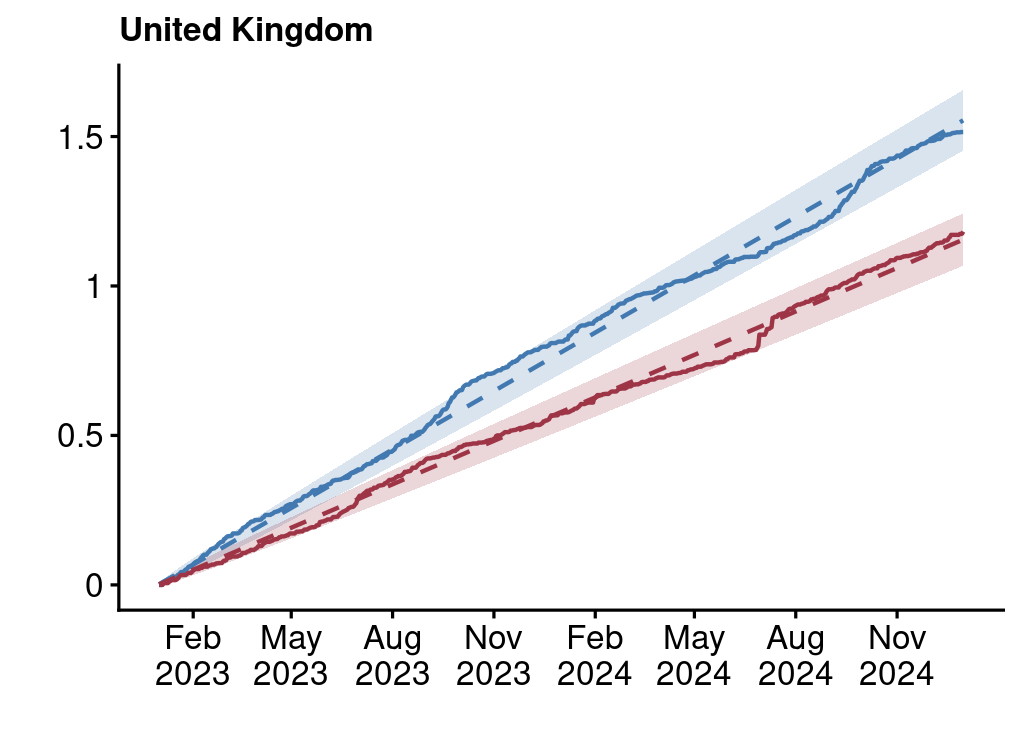} &
& \\
\end{longtable}
}
\subsection{By Age-Cohort} \label{appendix:parameter_age}
%%% Includes all ... age cohorts are...
{
Individual ages have been grouped into five age cohorts: children (ages 2–17), young adults (18–34), middle-aged adults (35–54), older adults (55–69), and seniors (70+). The following tables and plots summarise the estimated arrival and exit rates for each cohort.
}
%%% TABLE
{
\begin{table}[H]
\centering
\renewcommand{\arraystretch}{1.25} 
\begin{tabular}{| p{0.28\textwidth} | c | c | c |}
\hline
\textbf{Age Cohort} & \textbf{Daily Arrivals} \boldmath$\lambda_i$ & \textbf{Daily Exits} \boldmath$\mu_i$ & \textbf{Net Migration} \\ \hline \hline
Children & 84.20 & 30.58 & 53.62 \\ \hline
Young Adults & 222.18 & 97.93 & 124.24 \\ \hline
Middle-Aged Adults & 146.86 & 72.63 & 74.23 \\ \hline
Older Adults & 34.88 & 21.75 & 13.13 \\ \hline
Seniors & 9.15 & 14.17 & -5.03 \\ \hline
\end{tabular}
\vspace{5pt}
\caption{Estimated daily arrival and exit rates, and net migration by age cohort, based on Poisson parameters.}
\label{tab:daily_flows_by_age}
\end{table}
}
%%% PLOTS
{
\begin{figure}[H]
\renewcommand{\arraystretch}{1.0}
\setlength{\tabcolsep}{4pt}
\begin{tabular}{rrr}
\includegraphics[width=0.31\textwidth]{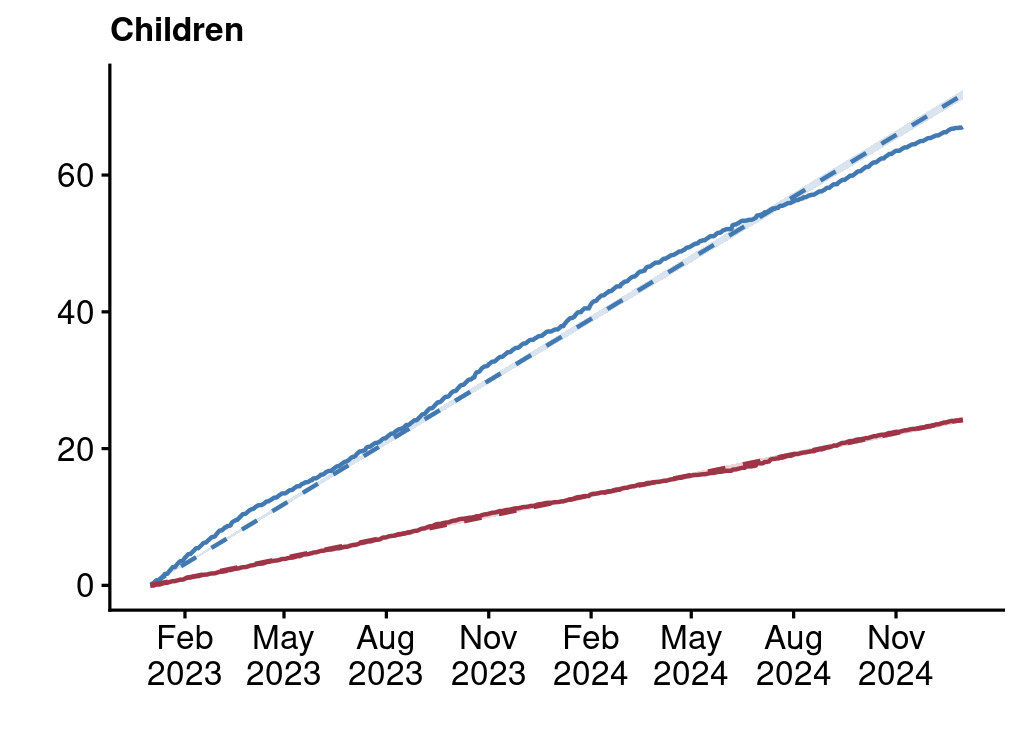} &
\includegraphics[width=0.31\textwidth]{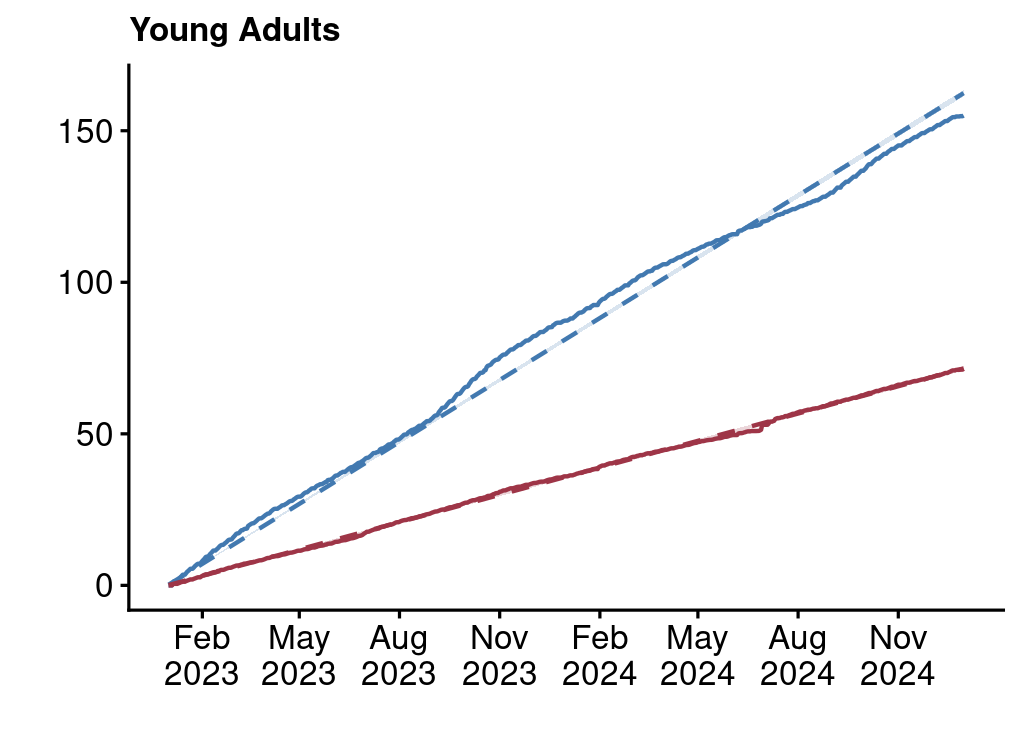} &
\includegraphics[width=0.31\textwidth]{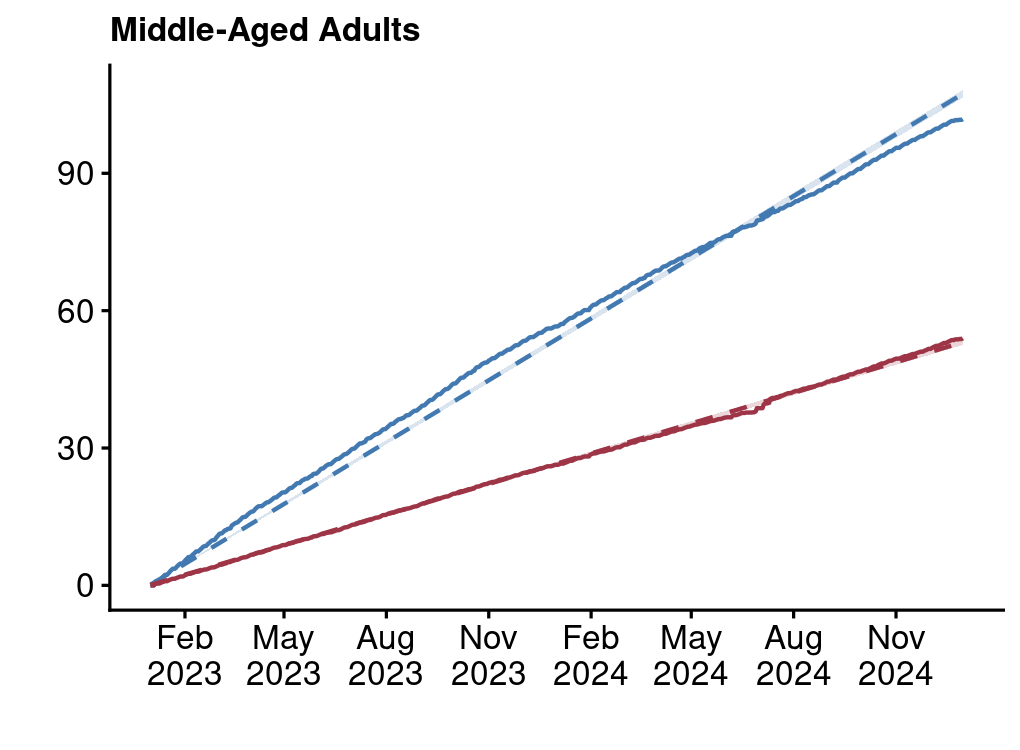} \\
\includegraphics[width=0.31\textwidth]{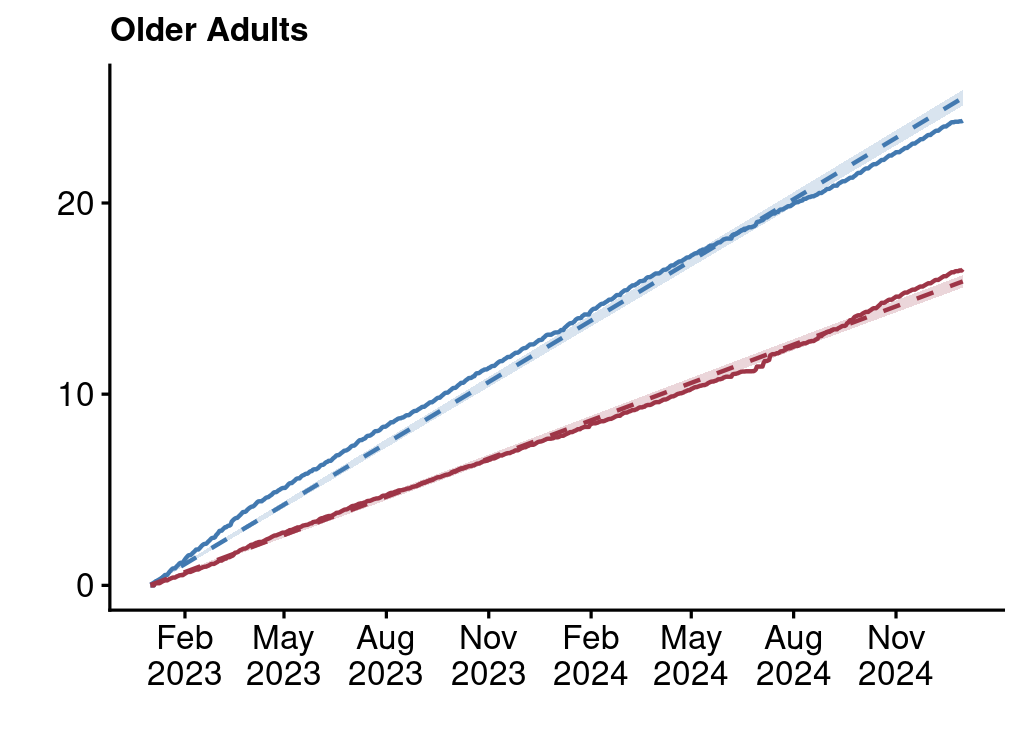} &
\includegraphics[width=0.31\textwidth]{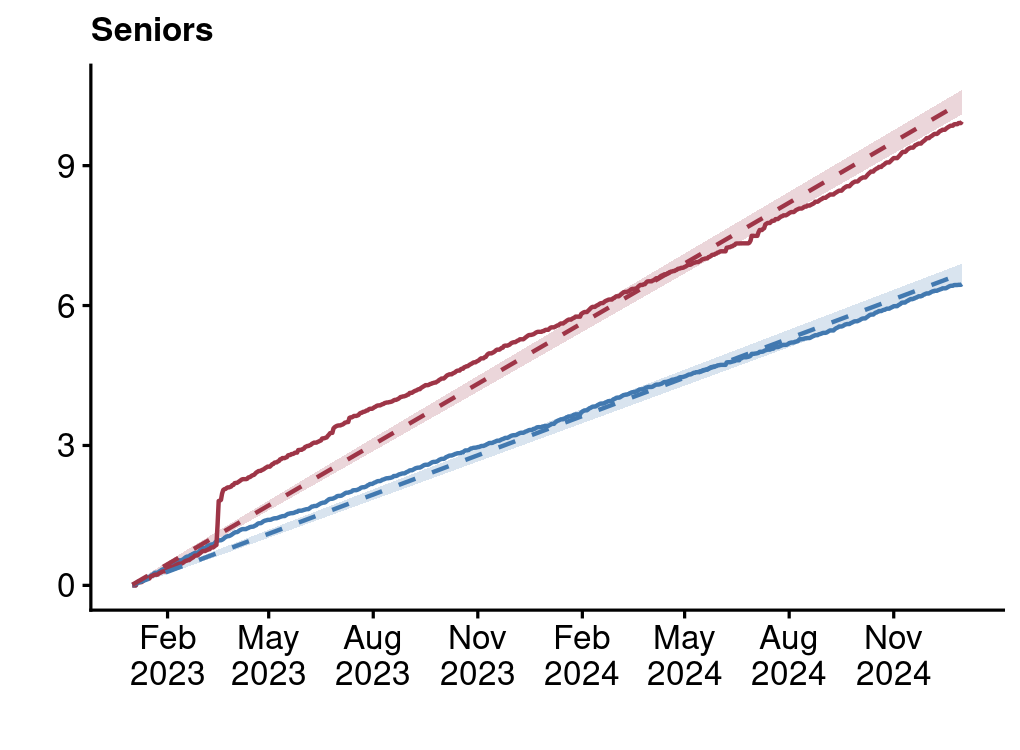} \\
 & \\
\end{tabular}
\caption{\textbf{Flow Intensities by Age-Cohort:} Cumulative number of arrivals (blue solid line) and exits (red solid line). Overlaid dashed lines represent the modelled daily rates, derived from a Poisson process, for both arrivals and exits. The shaded areas surrounding these dashed lines indicate the 99\% confidence intervals of the modelled rates. Ordered by ages of the corresponding age-cohorts: Children (2-17), Young Adults (18-34), Middle-Aged Adults (35-54), Older Adults (55-69), and Seniors (70+).}
\label{fig:appendix_age_flows}
\end{figure}
}

\section{Model Fitness} \label{appendix:fitness}
%%% Description of Fitness Metrics
{
To assess the suitability of a Poisson model for migration arrival and exit flows, we calculate the Mean Absolute Error (MAE) between the observed daily counts, $A_i(t)$ and $E_i(t)$, and the model’s expected counts, which are defined as $\lambda_it$ and $\mu_it$ for arrivals and exits, respectively, within a period of $t$ days. This metric quantifies the average absolute deviation between observed and expected values, providing a direct and interpretable measure of model performance in terms of expected error per day.
\[MAE_i=\tfrac{1}{T}\sum_{t=1}^T |A_i(t) - \lambda_it|.\]
To assess the model's performance for different migration counts over the entire time period of two years, we calculate the Mean Absolute Percentage Error (MAPE), which is given by: 
\[MAPE_i=\tfrac{100}{T}\sum_{t=1}^T |\tfrac{A_i(t) - \lambda_it}{A_i(t)}|.\]
Both metrics apply for exits with the observed count being $E_i(t)$ and the modelled count being $\mu_it$.
}
\subsection{Fitness of Poisson Parameter}
%%% INTRO
{
To assess how well the Poisson model represents actual migration flows, we compare observed and modelled weekly arrivals and exits across both citizenship groups and age cohorts. The comparison plot visualises the observed data alongside the modelled flows derived from the estimated Poisson parameters for each subgroup (see Figure \ref{fig:poisson_observed_modelled}). This visual check provides an overall sense of how accurately the model captures the dynamics of arrival and exit processes. The calculated errors are presented in the following two subsections for citizenships and age cohorts.
}
%%% PLOTS
{
\begin{figure}[H]
    \centering
    \includegraphics[width=\linewidth]{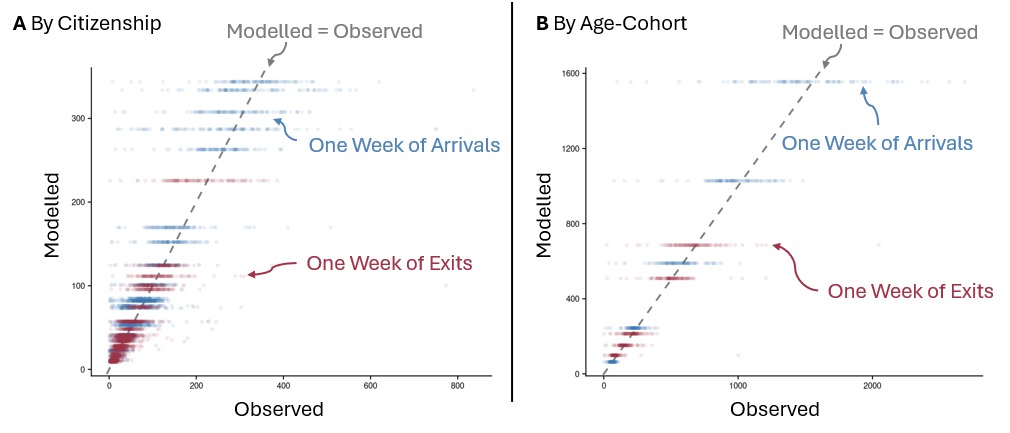}
    \caption{\textbf{Observed versus Modelled:} Arrival and exit flow intensities of \textbf{(A) citizenships} and \textbf{(B) age-cohorts} are modelled using a Poisson model. Points lying on the diagonal line indicate perfect agreement between observed and modelled values.}
    \label{fig:poisson_observed_modelled}
\end{figure}
}
\subsubsection{By Citizenship}
%%% TABLE
{
\begin{table}[H]
\centering
\renewcommand{\arraystretch}{1.25} 
\begin{tabular}{| p{0.28\textwidth} | c | c | c | c |}
\hline
\multirow{2}{*}{\textbf{Citizenship}} & \multicolumn{2}{c|}{\textbf{Arrivals}} & \multicolumn{2}{c|}{\textbf{Exits}} \\ \cline{2-5}
 & $\mathbf{MAE_i}$ & $\mathbf{MAPE_i}$ & $\mathbf{MAE_i}$ & $\mathbf{MAPE_i}$ \\ \hline \hline
Serbia & 10.6205 & 8.52\% & 9.9806 & 9.60\% \\ \hline
Turkey & 15.4516 & 9.43\% & 9.4918 & 9.21\% \\ \hline 
Germany & 31.6472 & 8.98\% & 11.4492 & 9.31\% \\ \hline 
Bosnia \& Herzegovina & 8.7703 & 8.84\% & 6.7547 & 10.35\% \\ \hline 
Romania & 27.6341 & 9.24\% & 12.2694 & 9.02\% \\ \hline 
Croatia & 14.2041 & 9.56\% & 5.2705 & 9.89\% \\ \hline 
Syria & 30.5799 & 11.12\% & 6.9822 & 11.65\% \\ \hline 
Ukraine & 31.2785 & 9.16\% & 22.2455 & 10.13\% \\ \hline 
Hungary & 23.2318 & 8.79\% & 7.9024 & 9.51\% \\ \hline 
Afghanistan & 7.9080 & 11.45\% & 4.5966 & 10.94\% \\ \hline 
Poland & 8.0494 & 9.75\% & 4.2439 & 10.38\% \\ \hline 
Russia & 7.7304 & 11.07\% & 4.4468 & 10.61\% \\ \hline 
Slovakia & 7.6053 & 9.28\% & 3.8948 & 10.52\% \\ \hline 
Bulgaria & 8.3555 & 10.43\% & 3.9393 & 10.25\% \\ \hline 
Kosovo & 4.1795 & 10.09\% & 2.8176 & 11.26\% \\ \hline 
North Macedonia & 3.9296 & 10.80\% & 2.0373 & 12.08\% \\ \hline 
Italy & 7.6756 & 9.13\% & 3.2812 & 10.69\% \\ \hline 
Iran & 6.5207 & 11.17\% & 3.5043 & 11.23\% \\ \hline 
Iraq & 1.4790 & 15.85\% & 2.4085 & 12.89\% \\ \hline 
China & 6.3879 & 10.74\% & 2.8774 & 11.40\% \\ \hline 
Slovenia & 3.6136 & 10.63\% & 1.5188 & 13.12\% \\ \hline 
India & 5.6635 & 10.54\% & 4.0946 & 10.48\% \\ \hline 
Somalia & 2.9528 & 15.28\% & 1.3484 & 13.69\% \\ \hline 
Czechia & 2.8406 & 10.75\% & 1.5854 & 12.62\% \\ \hline 
United Kingdom & 1.7861 & 11.79\% & 1.5649 & 13.24\% \\ \hline
\end{tabular}
\vspace{5pt}
\caption{Mean absolute error (MAE) and mean absolute percentage error (MAPE) for arrivals and exits, by citizenship.}
\label{tab:flow_errors_citizenship}
\end{table}
}
\subsubsection{By Age-Cohort}
%%% TABLE
{
\begin{table}[H]
\centering
\renewcommand{\arraystretch}{1.25} 
\begin{tabular}{| p{0.28\textwidth} | c | c | c | c |}
\hline
\multirow{2}{*}{\textbf{Age Cohort}} & \multicolumn{2}{c|}{\textbf{Arrivals}} & \multicolumn{2}{c|}{\textbf{Exits}} \\ \cline{2-5}
 & $\mathbf{MAE_i}$ & $\mathbf{MAPE_i}$ & $\mathbf{MAE_i}$ & $\mathbf{MAPE_i}$ \\ \hline \hline
Children & 49.7809 & 8.51\% & 20.2236 & 9.12\% \\ \hline
Young Adults & 129.4769 & 8.35\% & 60.3494 & 8.45\% \\ \hline
Middle-Aged Adults & 80.9622 & 7.95\% & 45.2912 & 8.39\% \\ \hline
Older Adults & 19.9970 & 8.23\% & 14.6653 & 8.87\% \\ \hline
Seniors & 6.1100 & 9.46\% & 10.4519 & 10.51\% \\ \hline
\end{tabular}
\vspace{5pt}
\caption{Mean absolute error (MAE) and mean absolute percentage error (MAPE) for arrivals and exits, by age cohort.}
\label{tab:flow_errors_age}
\end{table}
}
\subsection{Fitness of Diaspora dependent Flow Intensities}
%%% INTRO
{
To assess how well the diaspora-size-dependent model captures observed migration patterns, we compare weekly arrivals and exits derived from the estimated pull ($\rho$) and push ($\psi$) rates with observed flows across citizenships and age cohorts. The comparison plot shows modelled flows alongside the actual data, with each subgroup's arrivals and exits estimated based on the size of their respective diaspora. The following subsections provide calculated error metrics for both citizenships and age cohorts.
}
\subsubsection{By Citizenship}
%%% TABLE
{
\begin{table}[H]
\centering
\renewcommand{\arraystretch}{1.25} 
\begin{tabular}{| p{0.28\textwidth} | c | c | c | c |}
\hline
\multirow{2}{*}{\textbf{Citizenship}} & \multicolumn{2}{c|}{\textbf{Arrivals}} & \multicolumn{2}{c|}{\textbf{Exits}} \\ \cline{2-5}
 & $\mathbf{MAE_i}$ & $\mathbf{MAPE_i}$ & $\mathbf{MAE_i}$ & $\mathbf{MAPE_i}$ \\ \hline \hline
Serbia & 23.5588 & 18.89\% & 11.3375 & 10.91\% \\ \hline
Turkey & 21.0995 & 12.87\% & 10.1720 & 9.87\% \\ \hline 
Germany & 35.2915 & 10.01\% & 11.4414 & 9.31\% \\ \hline 
Bosnia \& Herzegovina & 17.0849 & 17.23\% & 8.8985 & 13.64\% \\ \hline 
Romania & 30.2357 & 10.11\% & 12.9675 & 9.53\% \\ \hline 
Croatia & 14.4065 & 9.69\% & 6.8042 & 12.77\% \\ \hline 
Syria & 29.5995 & 10.77\% & 7.9424 & 13.26\% \\ \hline 
Ukraine & 35.6910 & 10.45\% & 25.2138 & 11.48\% \\ \hline 
Hungary & 27.8718 & 10.54\% & 8.0755 & 9.72\% \\ \hline 
Afghanistan & 9.3126 & 13.48\% & 4.8433 & 11.52\% \\ \hline 
Poland & 8.0021 & 9.70\% & 4.2663 & 10.43\% \\ \hline 
Russia & 7.9241 & 11.35\% & 4.3958 & 10.49\% \\ \hline 
Slovakia & 7.7532 & 9.46\% & 3.8576 & 10.42\% \\ \hline 
Bulgaria & 8.4758 & 10.58\% & 3.9681 & 10.32\% \\ \hline 
Kosovo & 4.5390 & 10.96\% & 2.8330 & 11.32\% \\ \hline 
North Macedonia & 4.4158 & 12.13\% & 2.4467 & 14.50\% \\ \hline 
Italy & 8.6750 & 10.32\% & 3.2409 & 10.56\% \\ \hline 
Iran & 6.3064 & 10.80\% & 3.5180 & 11.28\% \\ \hline 
Iraq & 3.1773 & 34.06\% & 2.2573 & 12.08\% \\ \hline 
China & 6.5207 & 10.96\% & 2.8075 & 11.13\% \\ \hline 
Slovenia & 3.5617 & 10.47\% & 1.5749 & 13.60\% \\ \hline 
India & 5.8491 & 10.88\% & 4.5444 & 11.63\% \\ \hline 
Somalia & 2.8231 & 14.61\% & 1.3682 & 13.89\% \\ \hline 
Czechia & 2.8020 & 10.60\% & 1.5188 & 12.09\% \\ \hline 
United Kingdom & 1.8710 & 12.35\% & 1.4800 & 12.52\% \\ \hline
\end{tabular}
\vspace{5pt}
\caption{Mean absolute error (MAE) and mean absolute percentage error (MAPE) for arrivals and exits, by citizenship.}
\label{tab:flow_errors_diaspora_citizenship}
\end{table}
}

\subsubsection{By Age-Cohort}
%%% TABLE
{
\begin{table}[H]
\centering
\renewcommand{\arraystretch}{1.25} 
\begin{tabular}{| p{0.28\textwidth} | c | c | c | c |}
\hline
\multirow{2}{*}{\textbf{Age Cohort}} & \multicolumn{2}{c|}{\textbf{Arrivals}} & \multicolumn{2}{c|}{\textbf{Exits}} \\ \cline{2-5}
 & $\mathbf{MAE_i}$ & $\mathbf{MAPE_i}$ & $\mathbf{MAE_i}$ & $\mathbf{MAPE_i}$ \\ \hline \hline
Children & 53.3164 & 9.12\% & 20.1900 & 9.11\% \\ \hline
Young Adults & 154.4692 & 9.97\% & 70.1375 & 9.82\% \\ \hline
Middle-Aged Adults & 80.7411 & 7.93\% & 45.5409 & 8.44\% \\ \hline
Older Adults & 25.6250 & 10.54\% & 14.7891 & 8.94\% \\ \hline
Seniors & 11.2412 & 17.40\% & 10.3905 & 10.45\% \\ \hline
\end{tabular}
\vspace{5pt}
\caption{Mean absolute error (MAE) and mean absolute percentage error (MAPE) for arrivals and exits, by age cohort.}
\label{tab:flow_errors_diaspora_age}
\end{table}
}

\subsection{Municipal-Level Allocation of Arrival and Exit Flows} \label{appendix:fitness_assortativity}
%%% Observed v Modelled Diaspora v Gravity: Scale insensitive error as in flow stability (?)
{
To evaluate the accuracy of the migration flow allocation, we assess how well the simulated distribution of arrivals and exits matches the observed data across municipalities. This is done by comparing two probabilistic allocation mechanisms: one based on the local share of each diaspora population and another based on total municipal population size, the latter commonly used in gravity models of migration. To quantify the accuracy of each approach, we calculate an error metric, denoted \( \varepsilon_i \), that captures the deviation between observed and modelled flows across all municipalities and time steps.

Let \( A_{ij}(t) \) and \( E_{ij}(t) \) denote the observed number of arrivals and exits for group \( i \) in municipality \( j \) on day \( t \). The corresponding modelled counts from the multinomial allocation are given by \( \hat{A}_{ij}(t) \) and \( \hat{E}_{ij}(t) \). For each flow type, the daily error is computed as the square root of the summed squared deviations between the observed and simulated values, normalised by the total number of municipalities and time steps:
\[
\varepsilon_{i, \text{arrivals}} = \frac{1}{nT} \sqrt{ \sum_{j=1}^{n} \sum_{t=1}^{T} \left( \hat{A}_{ij}(t) - A_{ij}(t) \right)^2} \quad \text{and} \quad
\varepsilon_{i,\text{exits}} = \frac{1}{nT} \sqrt{ \sum_{j=1}^{n} \sum_{t=1}^{T} \left( \hat{E}_{ij}(t) - E_{ij}(t) \right)^2},
\]
where \( n\) is the number of municipalities included and \( T = 731 \) is the number of days in the observation period. The modelled flows are generated by drawing 100 simulations for each day from a multinomial distribution, based on either diaspora or population shares. The resulting values of \( \varepsilon_i \) represent the average deviation between observed and simulated flows per municipality per day, with lower values indicating better model fit. To reduce noise from low-count observations, the analysis is limited to municipalities that recorded at least five arrivals over the observation period. This threshold ensures that flow comparisons are based on meaningful levels of migration activity.
}

\subsubsection{By Citizenship}
%%% INTRO
{
The table below presents the calculated daily allocation errors \( \varepsilon_i \) for each citizenship, comparing diaspora-based and population-based models for both arrivals and exits (see Table \ref{tab:allocation_errors_citizenship}). These values quantify how closely the simulated flows match the observed municipal-level distributions. Following the table, a series of diagnostic plots illustrates the allocation performance for each group. Each plot contrasts observed flows with those generated through 100 multinomial simulations based on population size (grey) and diaspora size (blue for arrivals, red for exits), offering a visual comparison of allocation accuracy across methods.
}
%%% TABLE Citizenship
{
\begin{table}[H]
\centering
\renewcommand{\arraystretch}{1.25}
\begin{tabular}{|p{0.25\textwidth}|c|c|c|c|}
\hline
\textbf{Citizenship} 
& \multicolumn{2}{c|}{\textbf{Diaspora}} 
& \multicolumn{2}{c|}{\textbf{Population}} \\ \cline{2-5}
& \boldmath$\varepsilon_{i, \text{arrivals}}$ 
& \boldmath$\varepsilon_{i, \text{exits}}$ 
& \boldmath$\varepsilon_{i, \text{arrivals}}$ 
& \boldmath$\varepsilon_{i, \text{exits}}$ \\ \hline \hline
Serbia & 0.00108 & 0.00160 & 0.01352 & 0.01942 \\ \hline
Turkey & 0.00185 & 0.00133 & 0.01441 & 0.01001 \\ \hline
Germany & 0.00186 & 0.00163 & 0.00572 & 0.00473 \\ \hline
Bosnia \& Herzegovina & 0.00094 & 0.00127 & 0.00820 & 0.01087 \\ \hline
Romania & 0.00185 & 0.00168 & 0.00635 & 0.00544 \\ \hline
Croatia & 0.00132 & 0.00105 & 0.00917 & 0.00664 \\ \hline
Syria & 0.00397 & 0.00223 & 0.03633 & 0.01819 \\ \hline
Ukraine & 0.00221 & 0.00217 & 0.00811 & 0.00768 \\ \hline
Hungary & 0.00182 & 0.00136 & 0.00540 & 0.00403 \\ \hline
Afghanistan & 0.00129 & 0.00101 & 0.02697 & 0.02093 \\ \hline
Poland & 0.00084 & 0.00080 & 0.01244 & 0.01113 \\ \hline
Russia & 0.00088 & 0.00082 & 0.02342 & 0.02119 \\ \hline
Slovakia & 0.00085 & 0.00071 & 0.00915 & 0.00754 \\ \hline
Bulgaria & 0.00089 & 0.00085 & 0.01571 & 0.01354 \\ \hline
Kosovo & 0.00059 & 0.00054 & 0.01014 & 0.00900 \\ \hline
North Macedonia & 0.00057 & 0.00049 & 0.01274 & 0.01075 \\ \hline
Italy & 0.00086 & 0.00076 & 0.01372 & 0.01147 \\ \hline
Iran & 0.00083 & 0.00073 & 0.02589 & 0.02206 \\ \hline
Iraq & 0.00031 & 0.00037 & 0.01670 & 0.02010 \\ \hline
China & 0.00080 & 0.00070 & 0.01585 & 0.01320 \\ \hline
Slovenia & 0.00053 & 0.00043 & 0.01219 & 0.00926 \\ \hline
India & 0.00075 & 0.00070 & 0.01727 & 0.01599 \\ \hline
Somalia & 0.00058 & 0.00039 & 0.02399 & 0.01614 \\ \hline
Czechia & 0.00042 & 0.00035 & 0.01154 & 0.00960 \\ \hline
United Kingdom & 0.00031 & 0.00031 & 0.01552 & 0.01533 \\ \hline
\end{tabular}
\vspace{5pt}
\caption{Daily error \( \varepsilon_i \) in arrival and exit flow allocations for the diaspora- and population-based models, by citizenship. Lower values indicate a closer match between observed and modelled flows.}
\label{tab:allocation_errors_citizenship}
\end{table}

}
%%% PLOTS Citizenship
{
\renewcommand{\arraystretch}{1.0}
\setlength{\tabcolsep}{4pt}
\begin{longtable}{ccc}
\caption{\textbf{Sub-National Allocation of Migration Flows by Citizenship:} Comparison of population size (grey) versus diaspora size for arrivals (blue) and exits (red) as determinant for migration flows on the municipal level. In each of the 100 bootstrap iterations, we draw samples from a multinomial distribution, using the municipality-level shares of the total population and the total diaspora as parameters. These shares define the probability vector for allocating individuals across municipalities within each group.} \label{fig:flow_allocation_citizenship} \\
\endfirsthead

\multicolumn{3}{r}%
{{\bfseries \tablename\ \thetable{} -- continued from previous page}} \\
\\
\endhead

\includegraphics[width=0.48\textwidth]{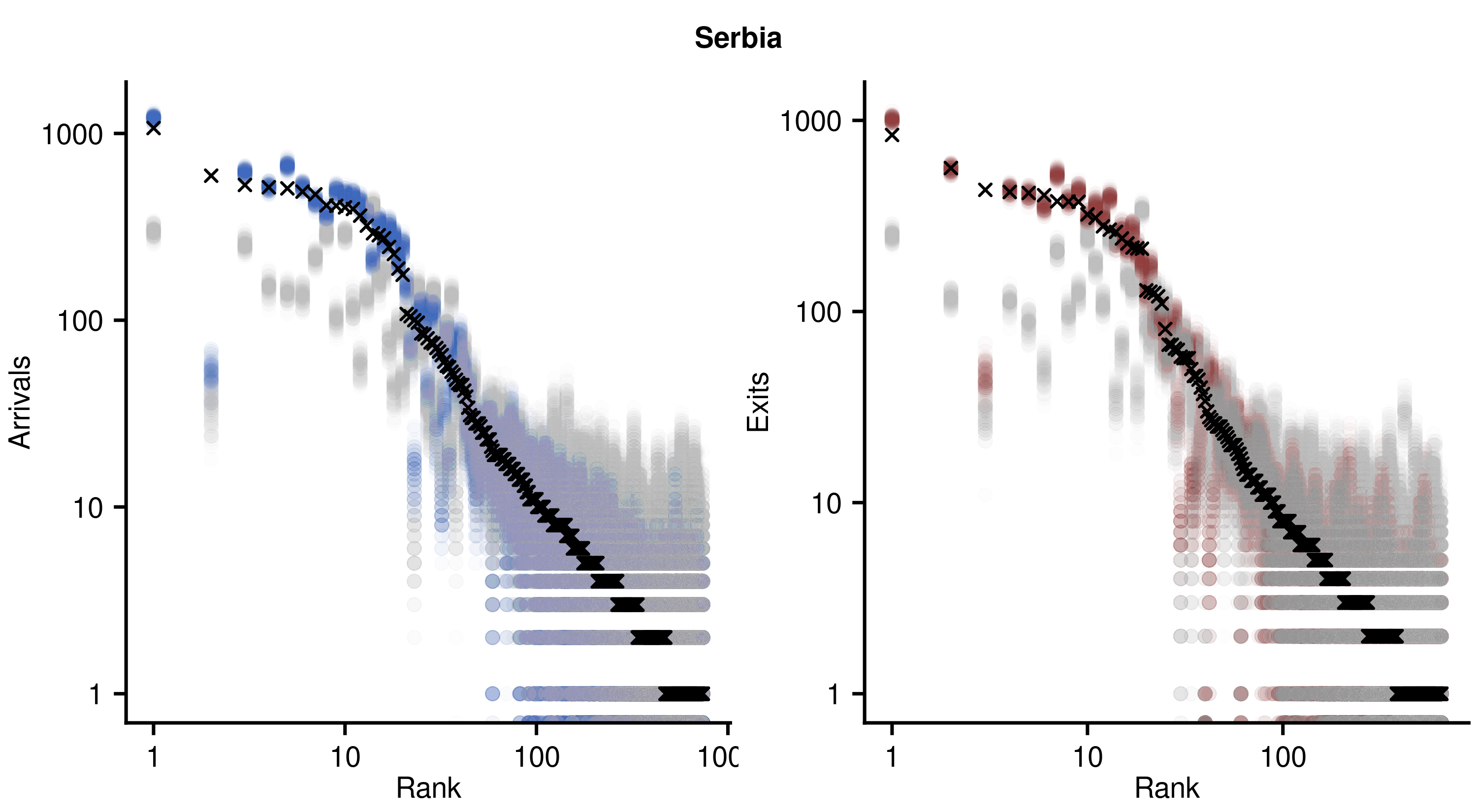} &
\includegraphics[width=0.48\textwidth]{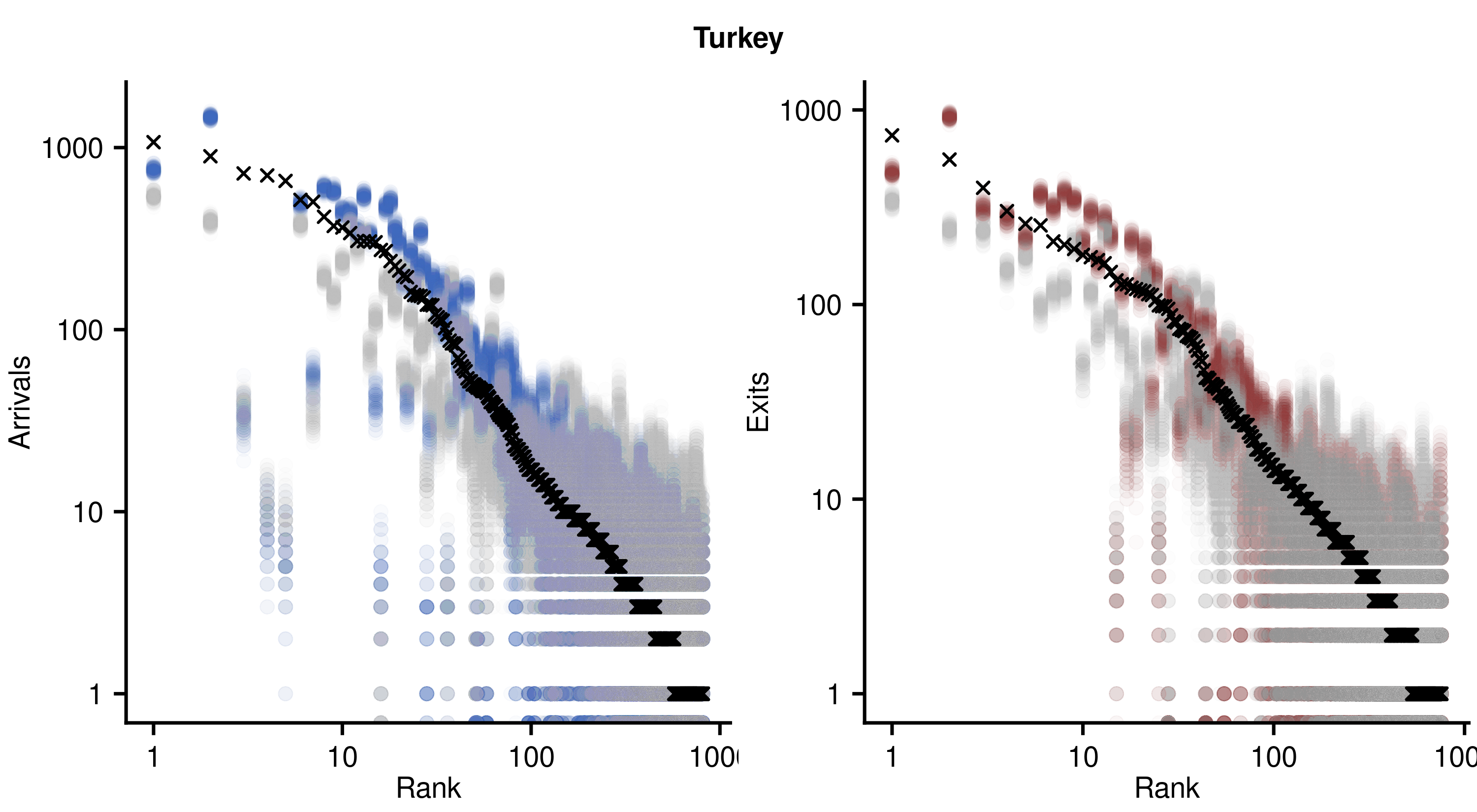} \\

\includegraphics[width=0.48\textwidth]{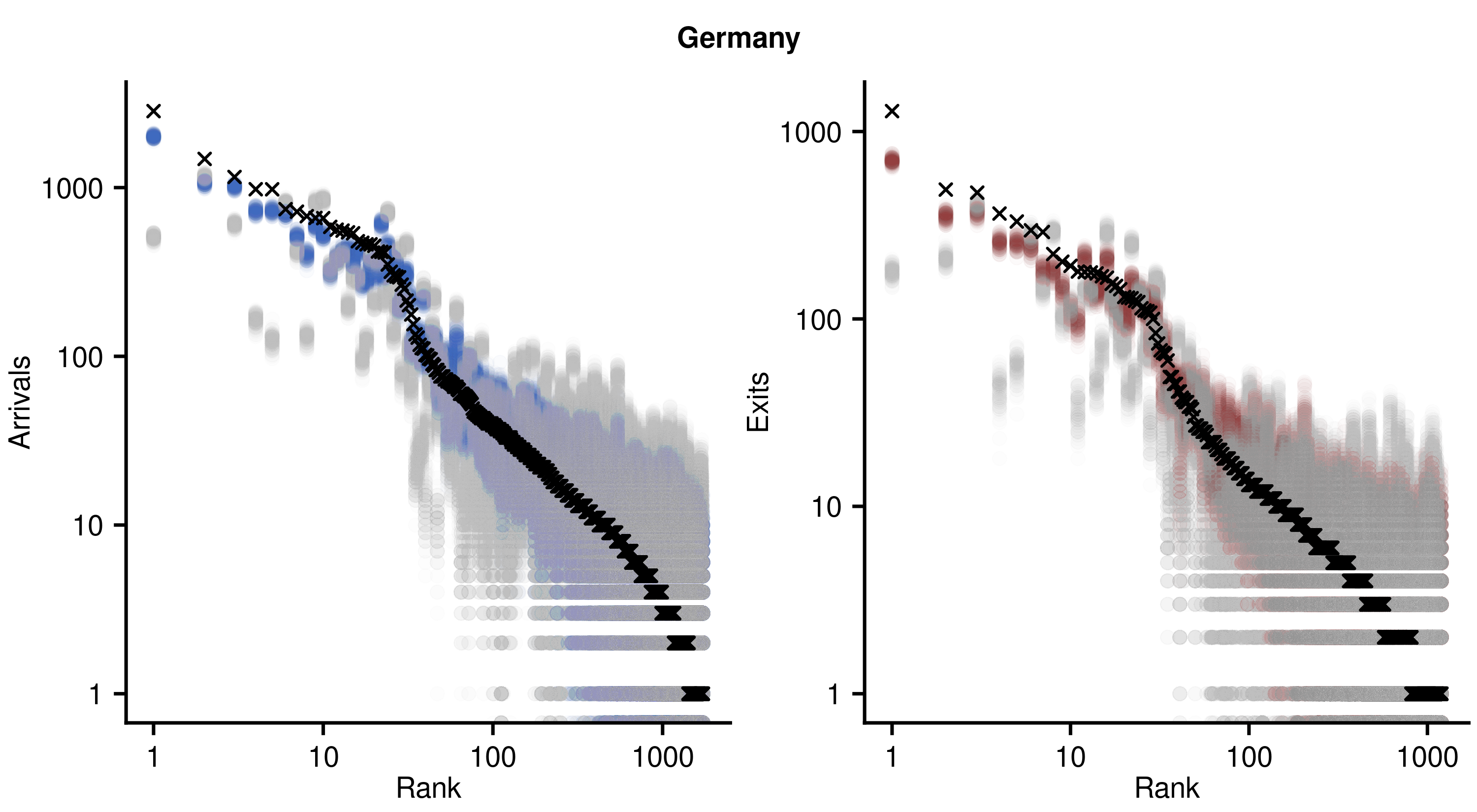} &
\includegraphics[width=0.48\textwidth]{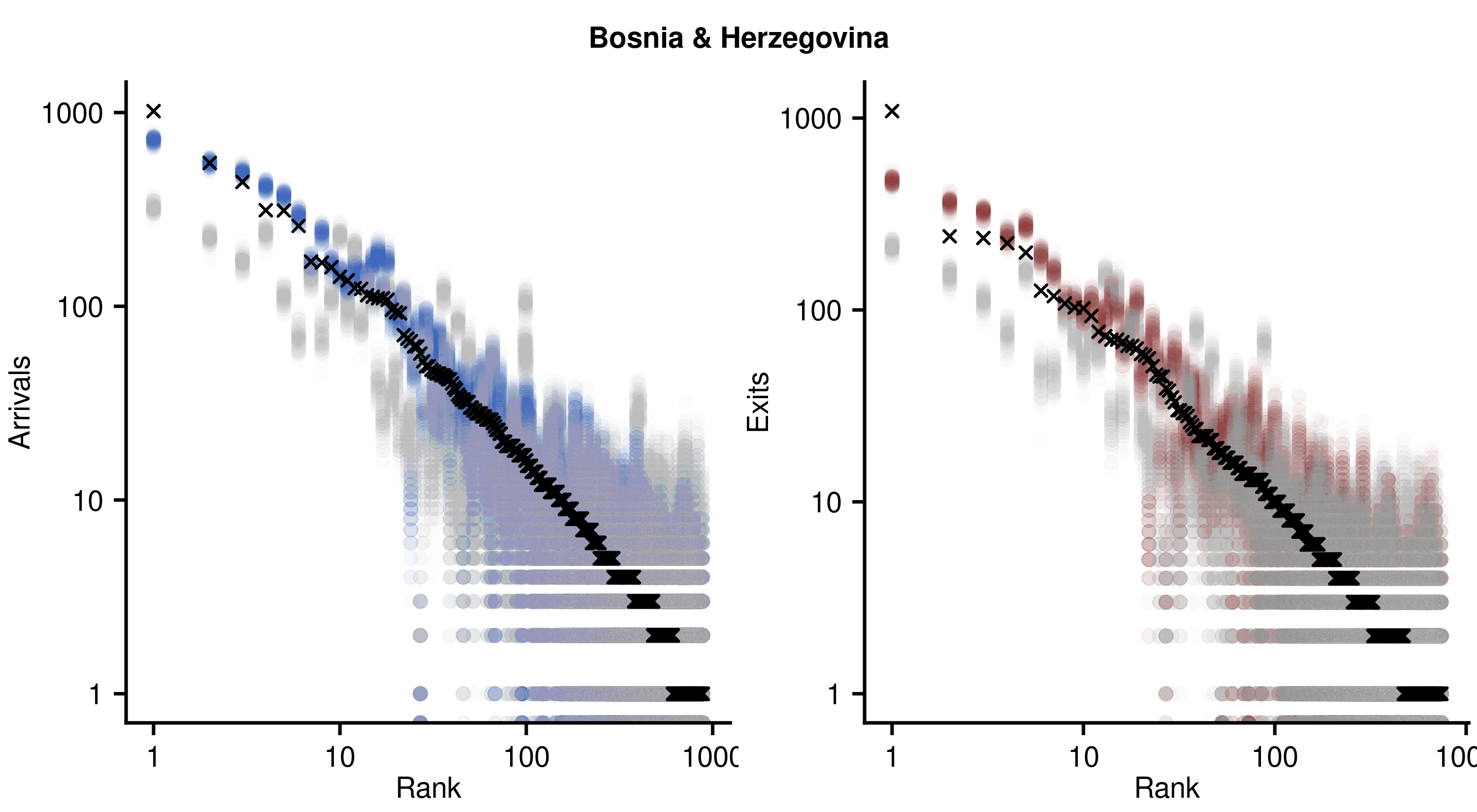} \\

\includegraphics[width=0.48\textwidth]{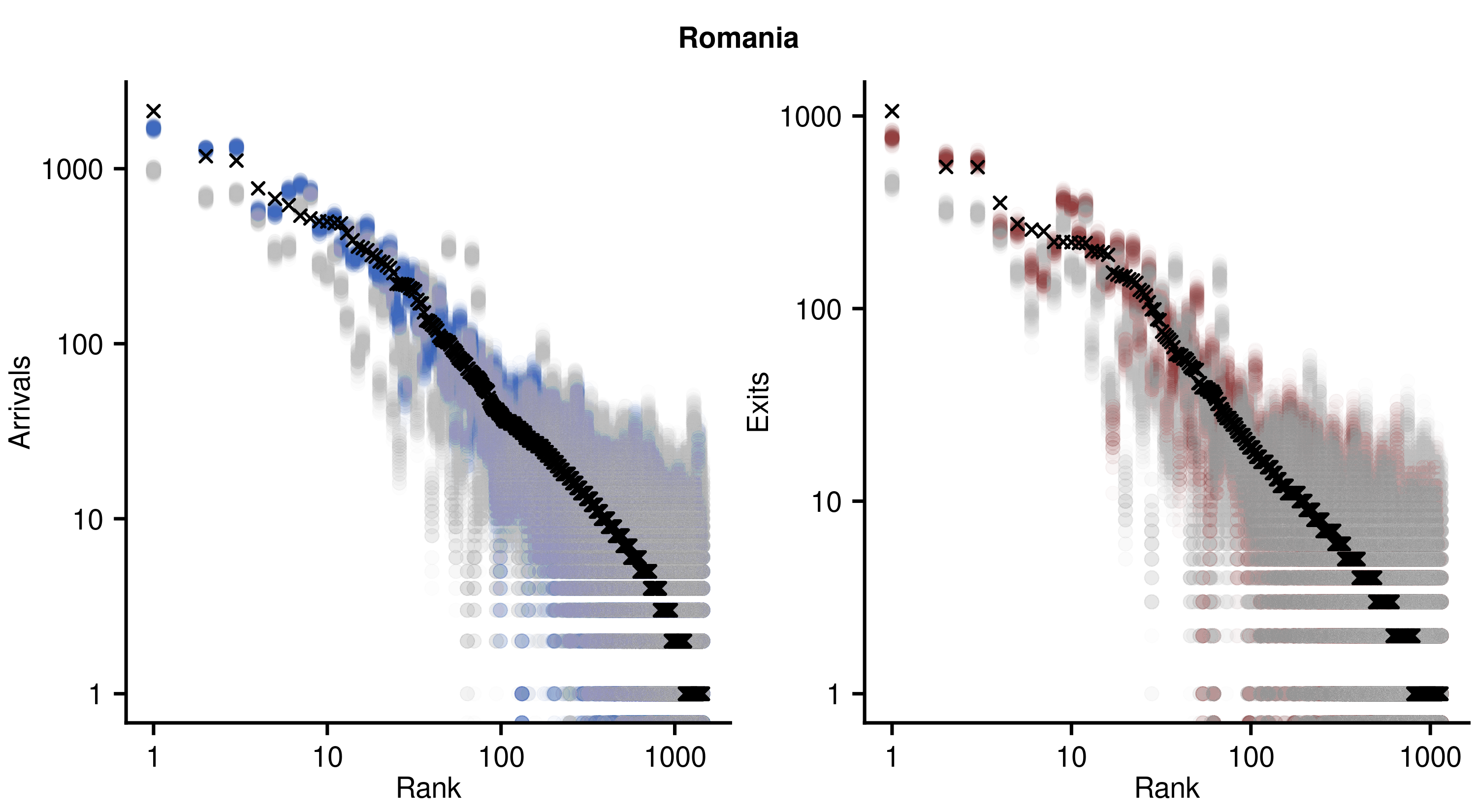} &
\includegraphics[width=0.48\textwidth]{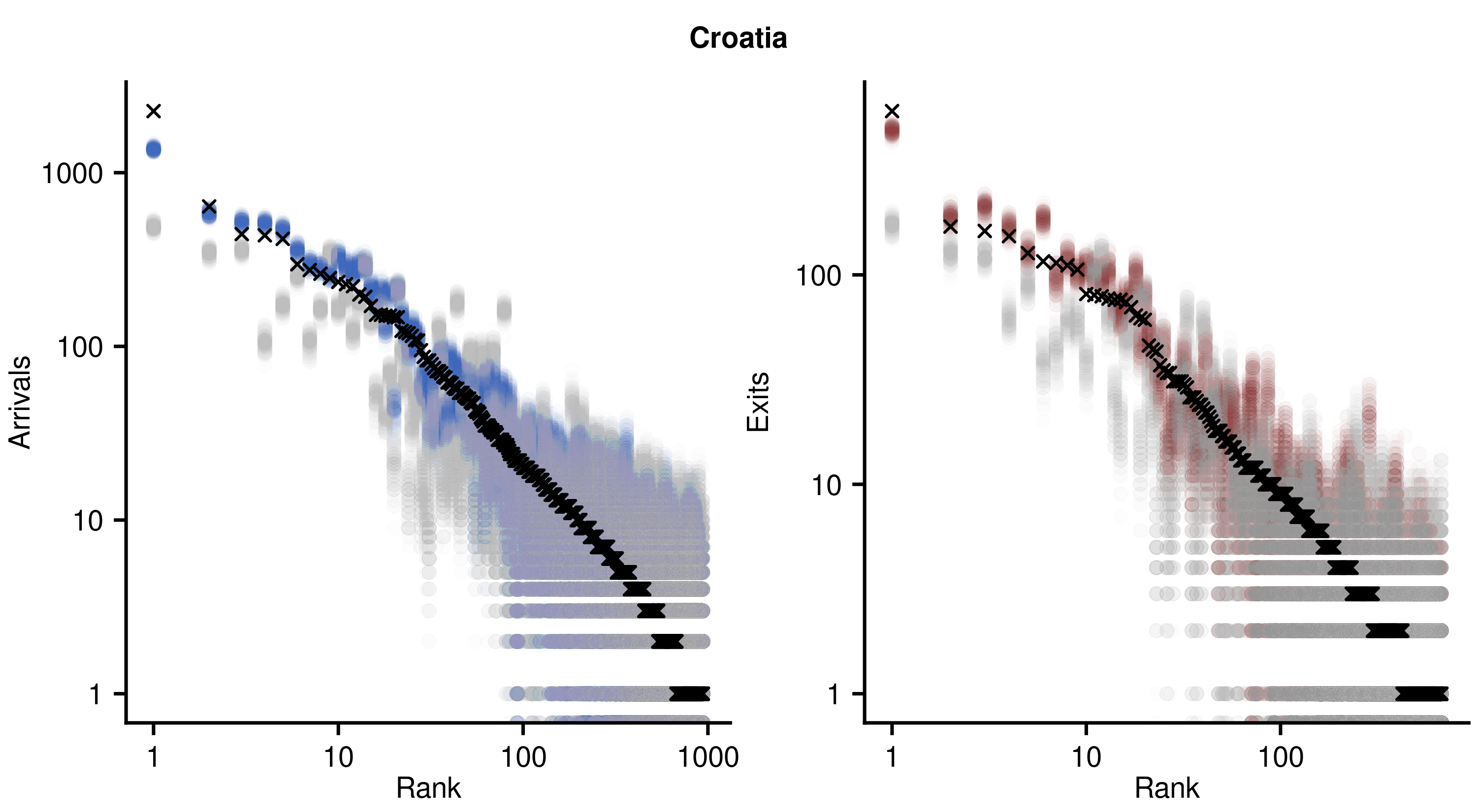} \\

\includegraphics[width=0.48\textwidth]{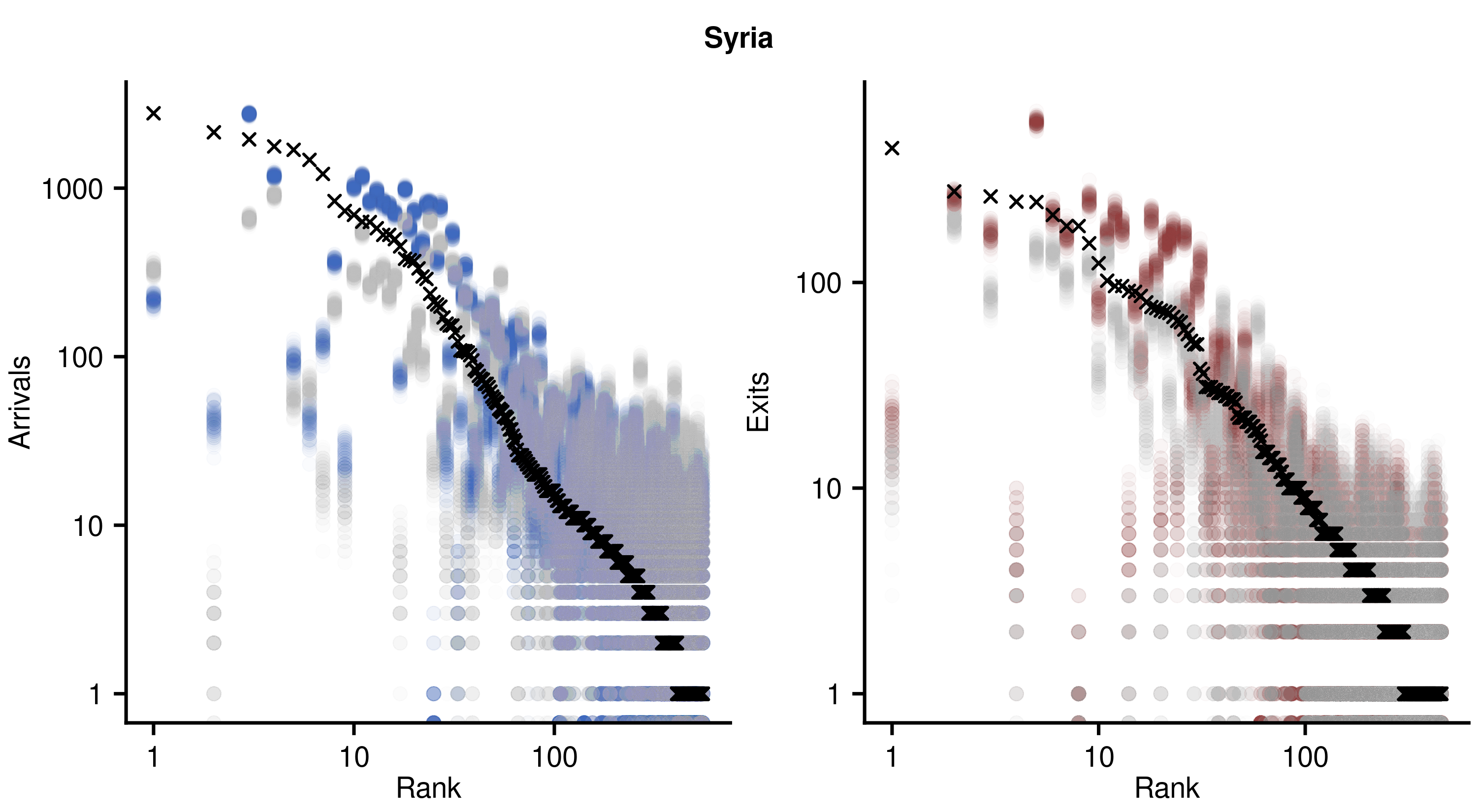} &
\includegraphics[width=0.48\textwidth]{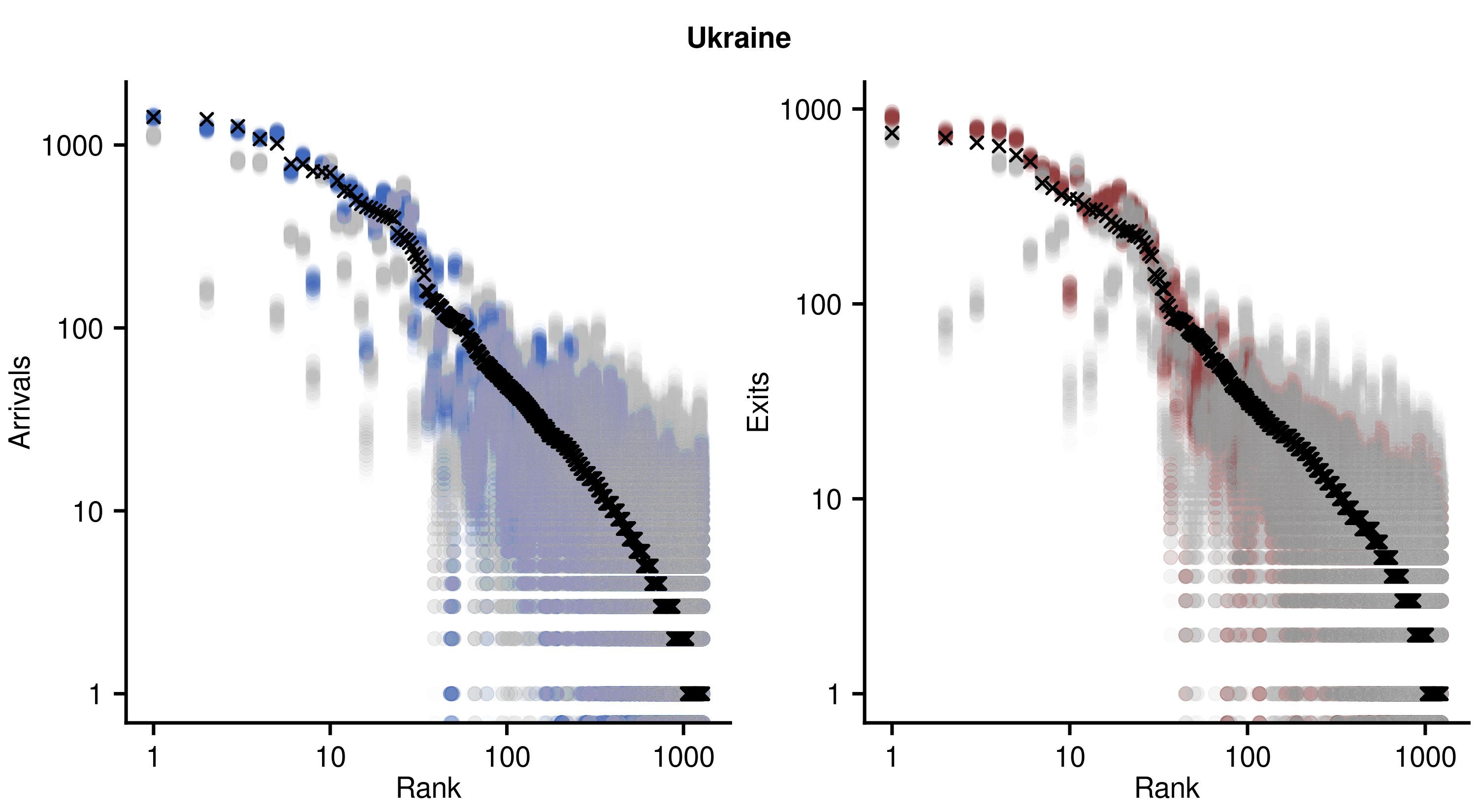} \\

\includegraphics[width=0.48\textwidth]{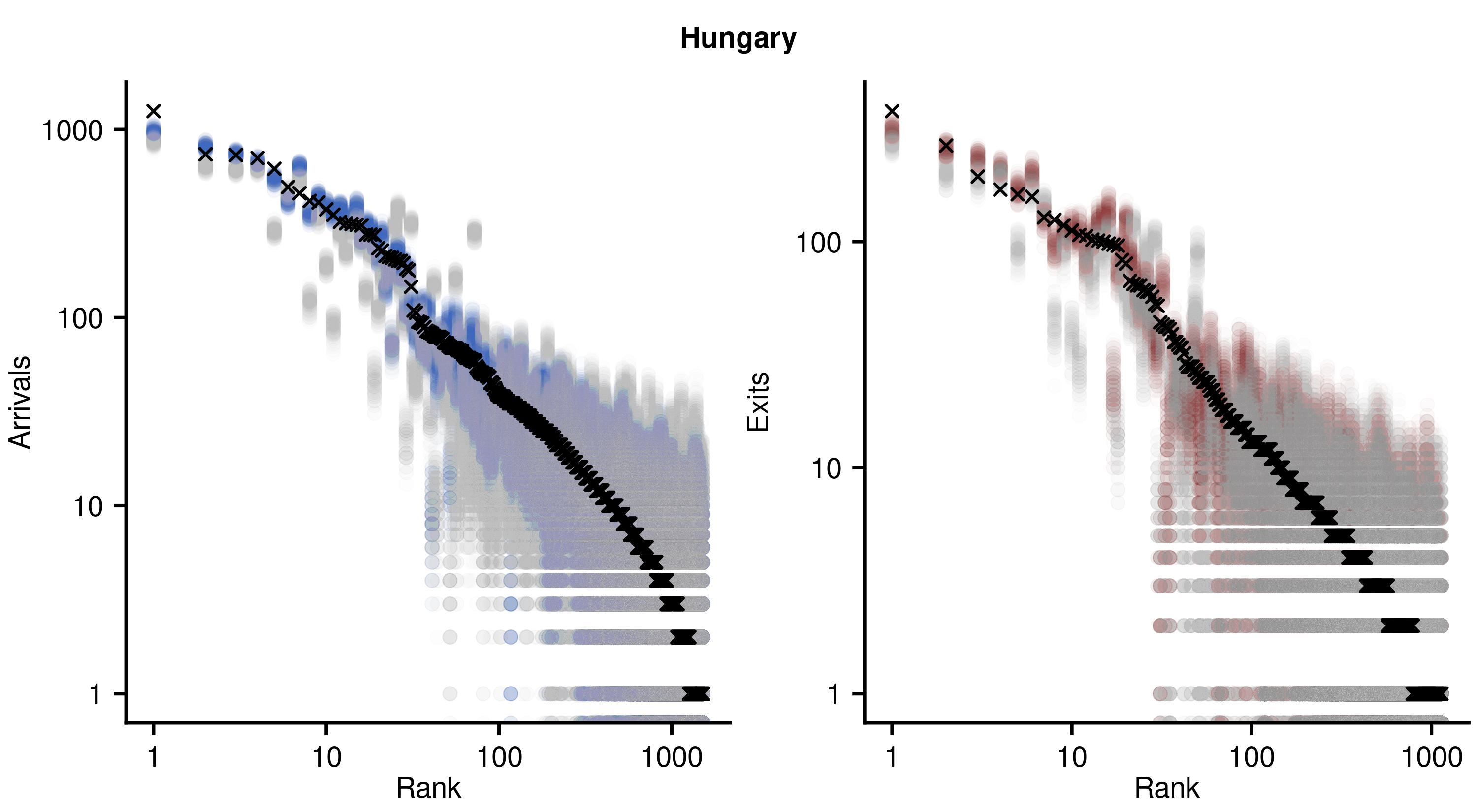} &
\includegraphics[width=0.48\textwidth]{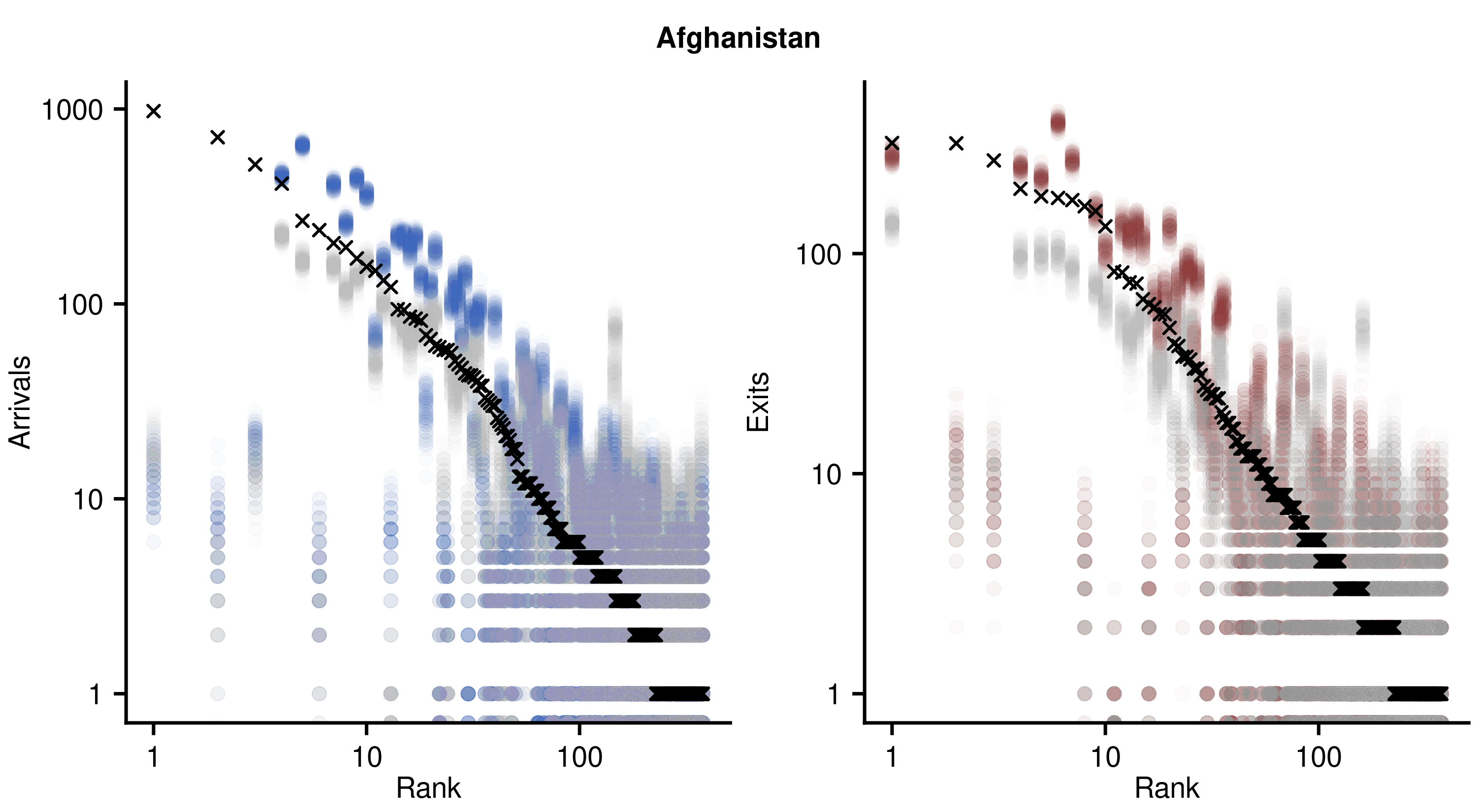} \\

\includegraphics[width=0.48\textwidth]{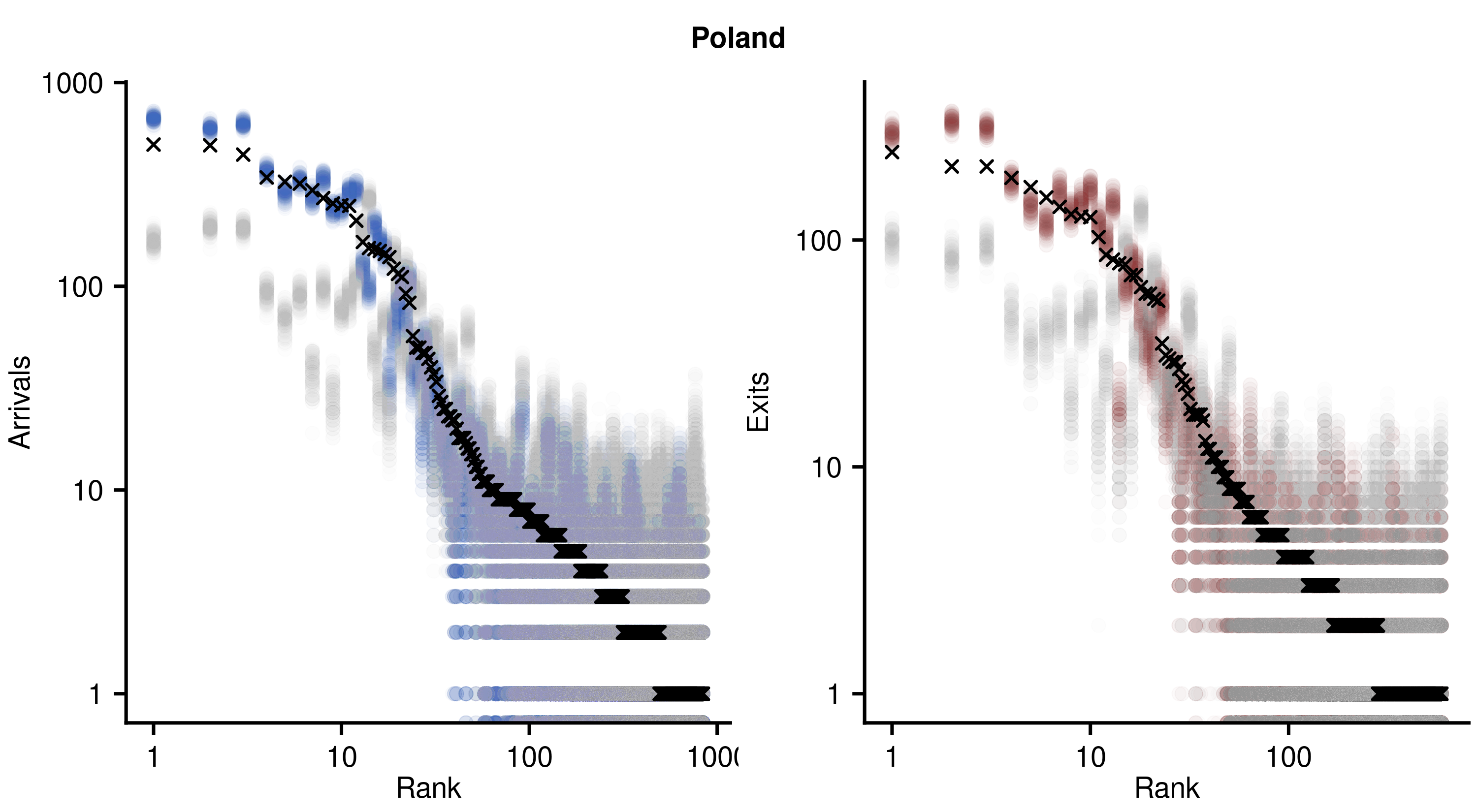} &
\includegraphics[width=0.48\textwidth]{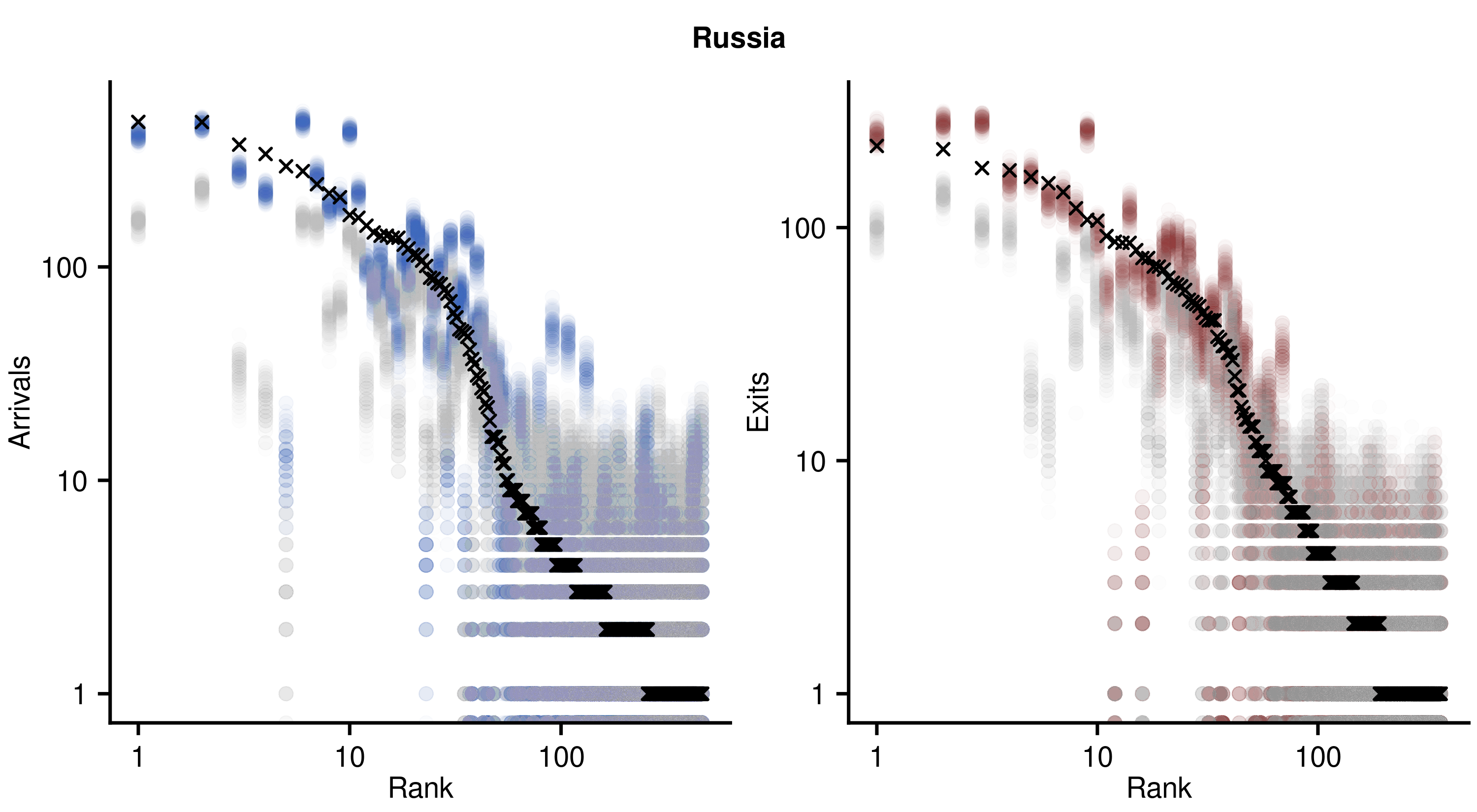} \\

\includegraphics[width=0.48\textwidth]{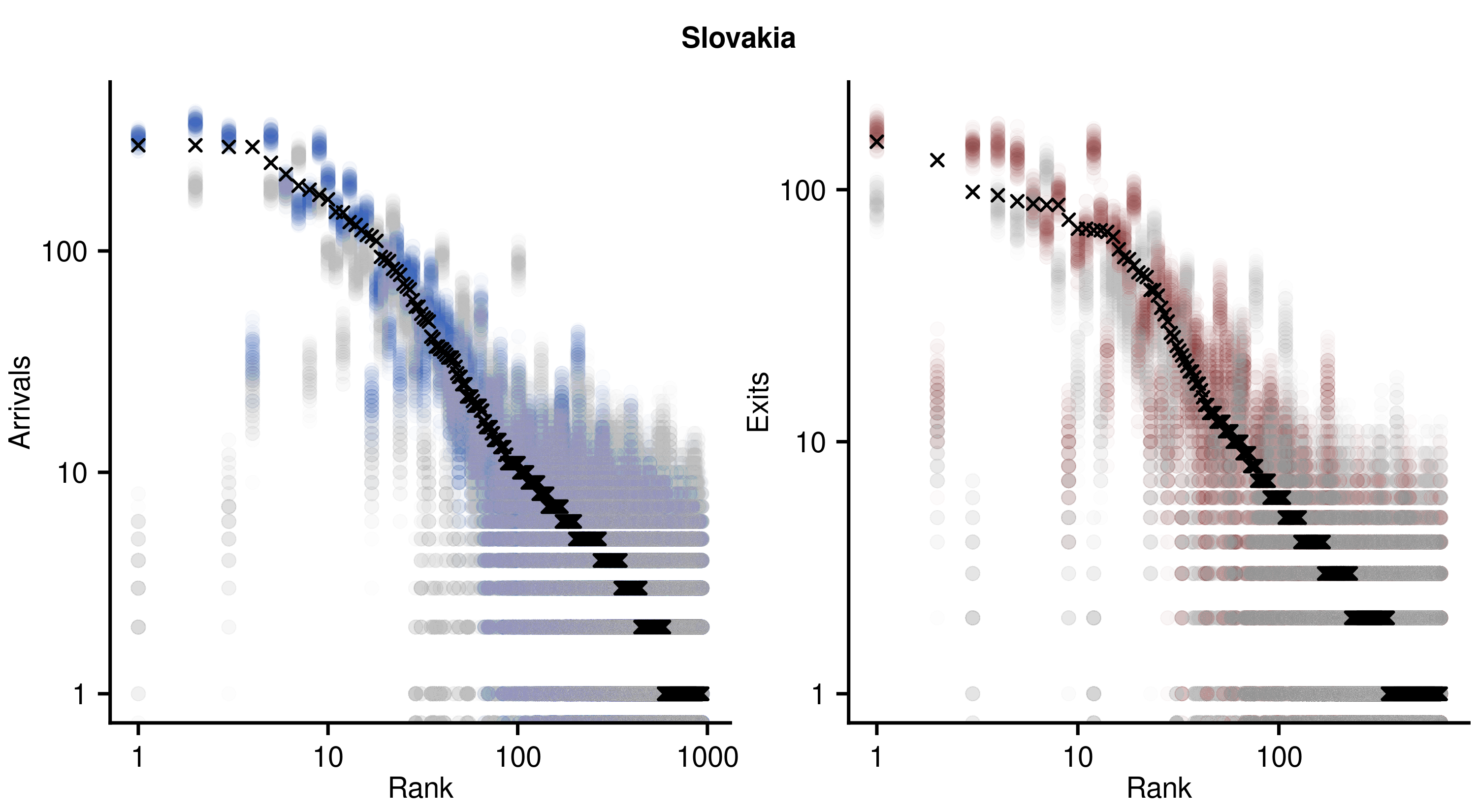} &
\includegraphics[width=0.48\textwidth]{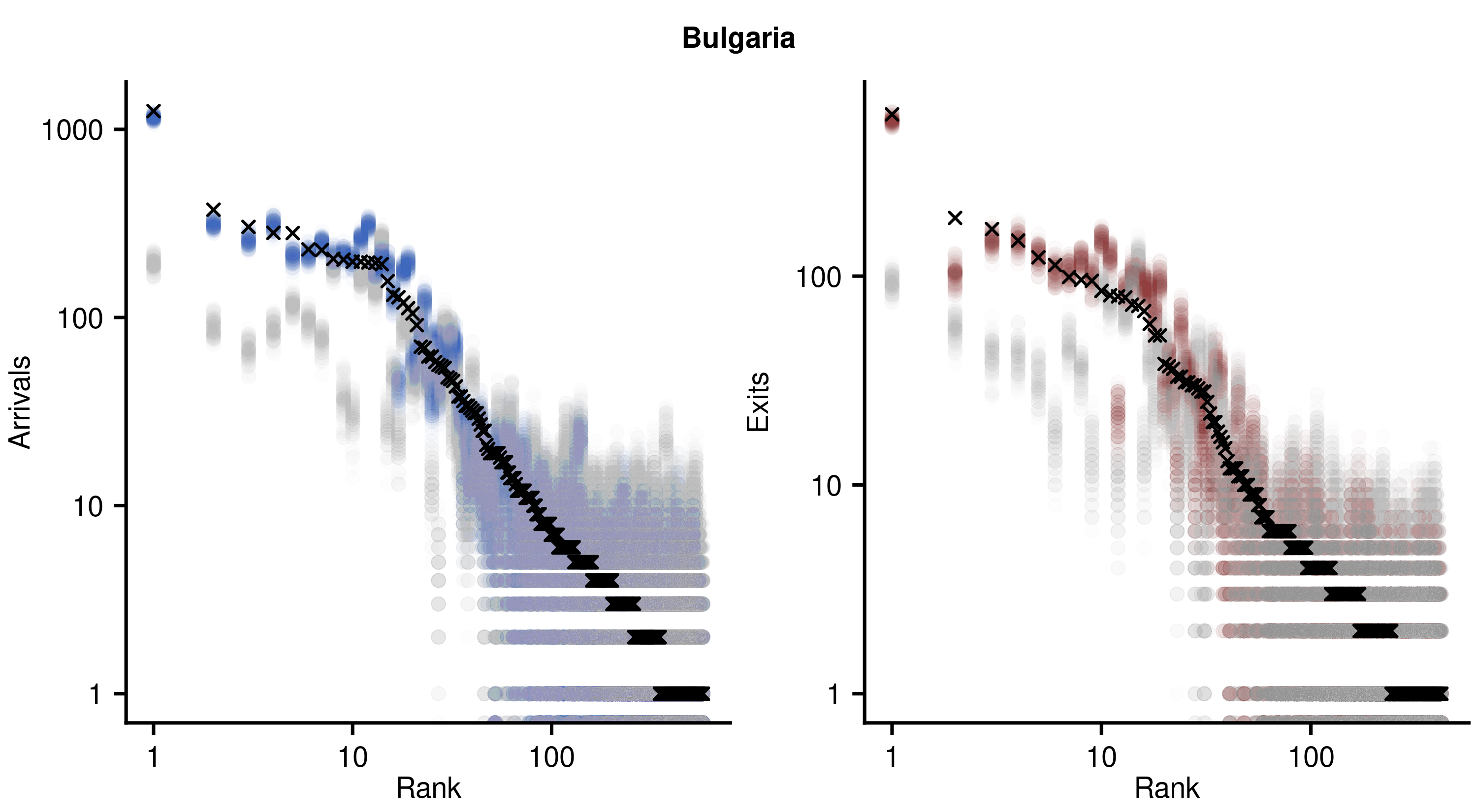} \\

\includegraphics[width=0.48\textwidth]{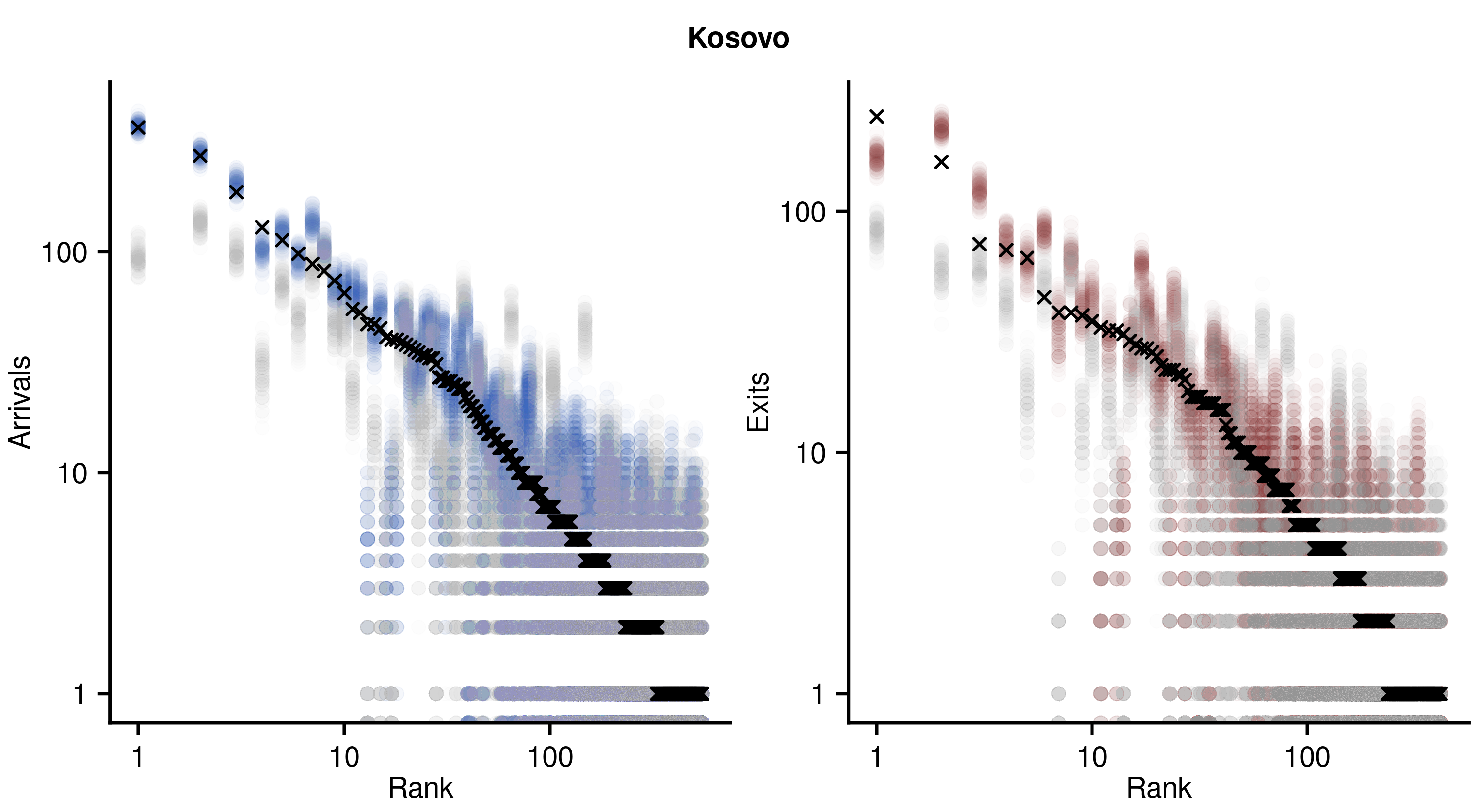} &
\includegraphics[width=0.48\textwidth]{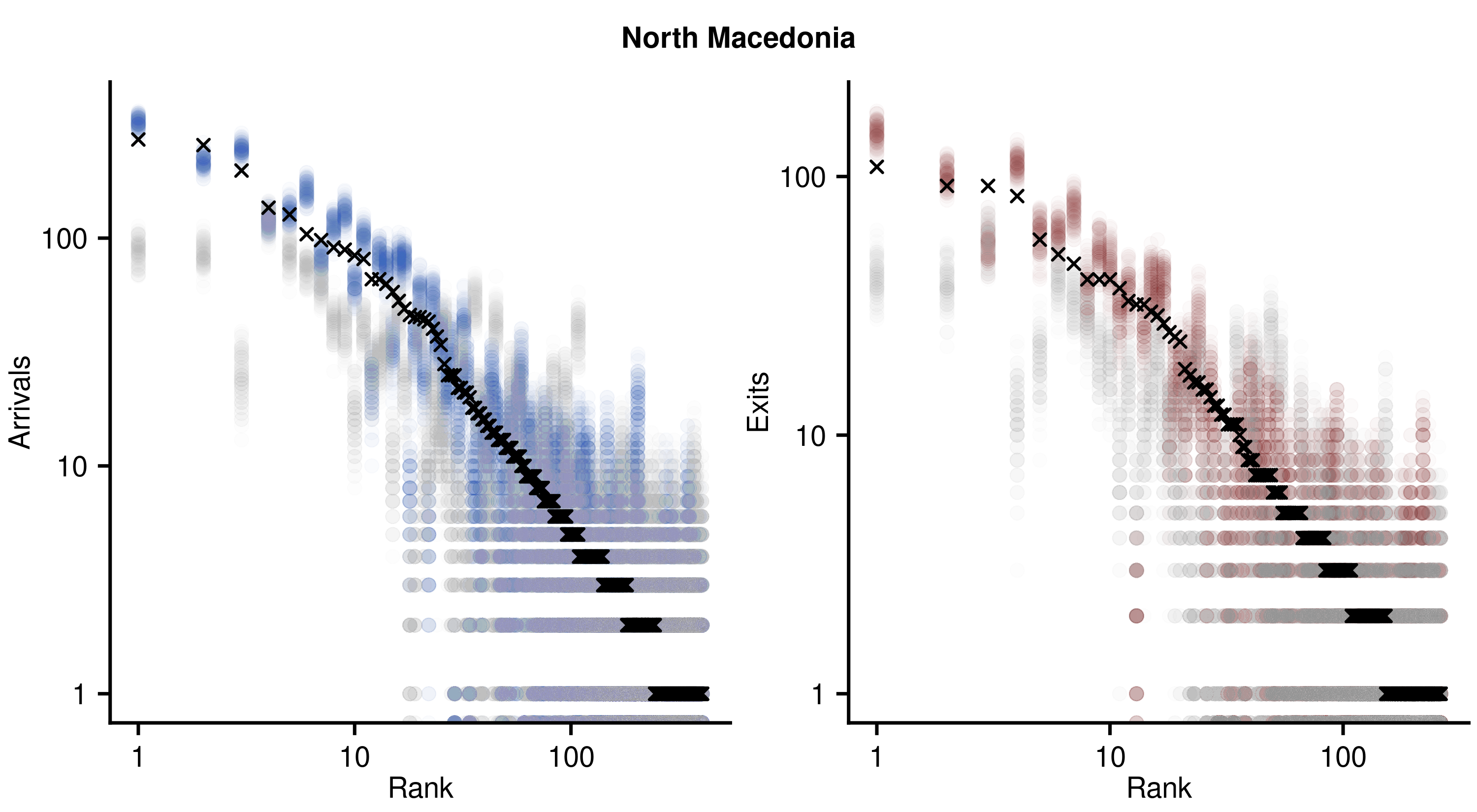} \\

\includegraphics[width=0.48\textwidth]{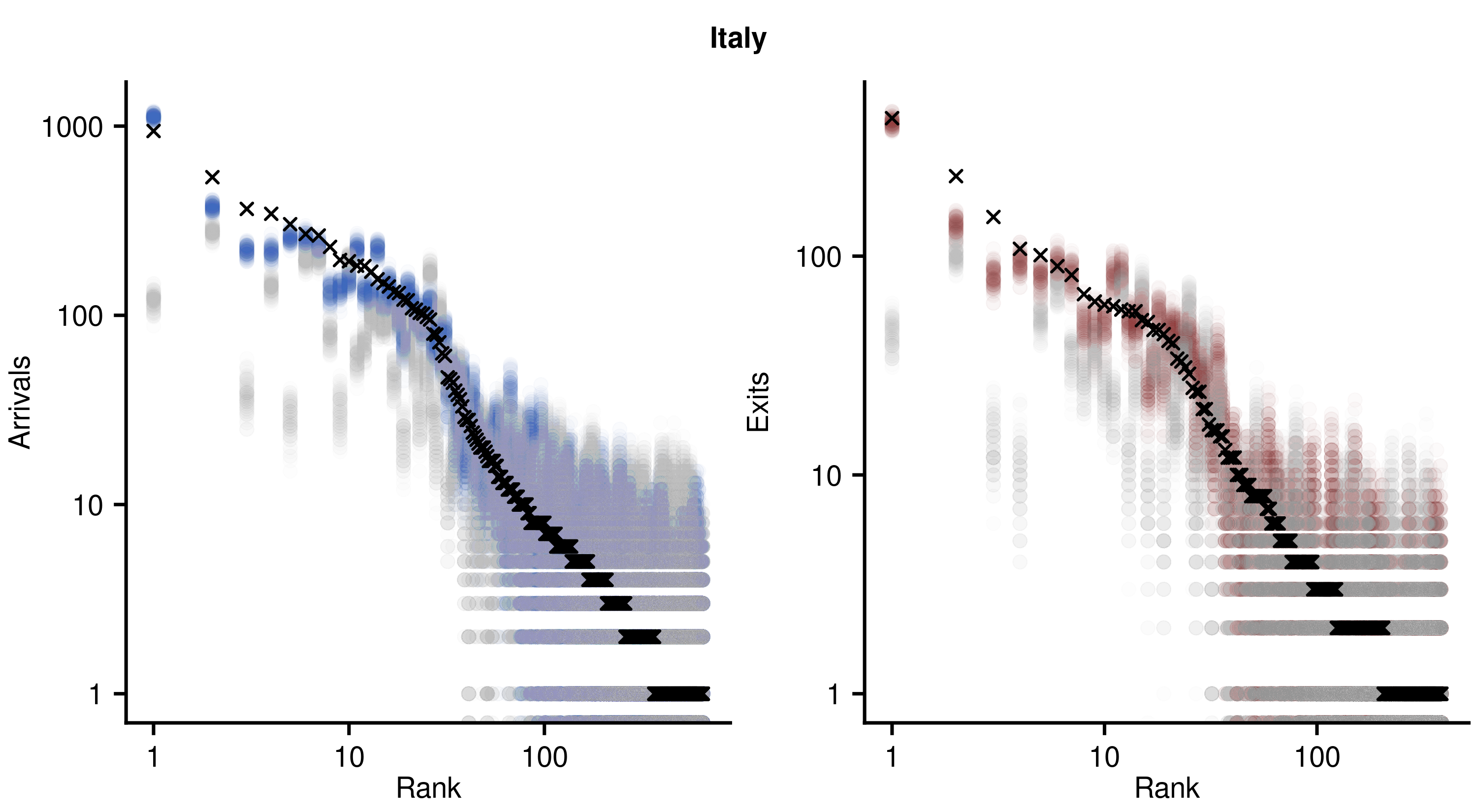} &
\includegraphics[width=0.48\textwidth]{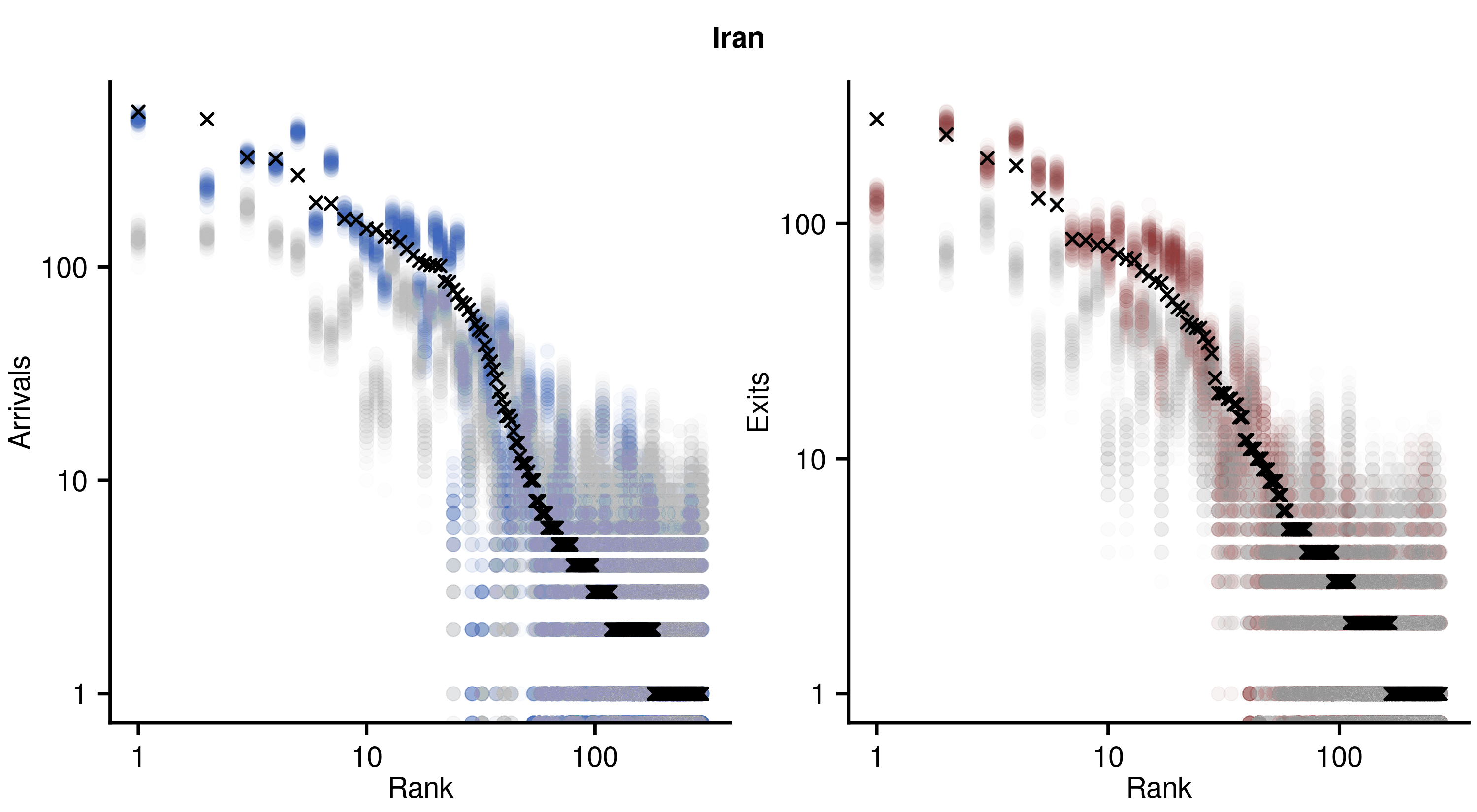} \\

\includegraphics[width=0.48\textwidth]{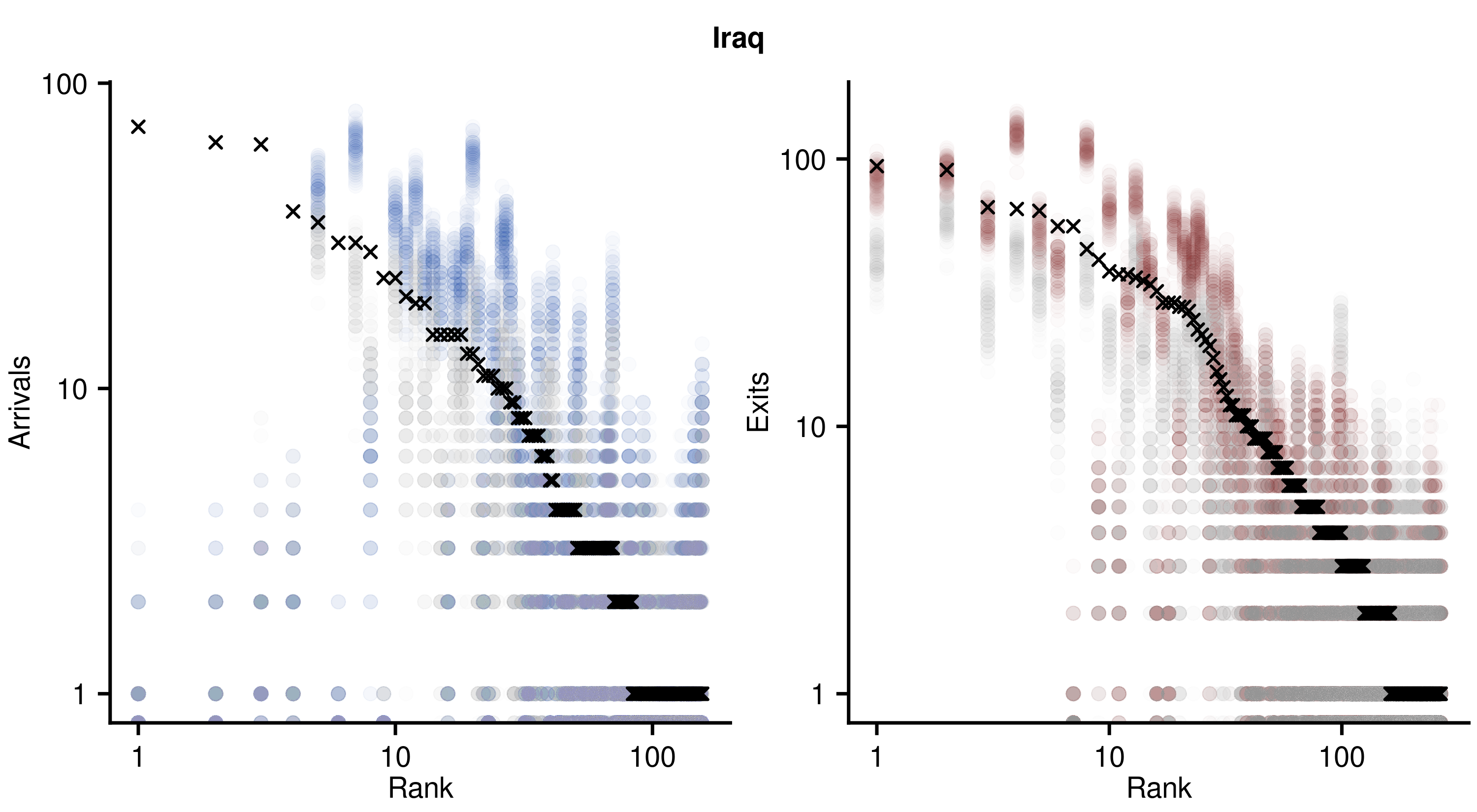} &
\includegraphics[width=0.48\textwidth]{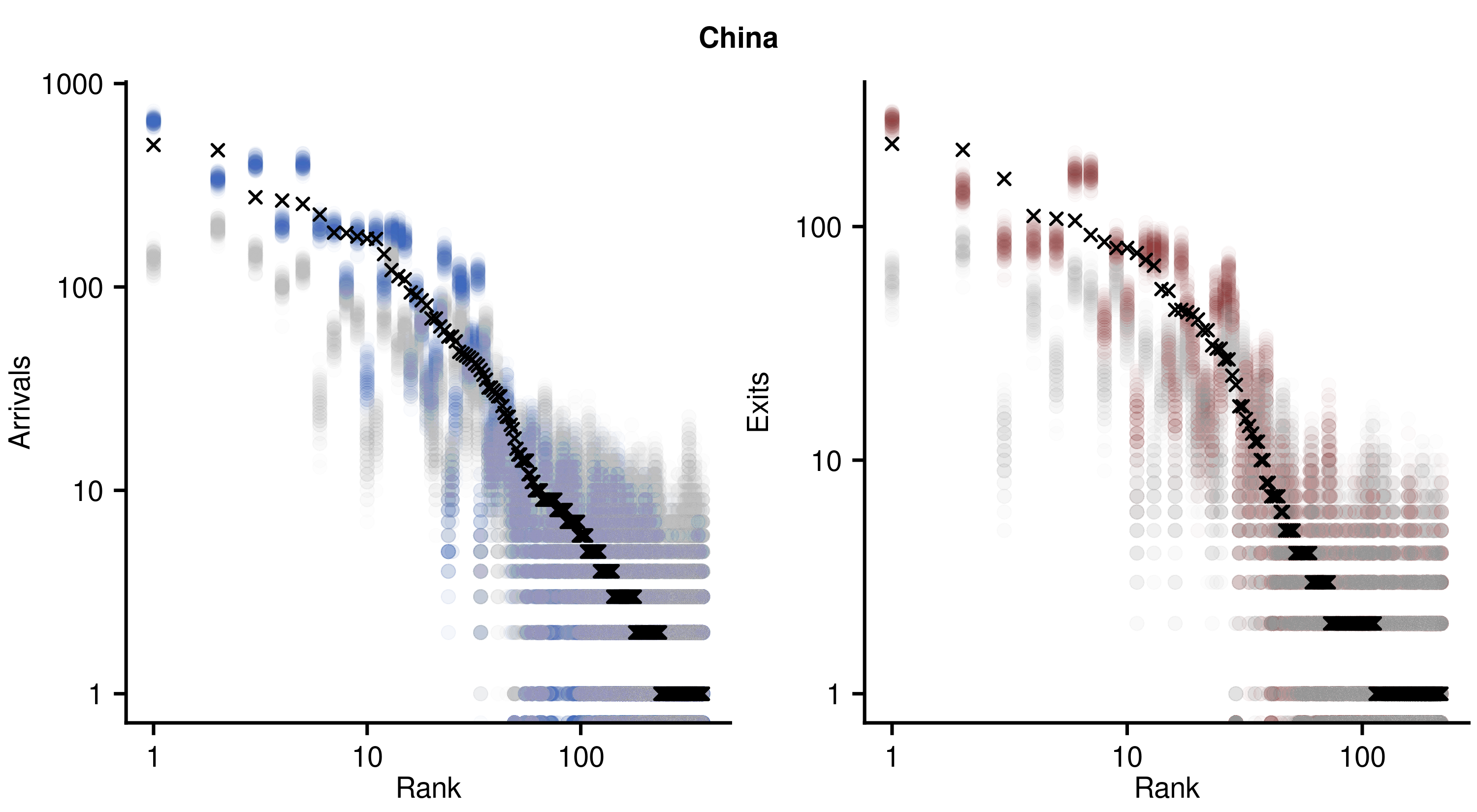} \\

\includegraphics[width=0.48\textwidth]{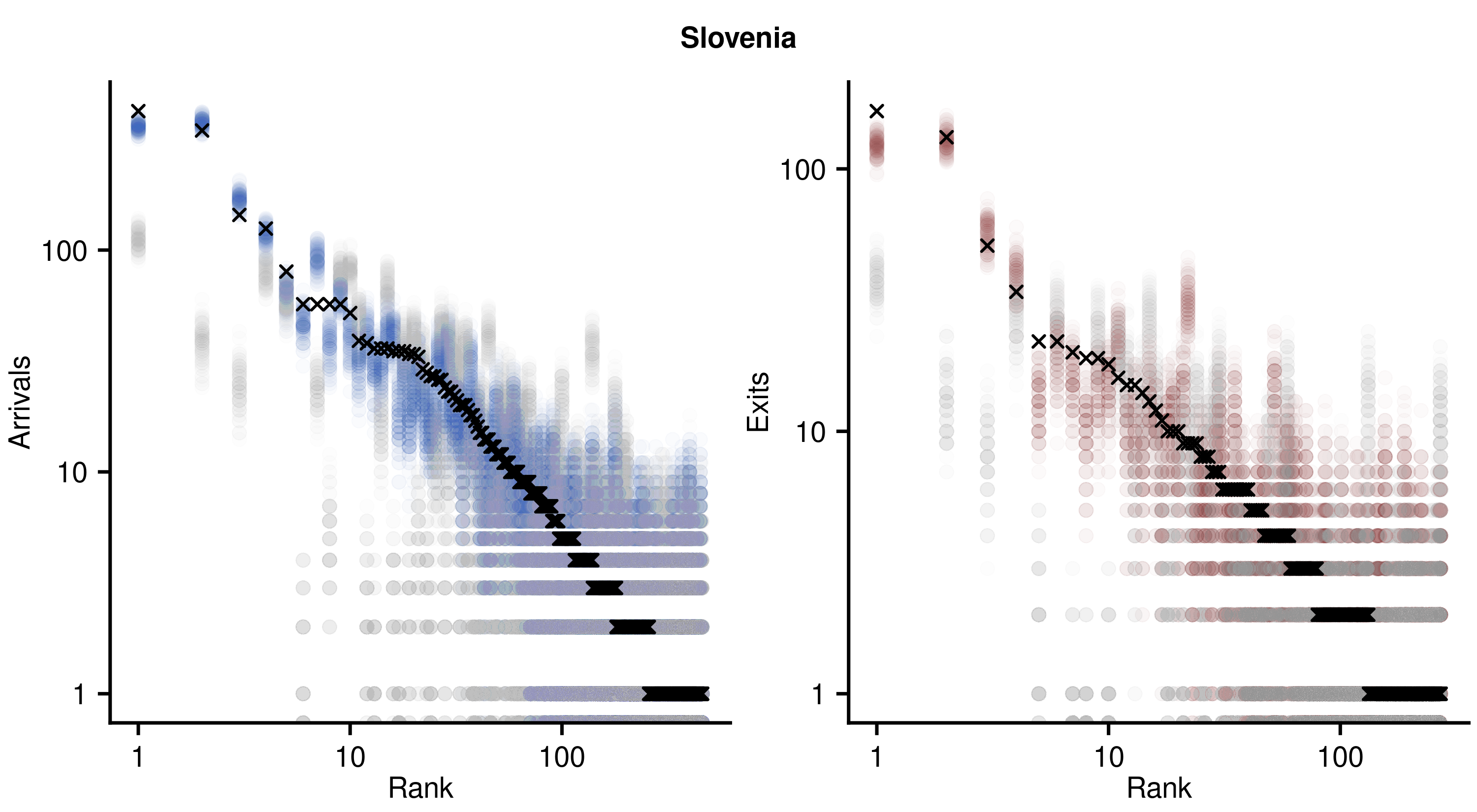} &
\includegraphics[width=0.48\textwidth]{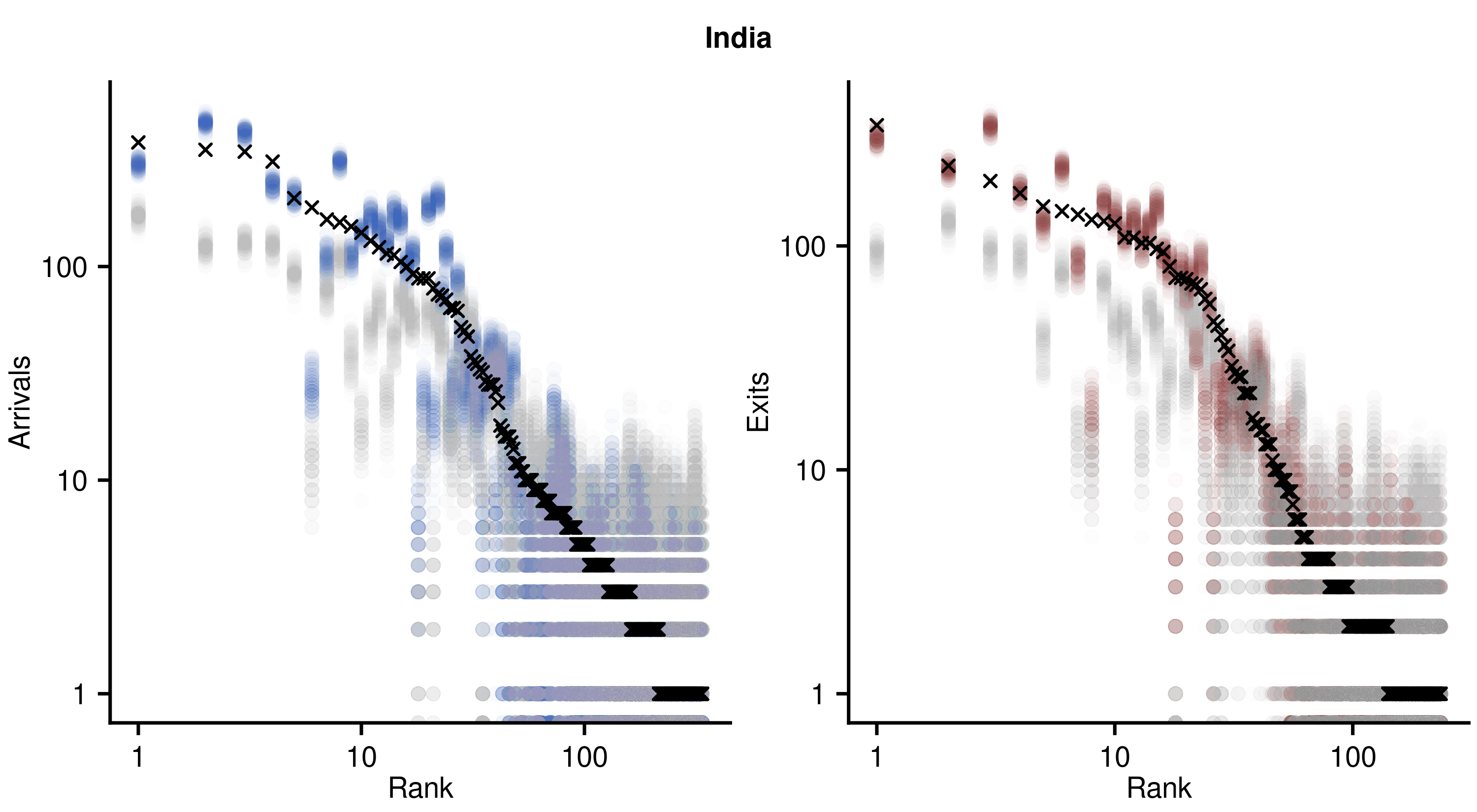} \\

\includegraphics[width=0.48\textwidth]{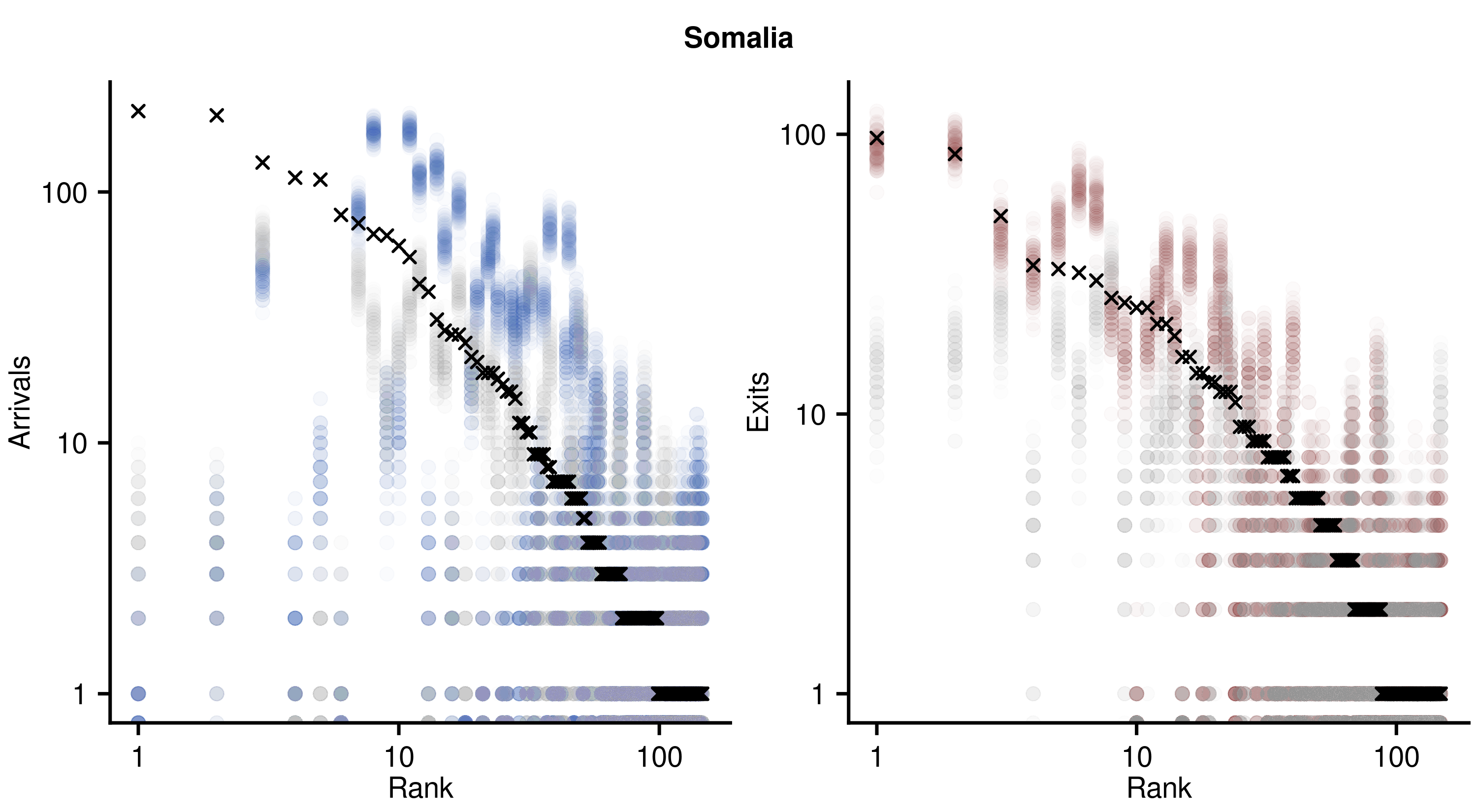} &
\includegraphics[width=0.48\textwidth]{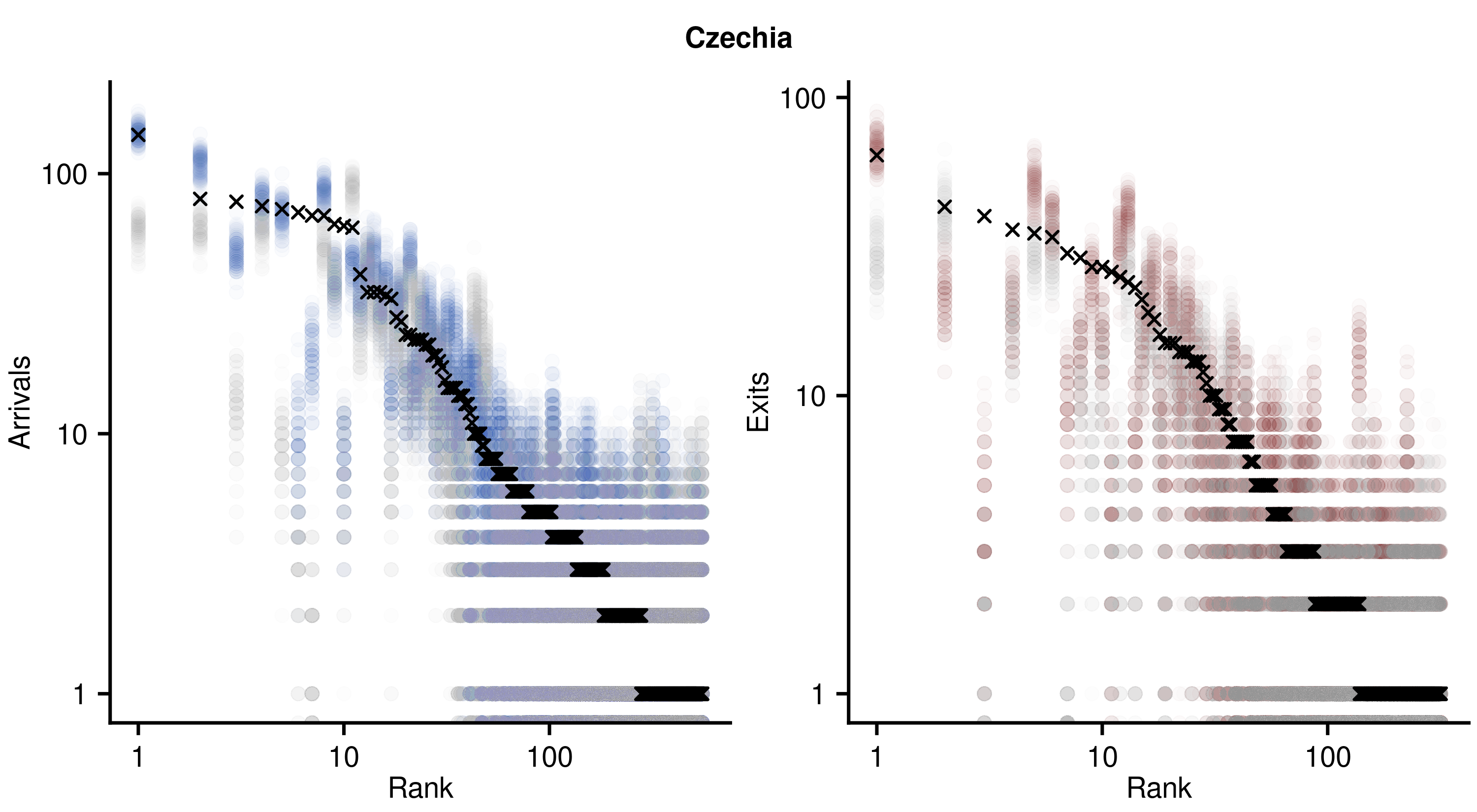} \\

\includegraphics[width=0.48\textwidth]{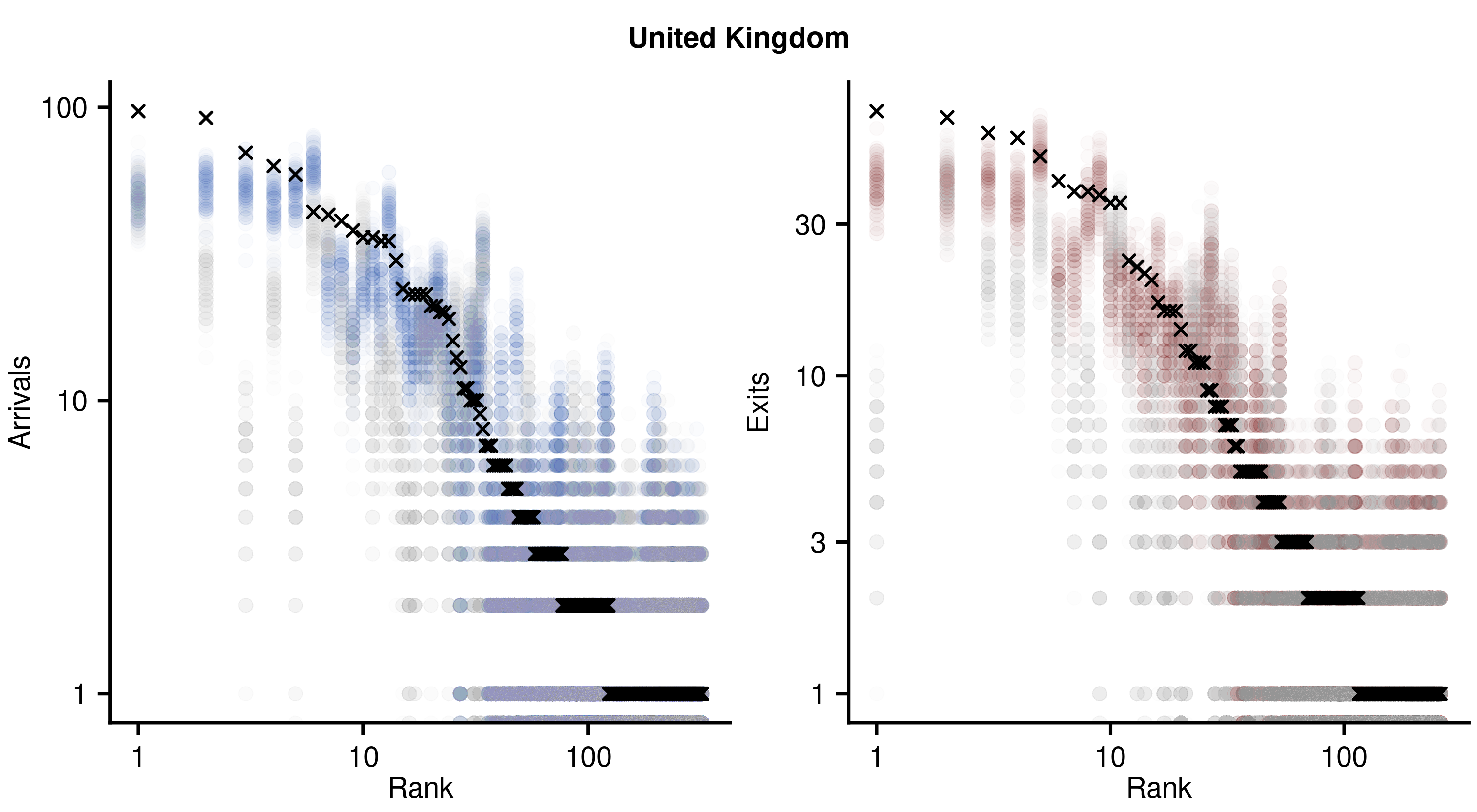} &
\\

\end{longtable}
}

\subsubsection{By Age-Cohort}
%%% INTRO
{
The table below presents the calculated daily allocation errors \( \varepsilon_i \) for each age cohort, comparing diaspora-based and population-based models for both arrivals and exits (see Table \ref{tab:allocation_errors_age}). These values quantify how closely the simulated flows match the observed municipal-level distributions across different life stages. Following the table, a series of diagnostic plots illustrates the allocation performance for each cohort. Each plot contrasts observed flows with those generated through 100 multinomial simulations based on population size (grey) and diaspora size (blue for arrivals, red for exits), providing a visual comparison of model accuracy across age groups.
}
%%% TABLE Age-Cohort
{
\begin{table}[H]
\centering
\renewcommand{\arraystretch}{1.25}
\begin{tabular}{|p{0.20\textwidth}|c|c|c|c|}
\hline
\textbf{Age-Cohort} 
& \multicolumn{2}{c|}{\textbf{Diaspora}} 
& \multicolumn{2}{c|}{\textbf{Population}} \\ \cline{2-5}
& \boldmath$\varepsilon_{i, \text{arrivals}}$ 
& \boldmath$\varepsilon_{i, \text{exits}}$ 
& \boldmath$\varepsilon_{i, \text{arrivals}}$ 
& \boldmath$\varepsilon_{i, \text{exits}}$ \\ \hline \hline
Children & 0.00361 & 0.00278 & 0.00884 & 0.00581 \\ \hline
Young Adults & 0.00598 & 0.00627 & 0.01061 & 0.00828 \\ \hline
Middle-Aged Adults & 0.00353 & 0.00381 & 0.00561 & 0.00510 \\ \hline
Older Adults & 0.00144 & 0.00148 & 0.00477 & 0.00479 \\ \hline
Seniors & 0.00066 & 0.00181 & 0.00835 & 0.02300 \\ \hline
\end{tabular}
\vspace{5pt}
\caption{Daily error \( \varepsilon_i \) in arrival and exit flow allocations for diaspora- and population-based models, by age cohort. Errors reflect the average deviation per municipality per day.}
\label{tab:allocation_errors_age}
\end{table}
}

%%% PLOTS Age-Cohort
{
\renewcommand{\arraystretch}{1.0}
\setlength{\tabcolsep}{4pt}
\begin{longtable}{ccc}
\caption{\textbf{Sub-National Allocation of Migration Flows by Age-Cohort:} Comparison of population size (grey) versus diaspora size for arrivals (blue) and exits (red) as determinant for migration flows on the municipal level. In each of the 100 bootstrap iterations, we draw samples from a multinomial distribution, using the municipality-level shares of the total population and the total diaspora as parameters. These shares define the probability vector for allocating individuals across municipalities within each group.} \label{fig:flow_allocation_age} \\
\endfirsthead

\multicolumn{3}{r}%
{{\bfseries \tablename\ \thetable{} -- continued from previous page}} \\
\\
\endhead
\includegraphics[width=0.48\textwidth]{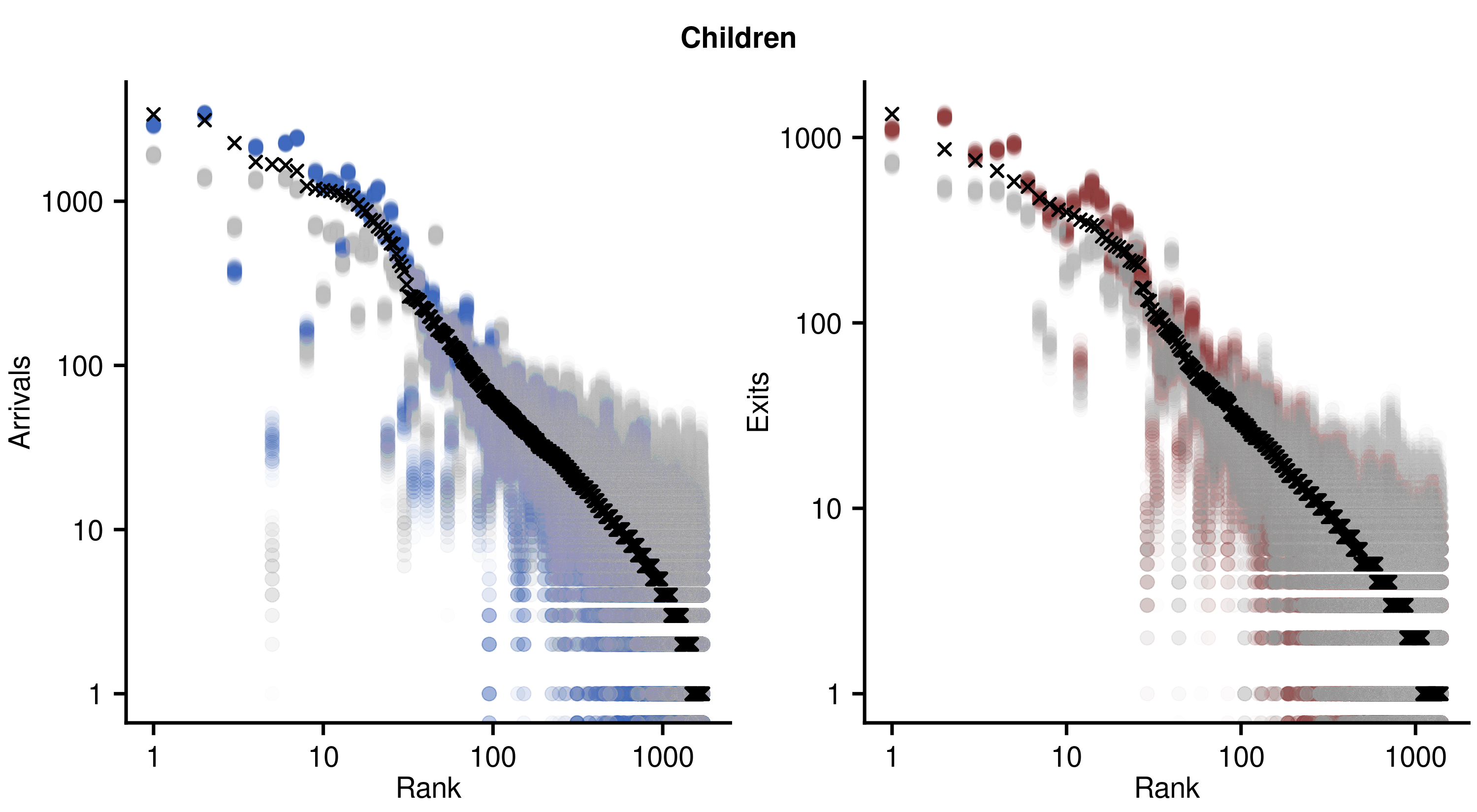} &
\includegraphics[width=0.48\textwidth]{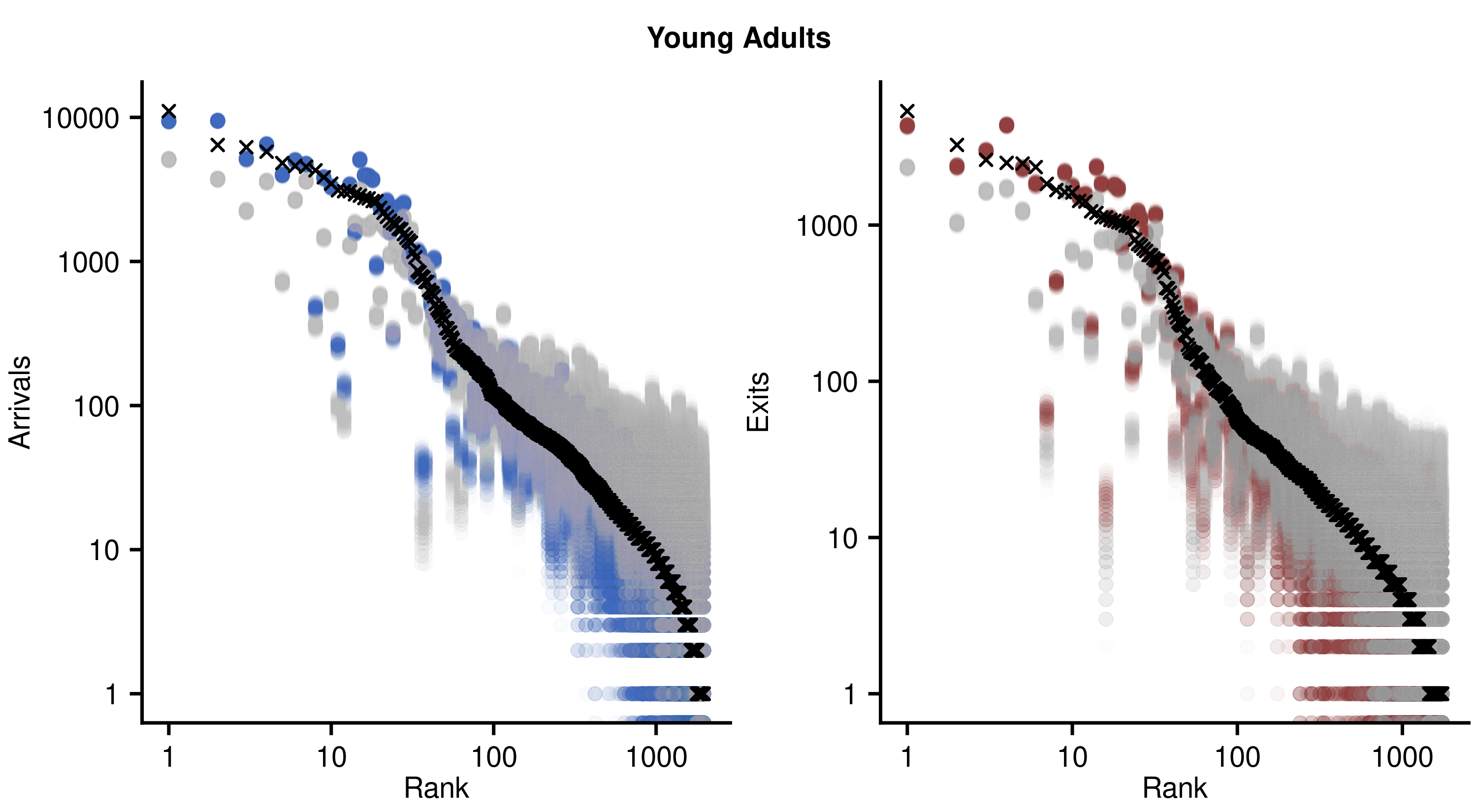} \\

\includegraphics[width=0.48\textwidth]{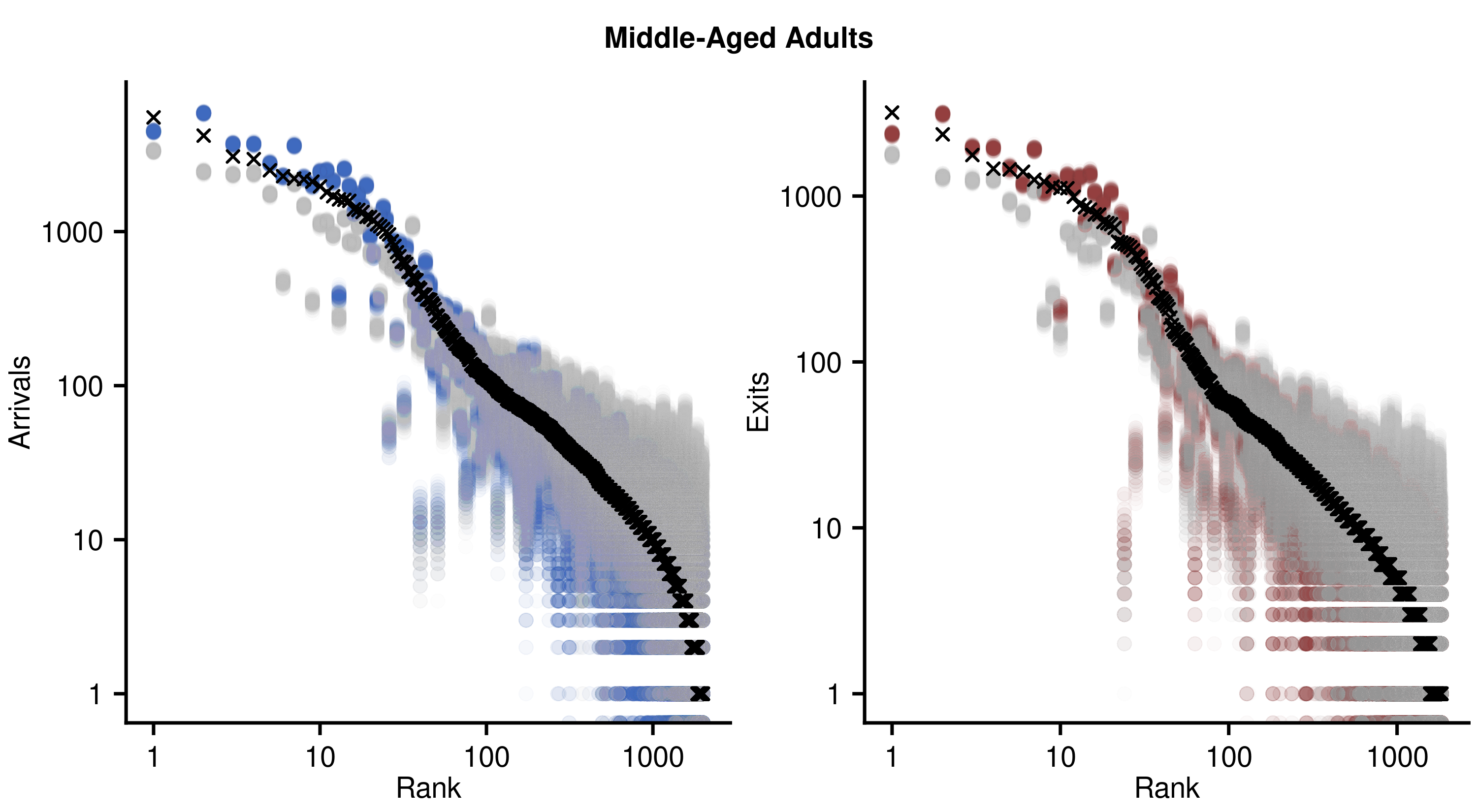} &
\includegraphics[width=0.48\textwidth]{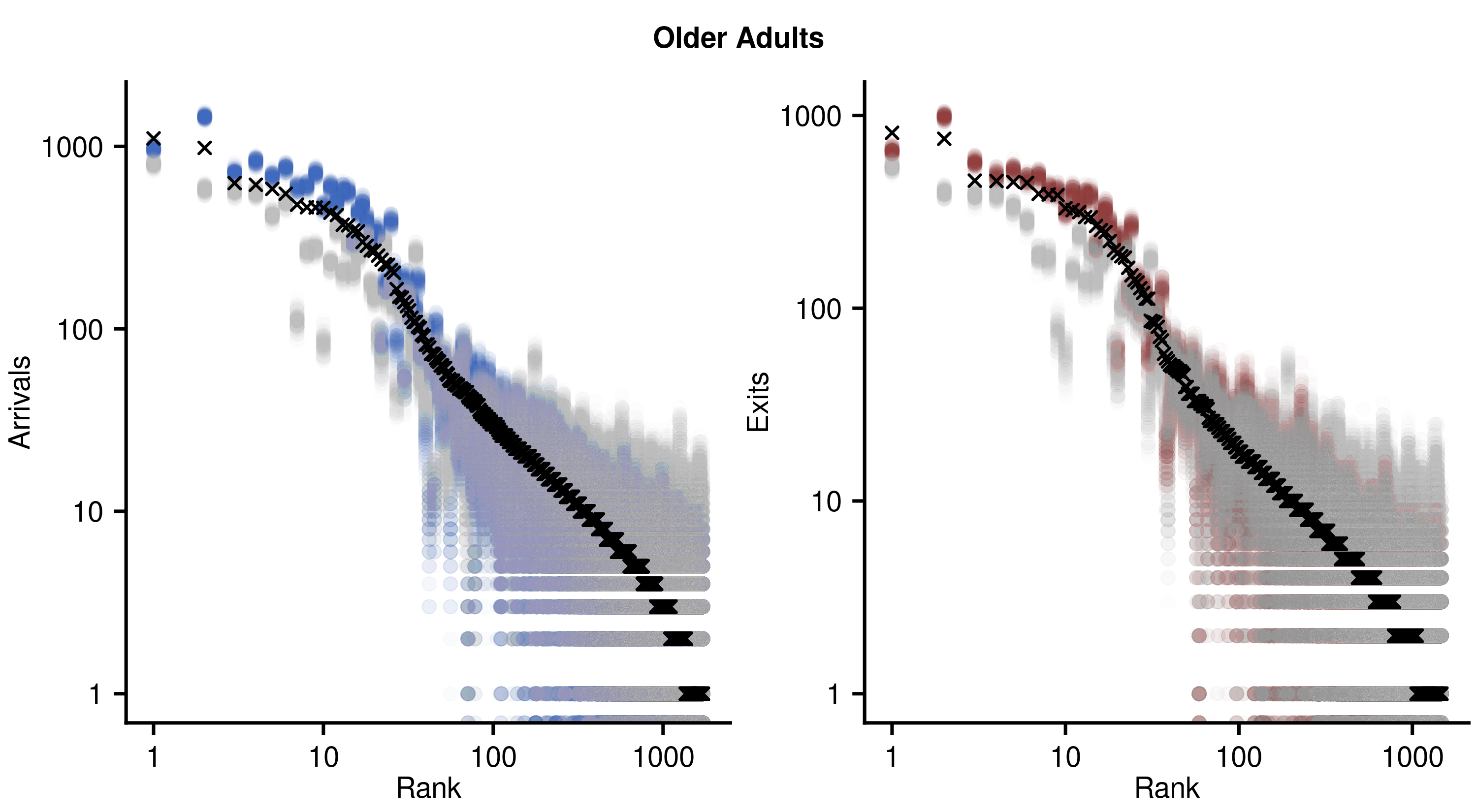} \\

\includegraphics[width=0.48\textwidth]{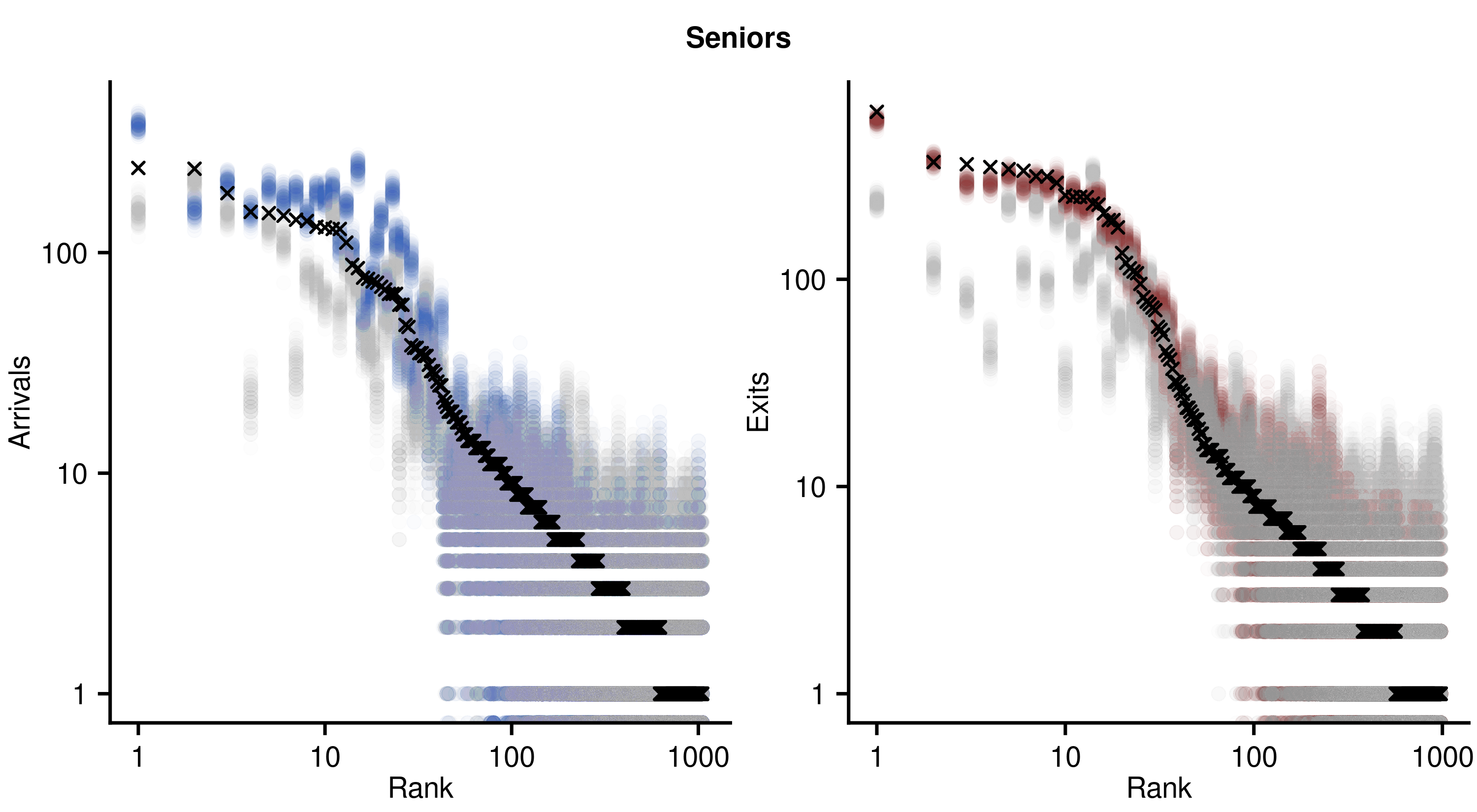} \\

\end{longtable}
}

\section{Diaspora Stability} \label{appendix:diaspora_stability}
%%% Diapsora stability by calculating years until 50% of replacement
{
When a fixed percentage of a diaspora $D_i$ exits a host country each year, due to return or onwards migration, the process can be modelled as an exponential decay of the diaspora population. Let $r$ represent the annual exit rate, defined as the proportion of the remaining diaspora that leaves in a given year. This assumes that the individual propensity to exit remains constant over time (i.e., each member of the diaspora faces the same probability of leaving regardless of how long they have stayed in the host country). The remaining diaspora population $D_i(t)$ after $t$ years can thus be modelled as: 
\[D_i(t) = D_{i,0}(1-r_i)^t,\]
where $D_{i,0}$ is the initial population in time $t = 0$ and $r$ is given by the percentage of migrants exiting in time $t$, given by $r_i=\tfrac{E_i(t)}{D_{i,0}}$. To calculate the time it takes for half the diaspora to exit the host country (i.e., the half-life $T_{i,1/2}$ of a diaspora), we solve for $t$ when $D_i(t)=\tfrac{1}{2}D_{i,0}$: 
\[\tfrac{1}{2} = (1-r)^{T_{1/2}}.\]
Taking the natural logarithm of both sides yields: 
\[T_{i,1/2}=\tfrac{\text{ln}(2)}{-\text{ln}(1-r_i)}.\]
Thus, \(T_{i,1/2}\) describes the number of years it would take for half of the initial diaspora population to leave the host country, assuming a constant individual exit probability over time.
}

\subsection{By Citizenship} \label{appendix:diaspora_stability_citizenship}
%%% INTRO
{
This section presents diaspora half-life estimates by citizenship, structured in two parts: a plot showing comparative values across groups (see Figure \ref{fig:half_life_citizenship}) and a table listing initial diaspora size, annual exit rates, and corresponding half-lives for the 25 largest diasporas (see Table \ref{tab:top25_diaspora_halflife}).
}

%%% PLOT
{
\begin{figure}[H]
\centering
    \includegraphics[width=0.5\textwidth]{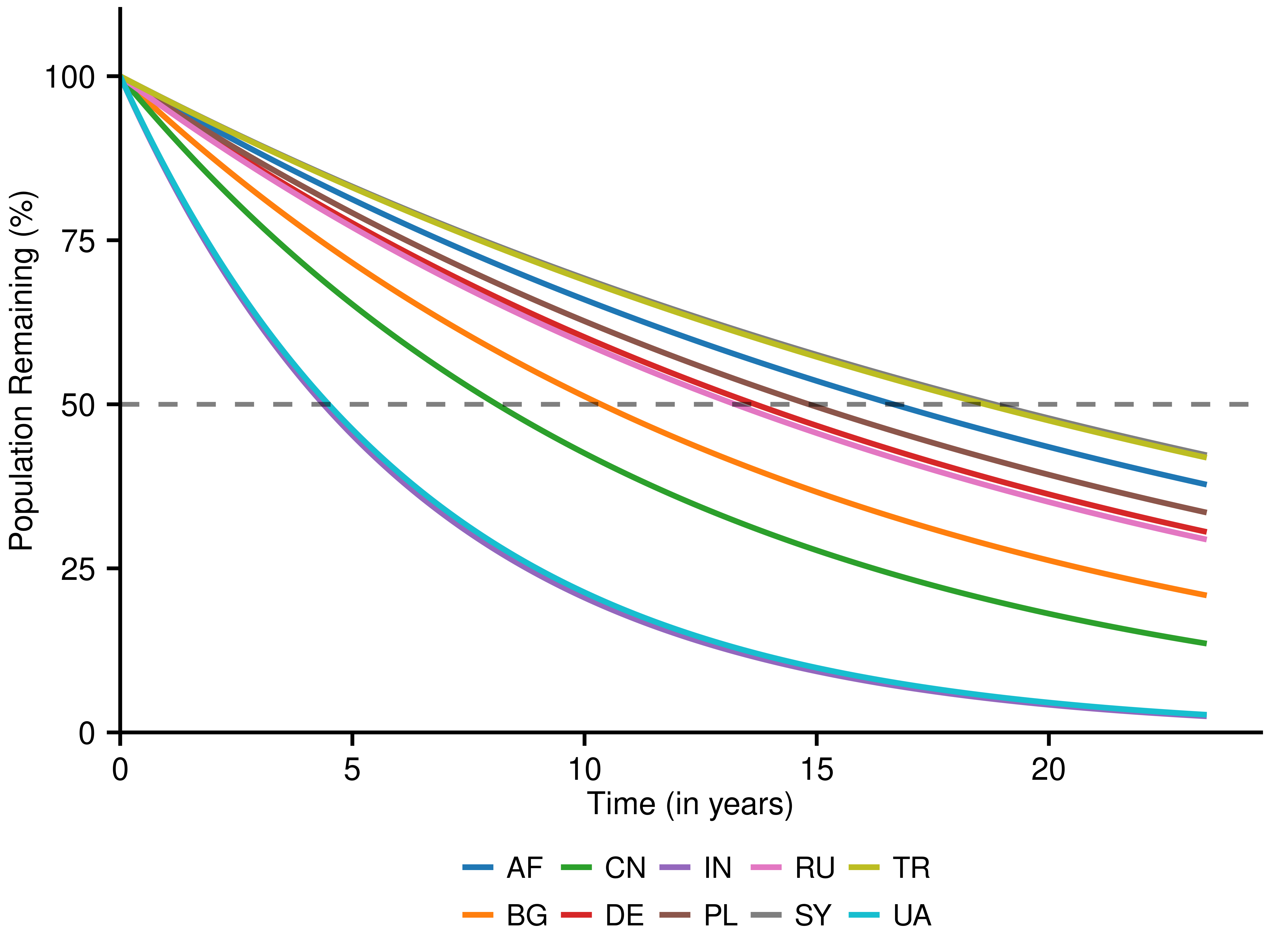}
    \caption{Half-life estimates by citizenship.}
    \label{fig:half_life_citizenship}
\end{figure}
}

%%% TABLE
{
\begin{table}[H]
\renewcommand{\arraystretch}{1.25} 
\centering
\begin{tabular}{| p{0.35\textwidth} | c | c | c |}
\hline
\textbf{Citizenship} & \boldmath$D_i$ & \boldmath$r_i$ & \boldmath$T_{i,1/2}$ \\ \hline \hline
Serbia & 143{,}348 & 0.0365 & 18.6320 \\ \hline
Turkey & 133{,}950 & 0.0372 & 18.2802 \\ \hline 
Germany & 114{,}615 & 0.0507 & 13.3272 \\ \hline 
Bosnia \& Herzegovina & 107{,}569 & 0.0276 & 24.7438 \\ \hline 
Romania & 99{,}683 & 0.0650 & 10.3165 \\ \hline 
Croatia & 88{,}239 & 0.0285 & 24.0010 \\ \hline 
Syria & 81{,}401 & 0.0368 & 18.4722 \\ \hline 
Ukraine & 76{,}159 & 0.1545 & 4.1306 \\ \hline 
Hungary & 63{,}355 & 0.0613 & 10.9589 \\ \hline 
Afghanistan & 51{,}421 & 0.0416 & 16.3035 \\ \hline 
Poland & 41{,}774 & 0.0467 & 14.4821 \\ \hline 
Russia & 40{,}400 & 0.0523 & 12.8948 \\ \hline 
Slovakia & 31{,}417 & 0.0554 & 12.1576 \\ \hline 
Bulgaria & 27{,}329 & 0.0669 & 10.0038 \\ \hline 
Kosovo & 26{,}017 & 0.0471 & 14.3790 \\ \hline 
North Macedonia & 25{,}739 & 0.0316 & 21.5677 \\ \hline 
Italy & 21{,}089 & 0.0703 & 9.5058 \\ \hline 
Iran & 17{,}720 & 0.0803 & 8.2811 \\ \hline 
Iraq & 14{,}594 & 0.0653 & 10.2632 \\ \hline 
China & 14{,}419 & 0.0855 & 7.7583 \\ \hline 
Slovenia & 13{,}890 & 0.0406 & 16.7052 \\ \hline 
India & 13{,}056 & 0.1583 & 4.0219 \\ \hline 
Somalia & 10{,}235 & 0.0457 & 14.8151 \\ \hline 
Czechia & 9{,}721 & 0.0618 & 10.8644 \\ \hline 
United Kingdom & 8{,}701 & 0.0663 & 10.1069 \\ \hline
\end{tabular}
\vspace{5pt}
\caption{Initial diaspora size $D_i$, exit rate $r_i$, and corresponding population half-life $T_{i,1/2}$ (in years) for the top 25 diasporas as of 1 January 2023. Values are based on revised daily exit rates and exponential decay assumptions.}
\label{tab:top25_diaspora_halflife}
\end{table}
}

\subsection{By Age-Cohort} \label{appendix:diaspora_stability_age}
%%% INTRO
{
This section presents half-life estimates for migrant populations by age-cohort. It includes a plot comparing half-life values across age-cohorts (see Figure \ref{fig:half_life_age}) and a table detailing initial diaspora sizes, annual exit rates, and the resulting half-life for each group (see Table \ref{tab:diaspora_by_age_halflife_updated}).
}

%%% PLOT
{
\begin{figure}[H]
    \centering
    \includegraphics[width=0.5\textwidth]{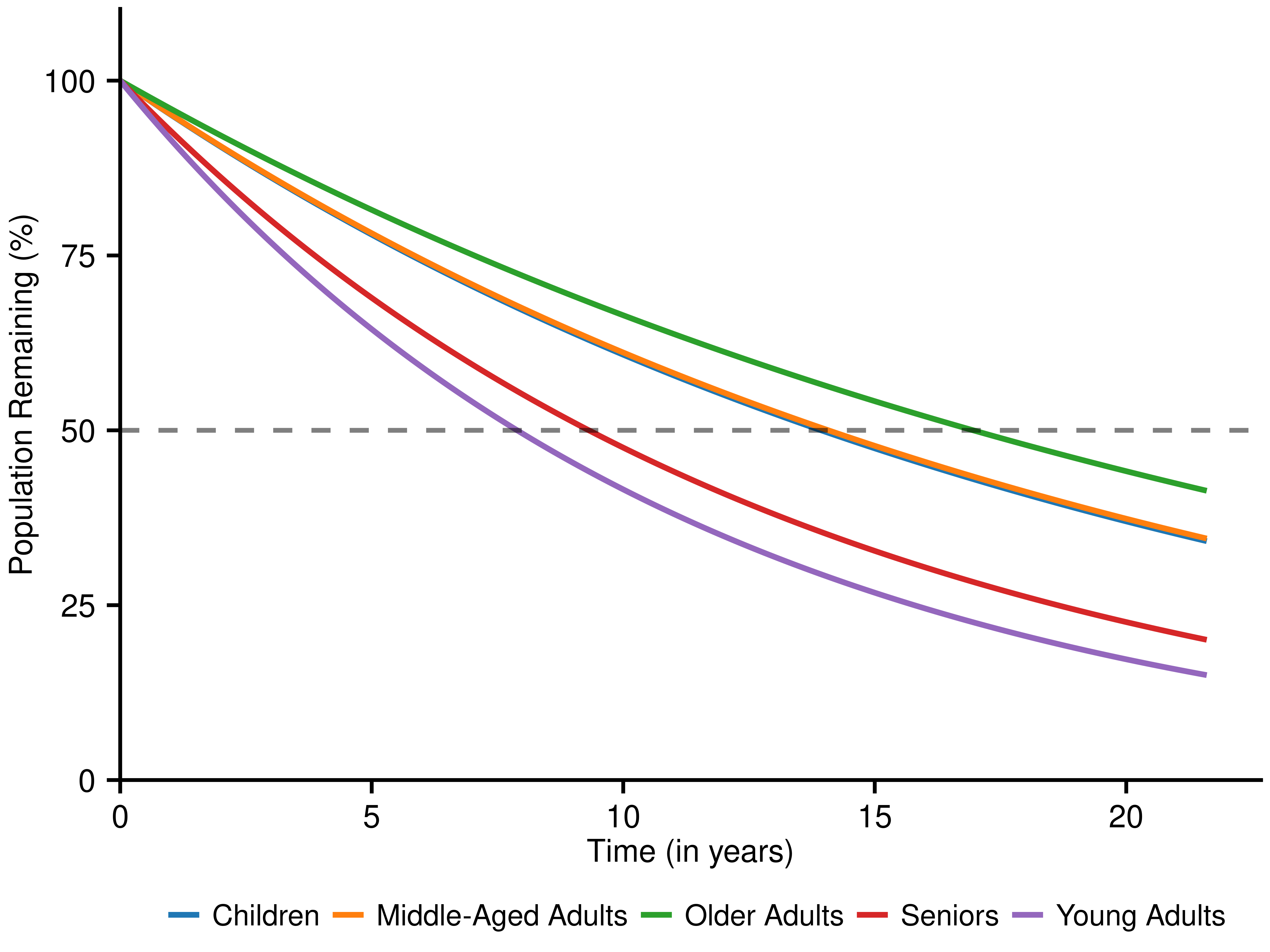}
    \caption{Half-life estimates by age-cohort.}
    \label{fig:half_life_age} 
\end
{figure}
}

{
\begin{table}[H]
\renewcommand{\arraystretch}{1.25} 
\centering
\begin{tabular}{| p{0.35\textwidth} | c | c | c |}
\hline
\textbf{Age Cohort} & \boldmath$D_i$ & \boldmath$r_i$ & \boldmath$T_{i, 1/2}$ \\ \hline \hline
Children & 224{,}409 & 0.0497 & 13.5867 \\ \hline
Young Adults & 406{,}986 & 0.0878 & 7.5402 \\ \hline
Middle-Aged Adults & 538{,}459 & 0.0492 & 13.7303 \\ \hline
Older Adults & 194{,}211 & 0.0409 & 16.6093 \\ \hline
Seniors & 69{,}531 & 0.0744 & 8.9661 \\ \hline
\end{tabular}
\vspace{5pt}
\caption{Initial diaspora size $D_i$, estimated exit rate $r_i$, and population half-life $T_{i, 1/2}$ (in years) by age cohort. Exit rates are based on observed daily exits and assume a constant exponential process.}
\label{tab:diaspora_by_age_halflife_updated}
\end{table}
}

\subsection{By Degree of Urbanisation} \label{appendix:diaspora_stability_urbanisation}
%%% INTRO
{
This section summarises diaspora half-life estimates across urban, intermediate, and rural areas. It includes a comparative plot of half-life values by degree of urbanisation (see Figure \ref{fig:half_life_urbanisation}) and a table listing the corresponding diaspora sizes, exit rates, and half-lives (see Table \ref{tab:diaspora_by_urbanisation}).
}
%%% PLOT
{
\begin{figure}[H]
    \centering
    \includegraphics[width=0.5\textwidth]{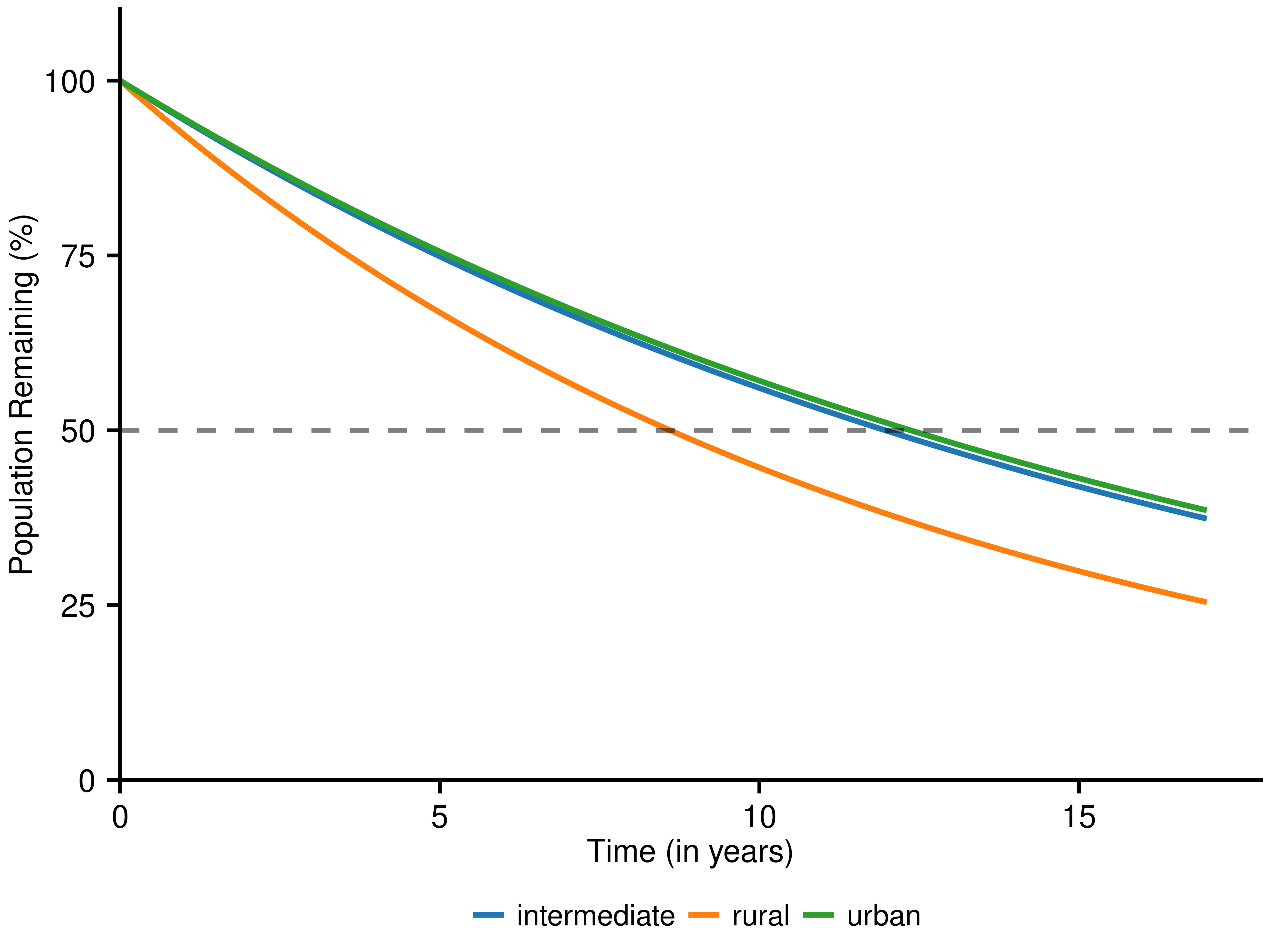}
    \caption{Half-life estimates by degree of urbanisation.}
    \label{fig:half_life_urbanisation}
\end{figure}
}

%%% TABLE
{
\begin{table}[H]
\renewcommand{\arraystretch}{1.25}
\centering
\begin{tabular}{|l|c|c|c|}
\hline
\textbf{Urbanisation Level} & \boldmath$D_i$ & \boldmath$r_i$ & \boldmath$T_{1/2}$ \\ \hline \hline
Urban & 811{,}437 & 0.0560 & 12.0167 \\ \hline
Intermediate & 400{,}680 & 0.0579 & 11.6208 \\ \hline
Rural & 221{,}479 & 0.0806 & 8.2520 \\ \hline
\end{tabular}
\vspace{5pt}
\caption{Initial diaspora size $D_i$, estimated exit rate $r_i$, and population half-life $T_{1/2}$ (in years) by urbanisation level. Values based on data from 2023 to 2024 and assume a constant exponential process.}
\label{tab:diaspora_by_urbanisation}
\end{table}
}

\subsection{By State and Degree of Urbanisation}
%%% INTRO
{
The table below provides half-life estimates by Austrian state, disaggregated by degree of urbanisation. It reports the initial diaspora size, estimated annual exit rate, and resulting half-life for each state–urban category combination (see Table \ref{tab:diaspora_by_state_urbanisation}).
}
%%% TABLE
{
\begin{table}[H]
\centering
\renewcommand{\arraystretch}{1.25}
\begin{tabular}{|l|l|c|c|c|}
\hline
\textbf{State} & \textbf{Urbanisation} & \boldmath$D_i$ & \boldmath$r_i$ & \boldmath$T_{1/2}$ \\ \hline \hline
Vienna & urban & 591{,}056 & 0.0518 & 13.0333 \\ \hline

\multirow{3}{*}{Lower Austria}
    & urban        & --        & --      & --      \\
    & intermediate & 114{,}792 & 0.0701  & 9.5439 \\
    & rural        & 51{,}518  & 0.0647  & 10.3582 \\ \hline

\multirow{3}{*}{Upper Austria}
    & urban        & 53{,}477  & 0.0565  & 11.9127 \\
    & intermediate & 99{,}251  & 0.0493  & 13.7171 \\
    & rural        & 47{,}842  & 0.0977  & 6.7447 \\ \hline

\multirow{3}{*}{Styria}
    & urban        & 68{,}869  & 0.0773  & 8.6206 \\
    & intermediate & 40{,}210  & 0.0593  & 11.3346 \\
    & rural        & 26{,}950  & 0.0977  & 6.7415 \\ \hline

\multirow{3}{*}{Salzburg}
    & urban        & 40{,}912  & 0.0623  & 10.7822 \\
    & intermediate & 23{,}005  & 0.0555  & 12.1505 \\
    & rural        & 23{,}039  & 0.0914  & 7.2306 \\ \hline

\multirow{3}{*}{Tyrol}
    & urban        & 35{,}312  & 0.0770  & 8.6476 \\
    & intermediate & 42{,}152  & 0.0564  & 11.9475 \\
    & rural        & 27{,}868  & 0.0763  & 8.7289 \\ \hline

\multirow{3}{*}{Carinthia}
    & urban        & 21{,}811  & 0.0576  & 11.6748 \\
    & intermediate & 20{,}905  & 0.0554  & 12.1634 \\
    & rural        & 20{,}027  & 0.0615  & 10.9239 \\ \hline

\multirow{3}{*}{Vorarlberg}
    & urban        & --        & --      & --      \\
    & intermediate & 52{,}144  & 0.0467  & 14.4979 \\
    & rural        & 6{,}763   & 0.0779  & 8.5435 \\ \hline

\multirow{3}{*}{Burgenland}
    & urban        & --        & --      & --      \\
    & intermediate & 8{,}221   & 0.0778  & 8.5622 \\
    & rural        & 17{,}472  & 0.0693  & 9.6447 \\ \hline \hline
\multirow{3}{*}{\textbf{All Combined}}
    & \textbf{urban}        & \textbf{811{,}437} & \textbf{0.0560}  & \textbf{12.0167} \\
    & \textbf{intermediate} & \textbf{400{,}680} & \textbf{0.0579}  & \textbf{11.6208} \\
    & \textbf{rural}        & \textbf{221{,}479} & \textbf{0.0805}  & \textbf{8.2520} \\ \hline 
\end{tabular}
\vspace{5pt}
\caption{Initial diaspora size $D_i$, estimated exit rate $r_i$, and population half-life $T_{1/2}$ (in years) by Austrian state and urbanisation level, based on 2024 estimates.}
\label{tab:diaspora_by_state_urbanisation}
\end{table}
}
\section{Flow Stability} \label{appendix:flow_stability}
\subsection{Assessing Stability of Migration Flows}
%%% Scale invariant error measure
{
To evaluate the stability of arrival and exit flows, we compare the observed values of cumulative migration over time $t$, denoted as $M_i(t)$, with the modelled cumulative migration, represented by $\lambda_it$ and $\mu_it$ respectively. The error term $\varepsilon_i$ captures the discrepancy in the flow of subpopulation $i$, and is defined as a scale-invariant measure of deviation:
\[\varepsilon_i=\tfrac{\sum_{t=1}^T\sqrt{(M_i(t)-\lambda_it})^2}{M_i(t)}.\]
This formulation allows direct comparison of migration flows of varying sizes, as it normalises deviations by the total observed migration. Intuitively, a lower value of $\varepsilon_i$ indicates that the model closely approximates the observed migration pattern for subpopulation $i$, suggesting a more stable and predictable flow. The modelled migration flows represent the most stable form, since the Poisson model assumes a constant flow per time period $t$.
}

%%% PLOT Flow Errors Citizenship & Age
{
\begin{figure}[H]
    \centering
    \includegraphics[width=\textwidth]{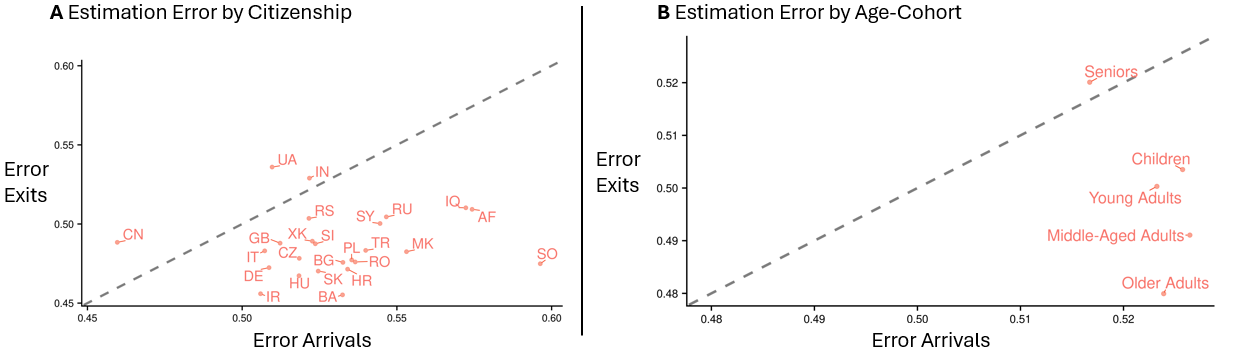}
    \caption{\textbf{Flow errors by (A) citizenship} of the top 25 diasporas in Austria and \textbf{(B) age-cohort} with Children (2-17), Young Adults (18-34), Middle-Aged Adults (35-54), Older Adults (55-69), and Seniors (70+). Analysis of both types of diaspora reveals the better predictability of exit flows compared to arrival flows due to lower errors across most diasporas.}
    \label{fig:flow_errors_combined}
\end{figure}
}

%%% Plots Flow Stability
\subsection{Temporal Distribution of Migration Flows}
%%% Explanation of plots and intention
{
This section presents plots illustrating the temporal distribution of migration flows. The x-axis denotes the fraction of time during the observation period, while the y-axis represents the cumulative fraction of arrivals and exits. This visualisation facilitates the identification of temporal patterns and the differences in flow dynamics between arrivals and exits. The first section lists the plots for the 25 largest diasporas in Austria, and the second section lists the age cohorts of Children (2-17), Young Adults (18-34), Middle-Aged Adults (35-54), Older Adults (55-69), and Seniors (70+).
}

\subsubsection{By Citizenship}
%%% PLOTS Citizenship
{
\renewcommand{\arraystretch}{1.0}
\setlength{\tabcolsep}{4pt}
\begin{longtable}{ccc}
\caption{\textbf{Flow Stability by Citizenship:} Cumulative distribution of arrivals (blue) and exits (red) over the observed period. The x-axis represents the fraction of time elapsed; the y-axis shows the cumulative share of migration. Dashed lines indicate a constant migration rate.} \label{fig:flow_stability_citizenship} \\
\endfirsthead

\multicolumn{3}{r}%
{{\bfseries \tablename\ \thetable{} -- continued from previous page}} \\
\\
\endhead

\includegraphics[width=0.31\textwidth]{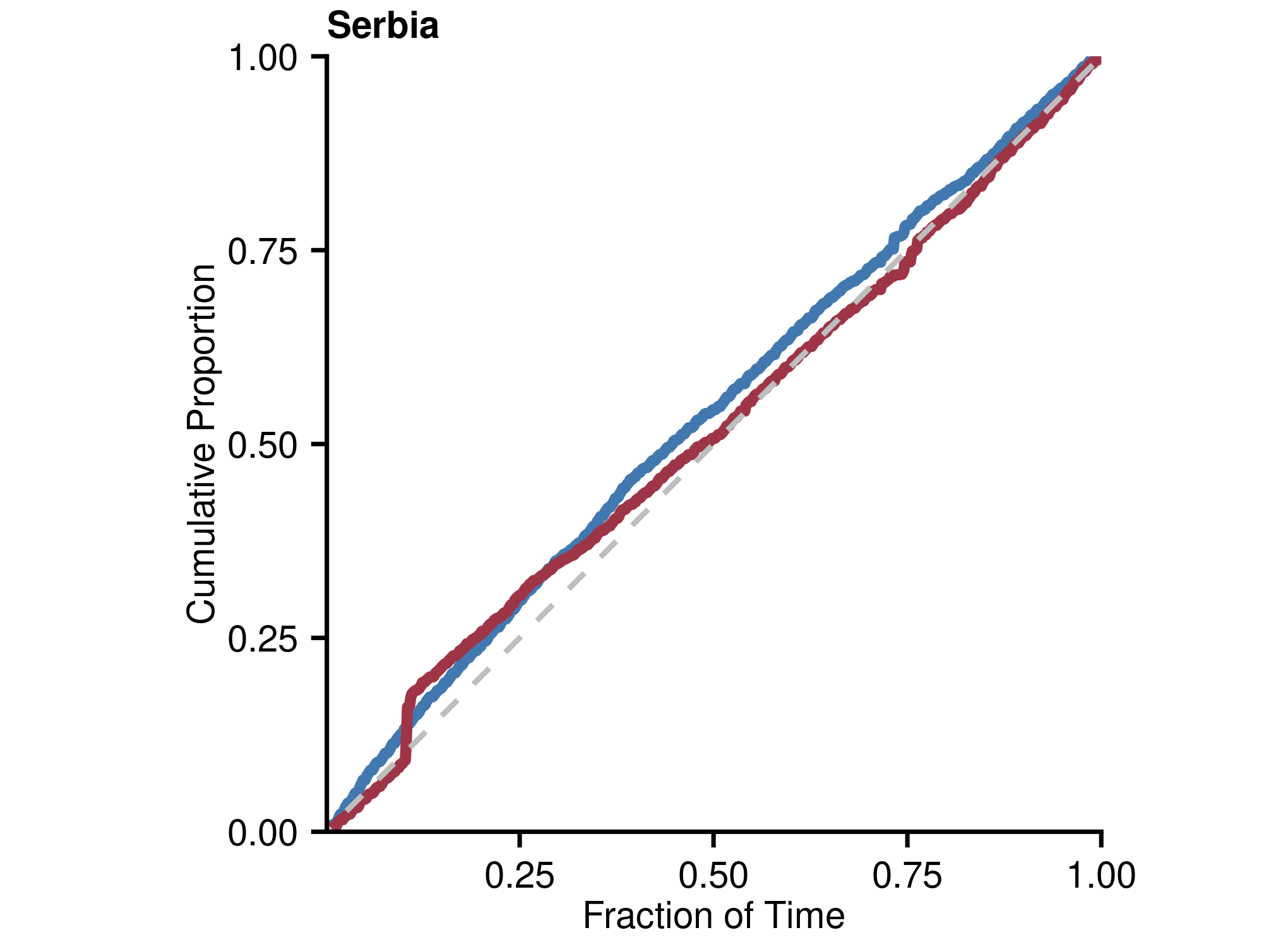} &
\includegraphics[width=0.31\textwidth]{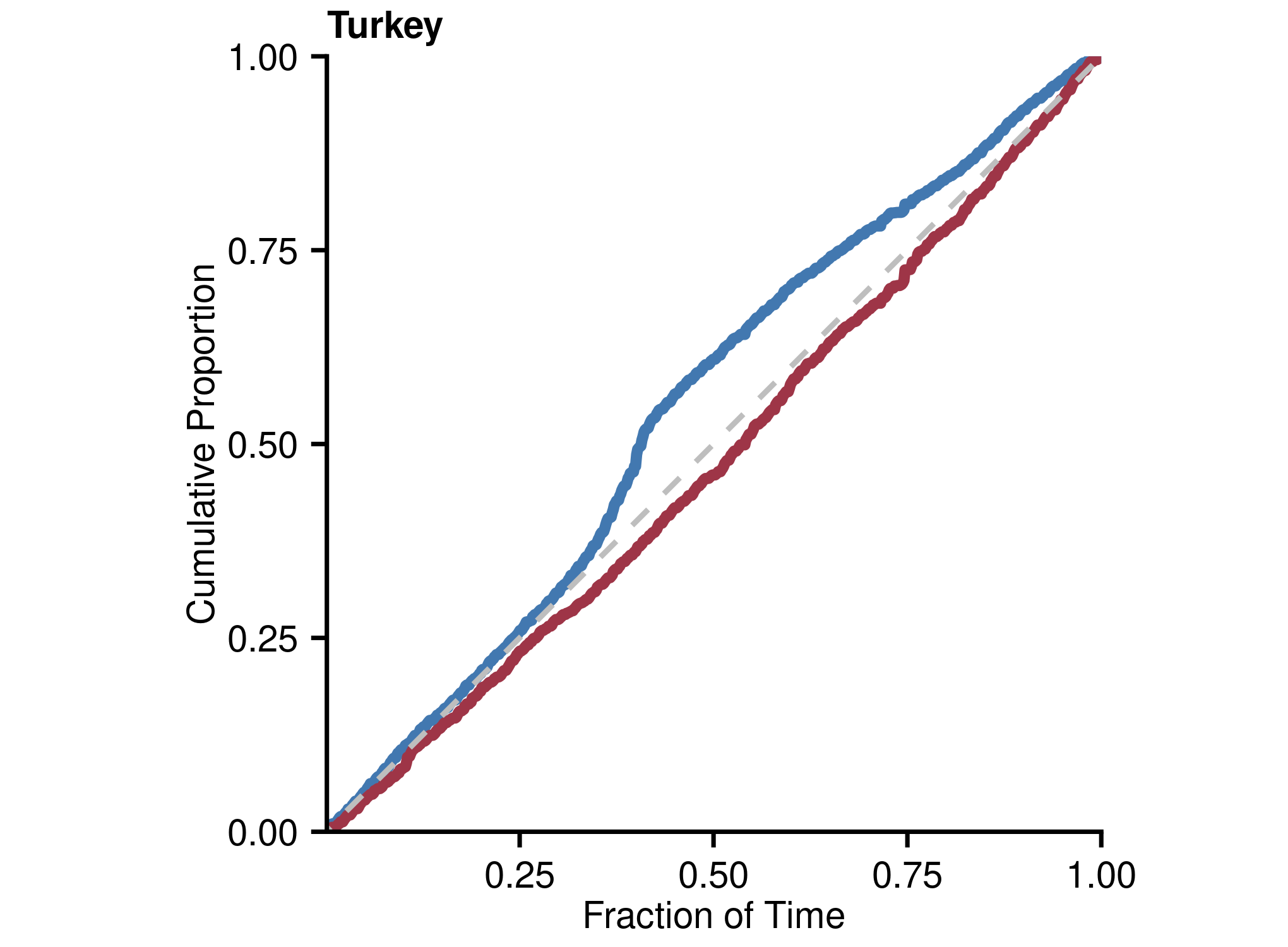} &
\includegraphics[width=0.31\textwidth]{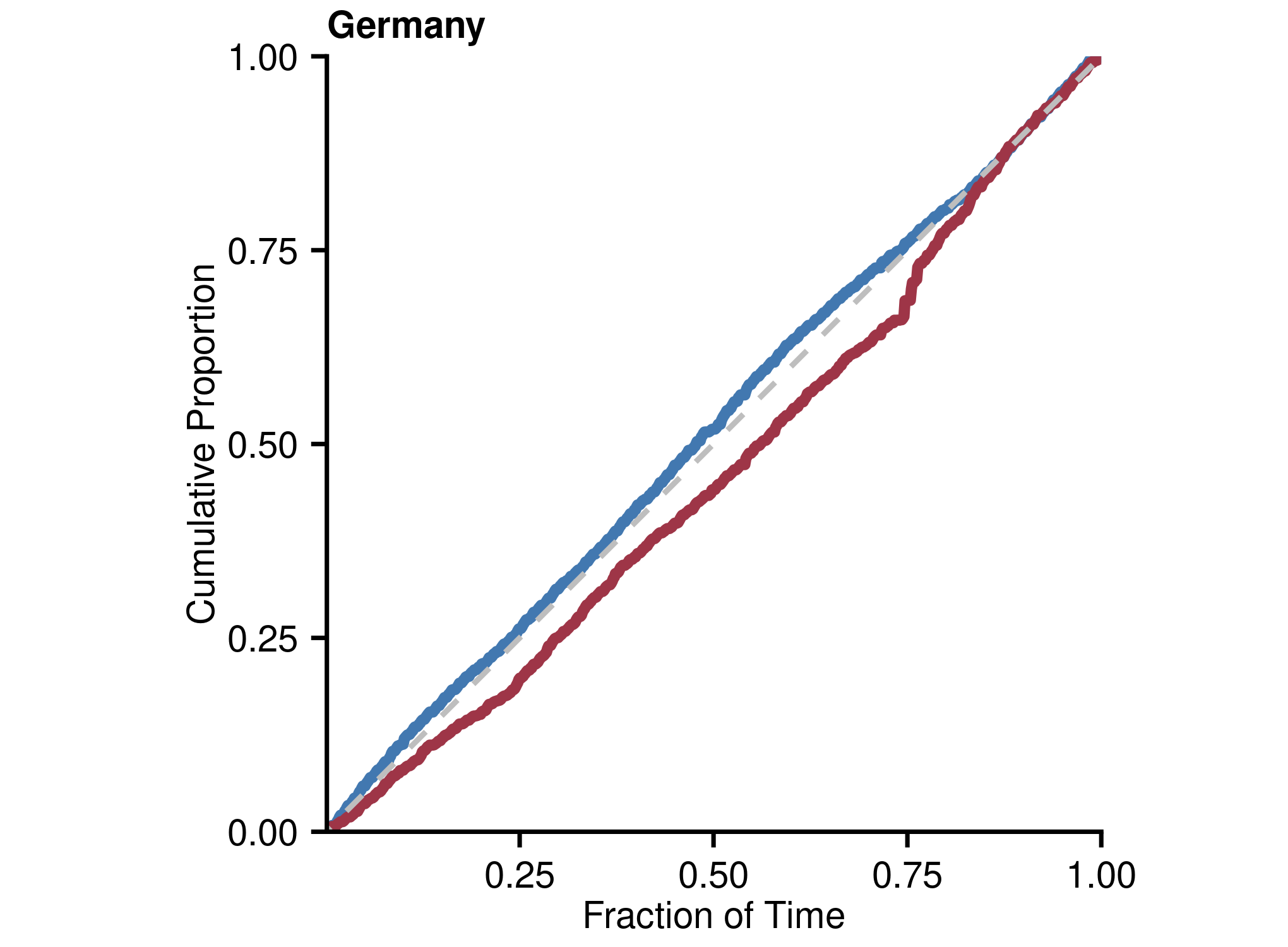} \\

\includegraphics[width=0.31\textwidth]{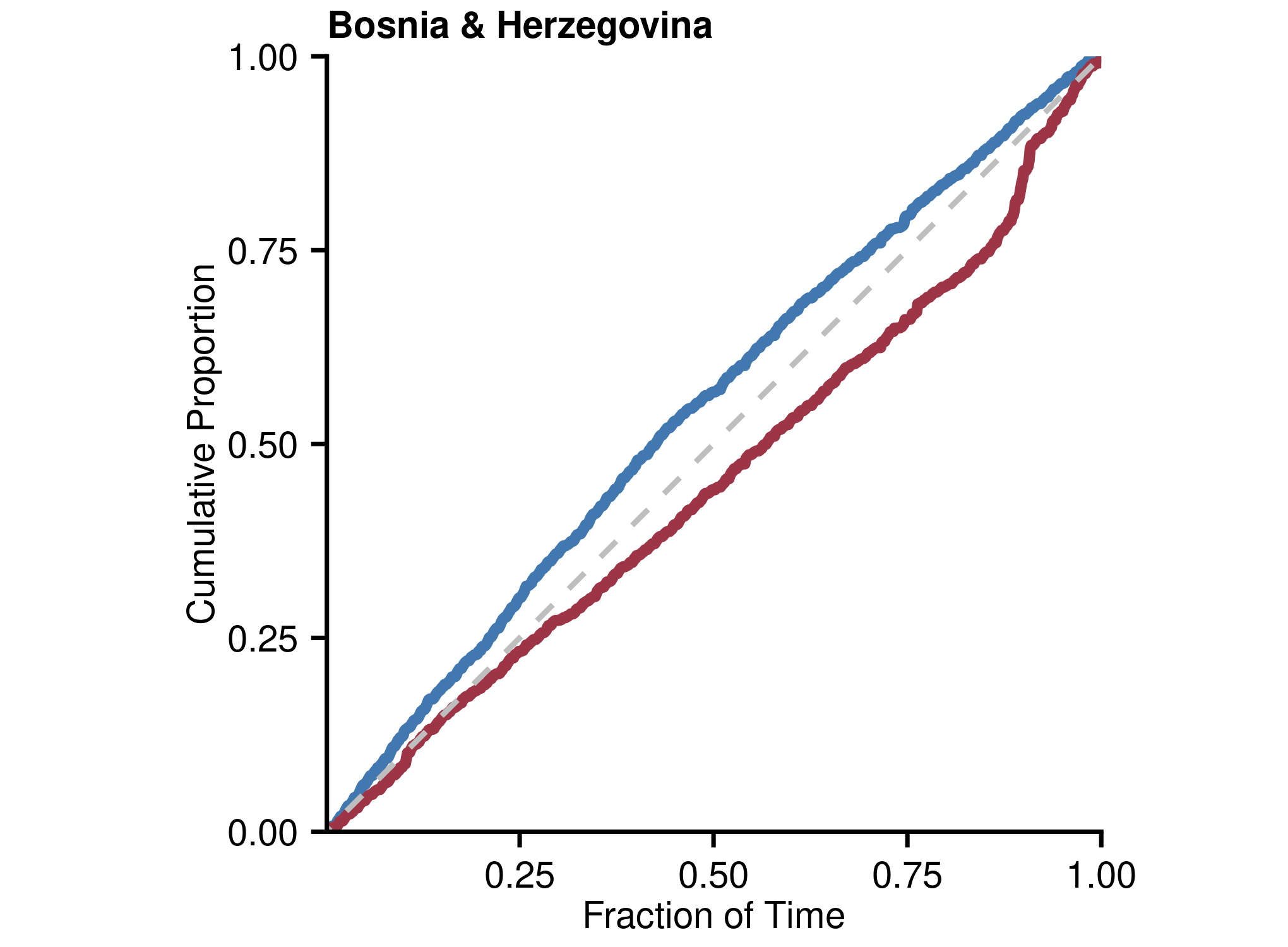} &
\includegraphics[width=0.31\textwidth]{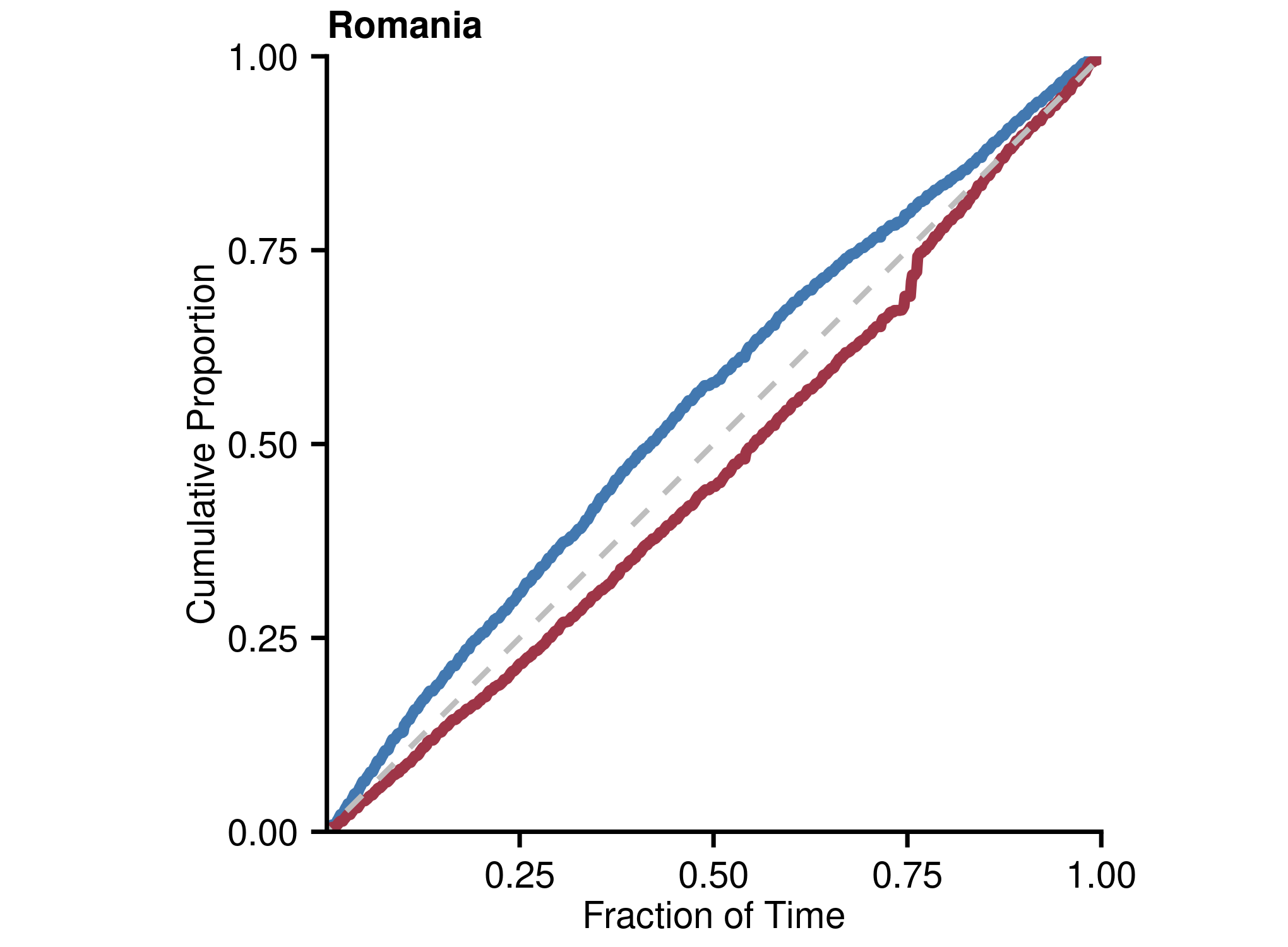} &
\includegraphics[width=0.31\textwidth]{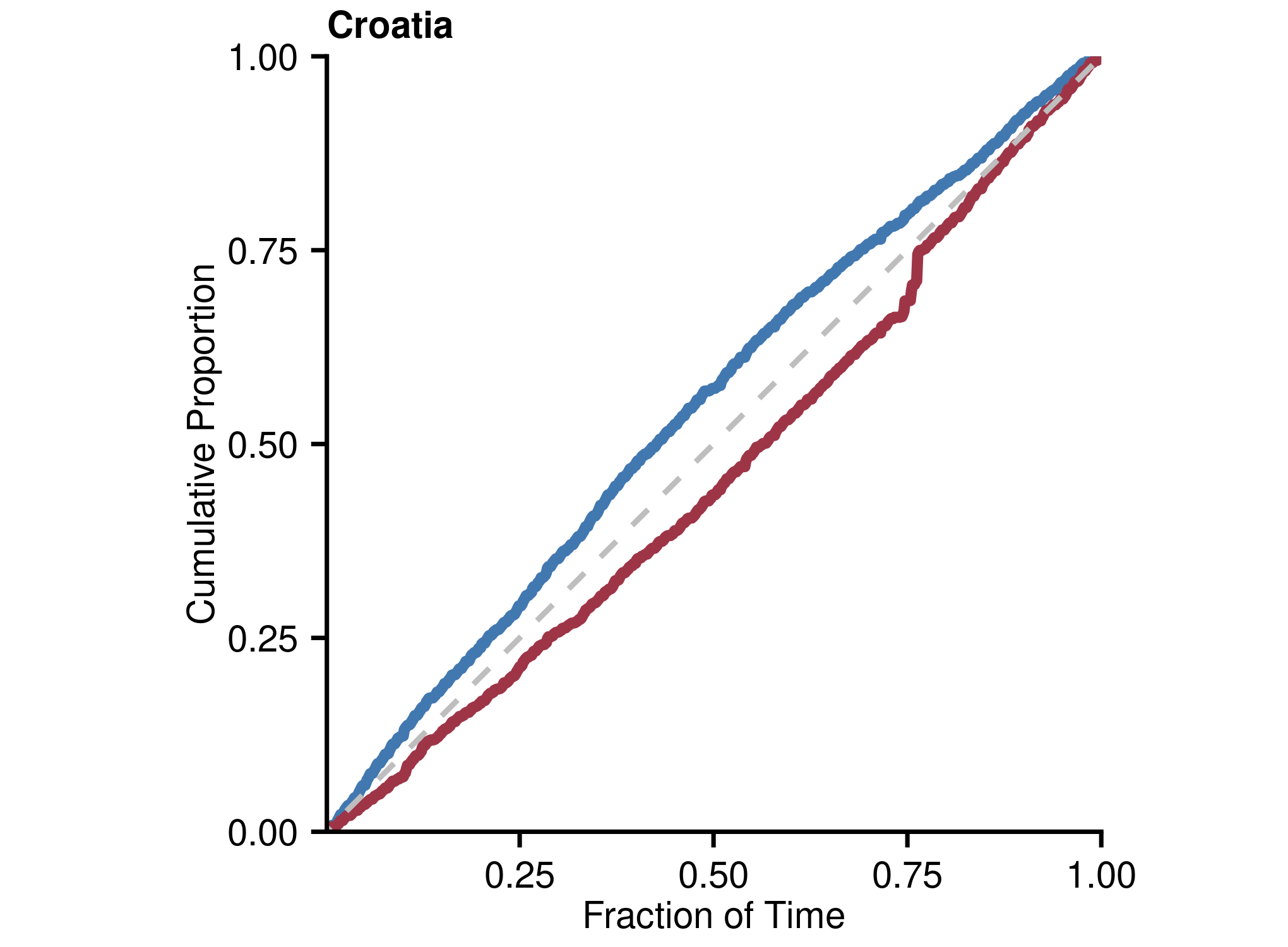} \\

\includegraphics[width=0.31\textwidth]{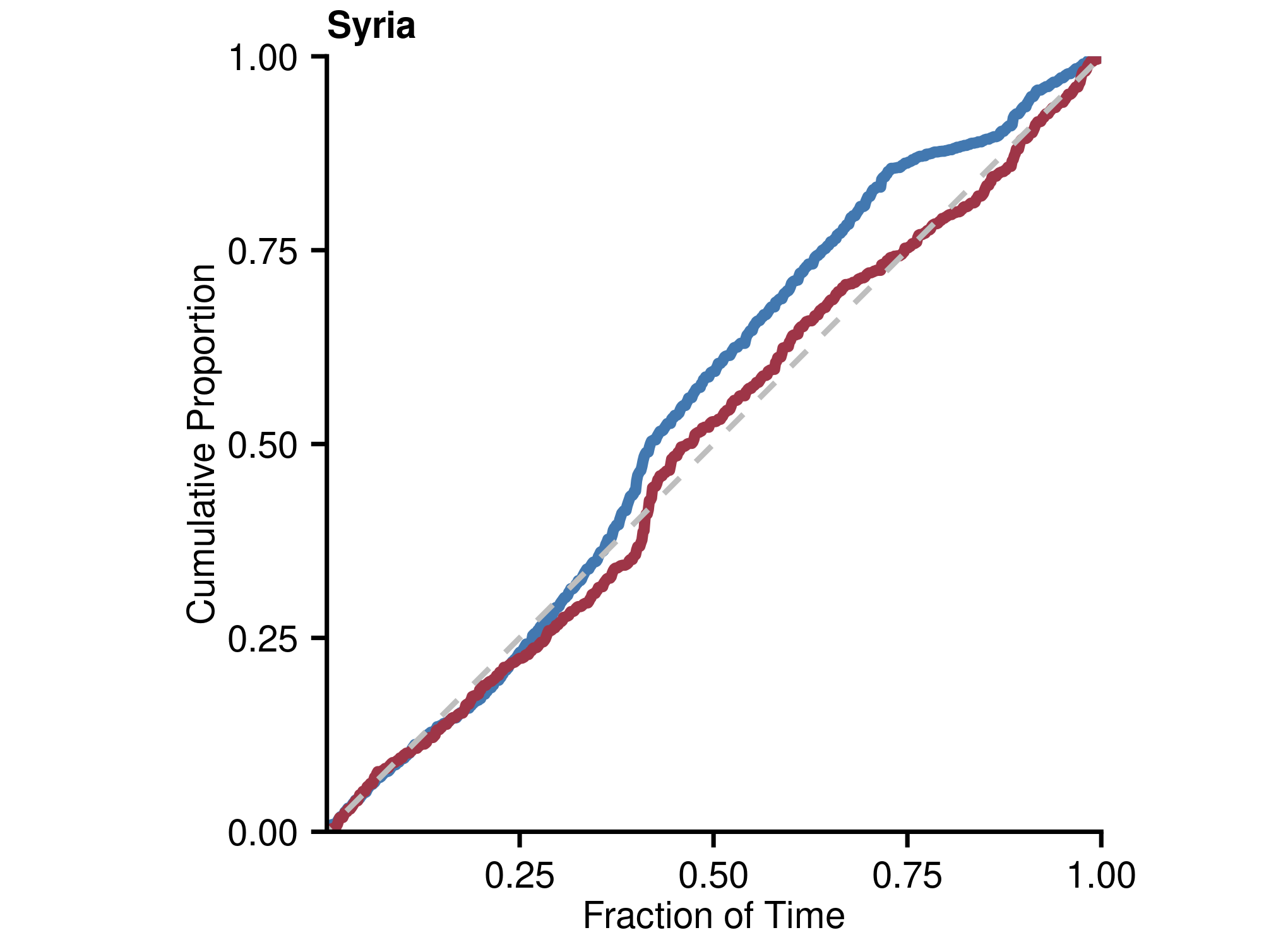} &
\includegraphics[width=0.31\textwidth]{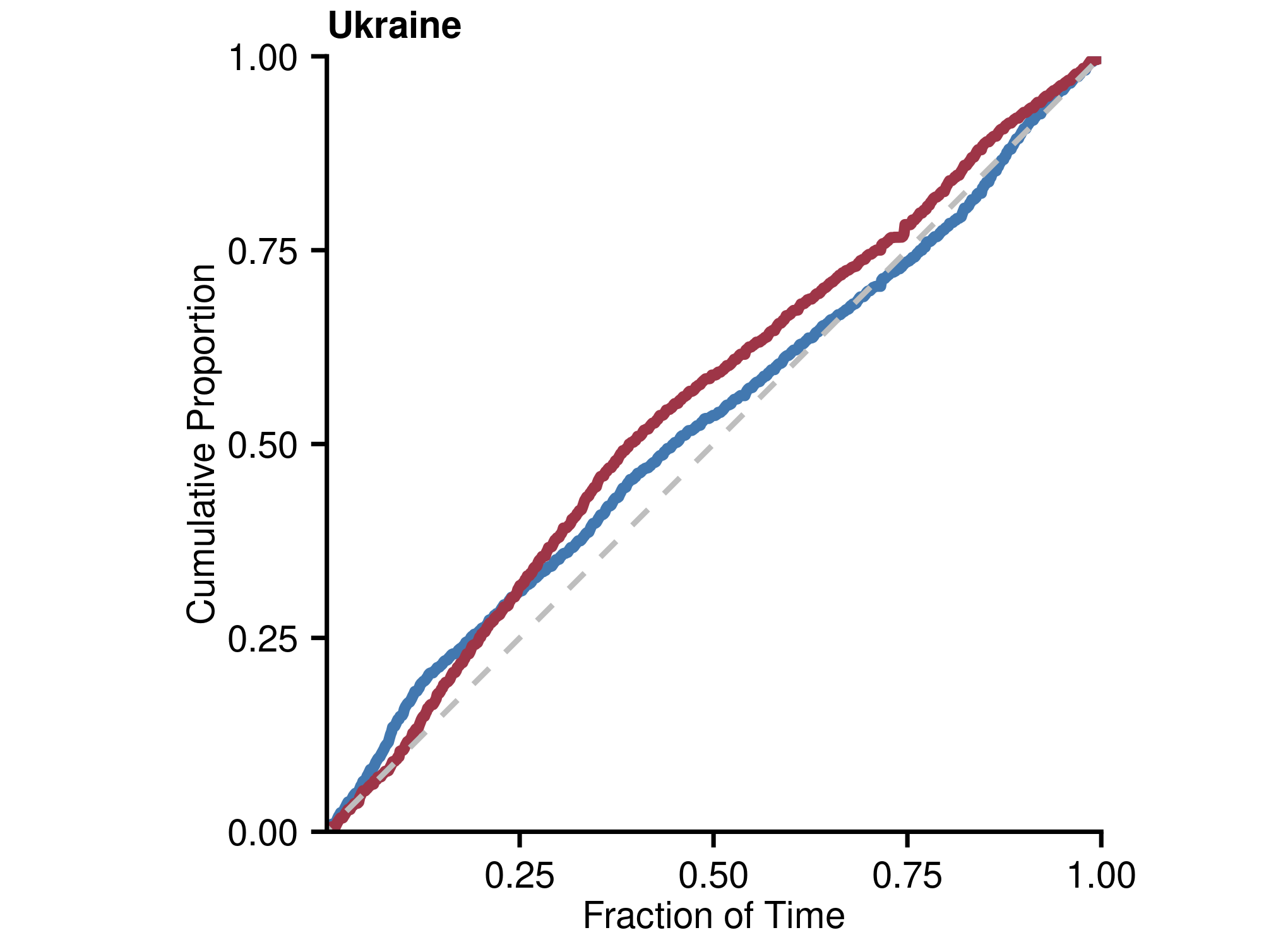} &
\includegraphics[width=0.31\textwidth]{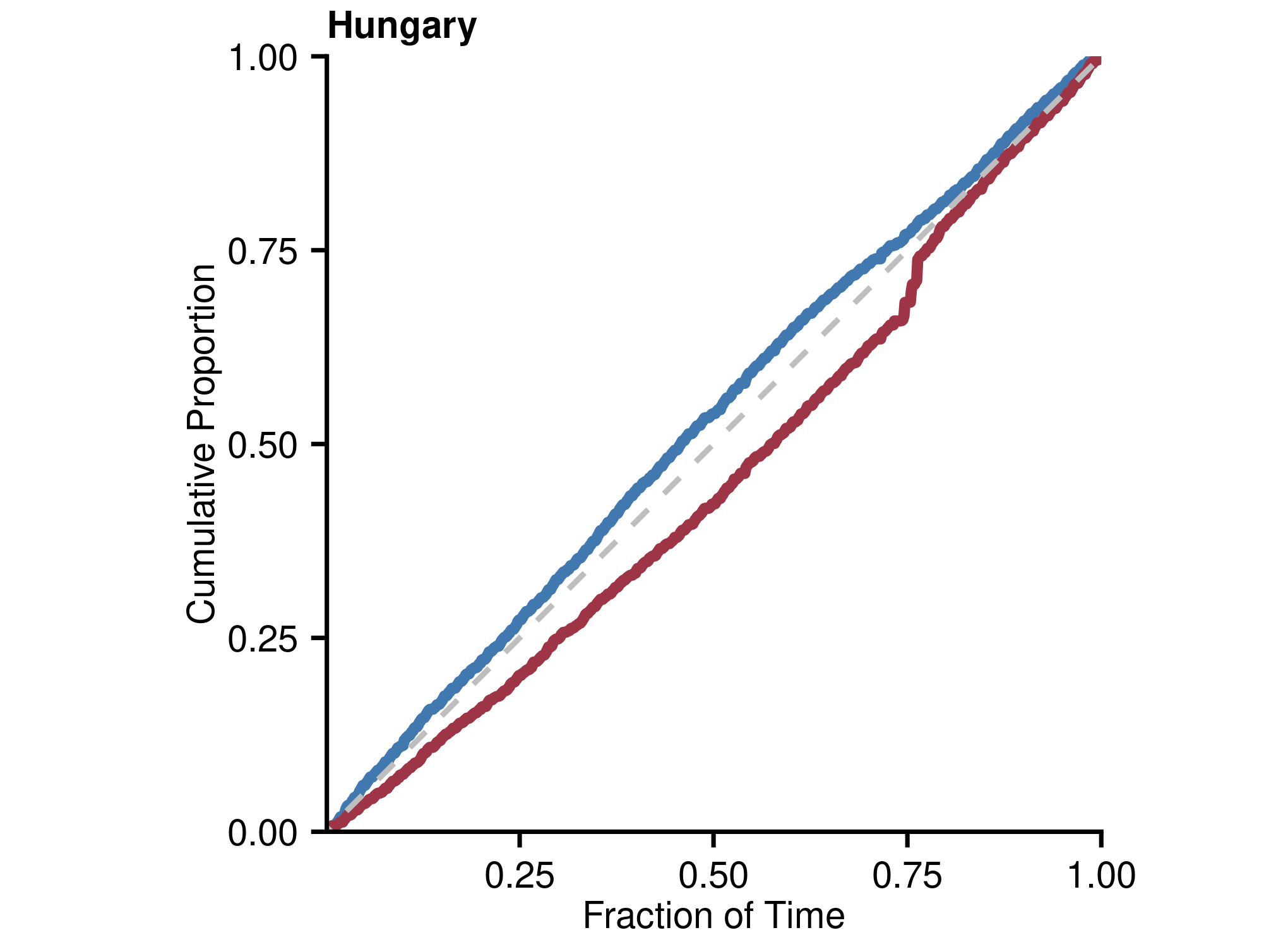} \\

\includegraphics[width=0.31\textwidth]{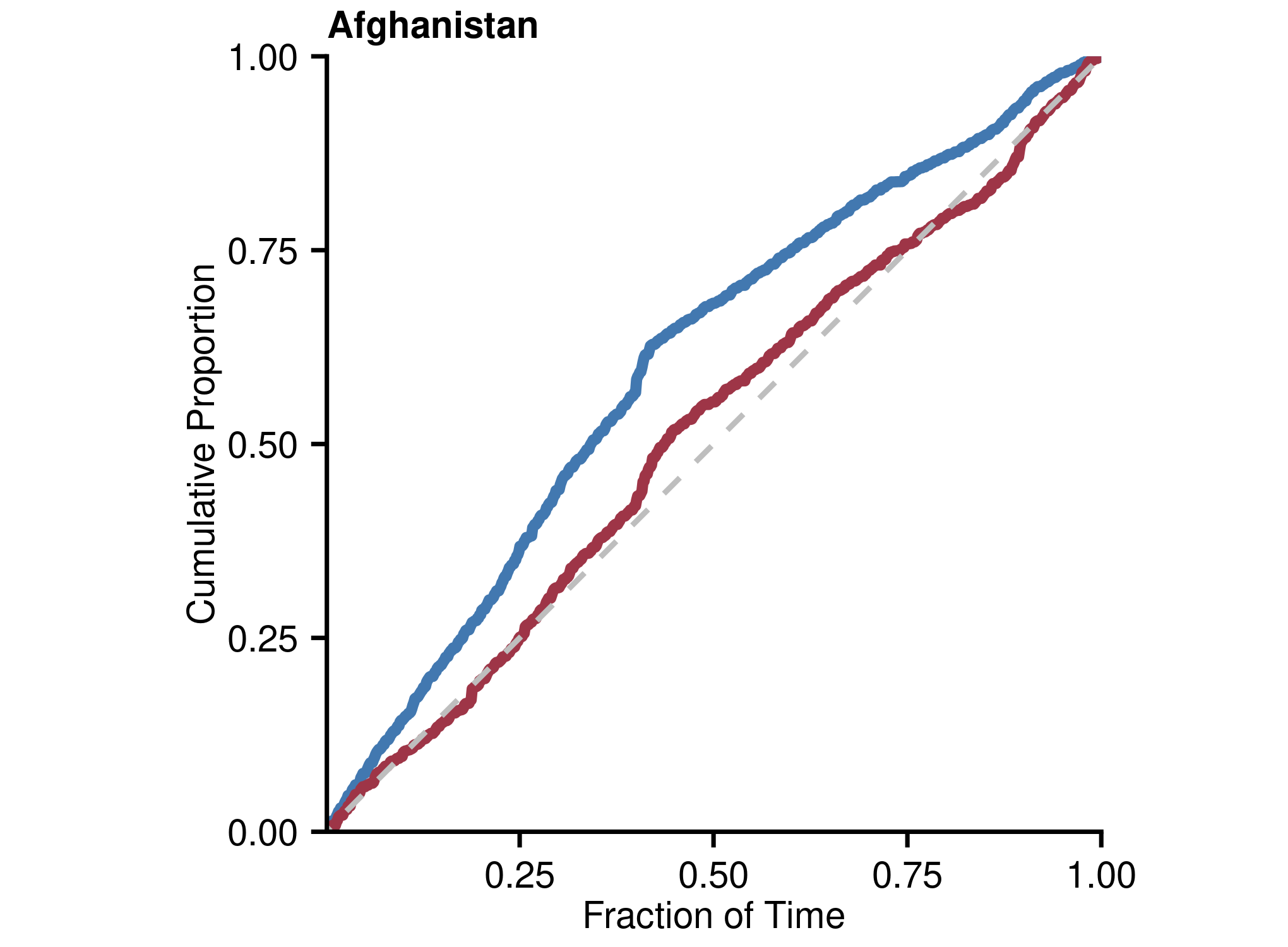} &
\includegraphics[width=0.31\textwidth]{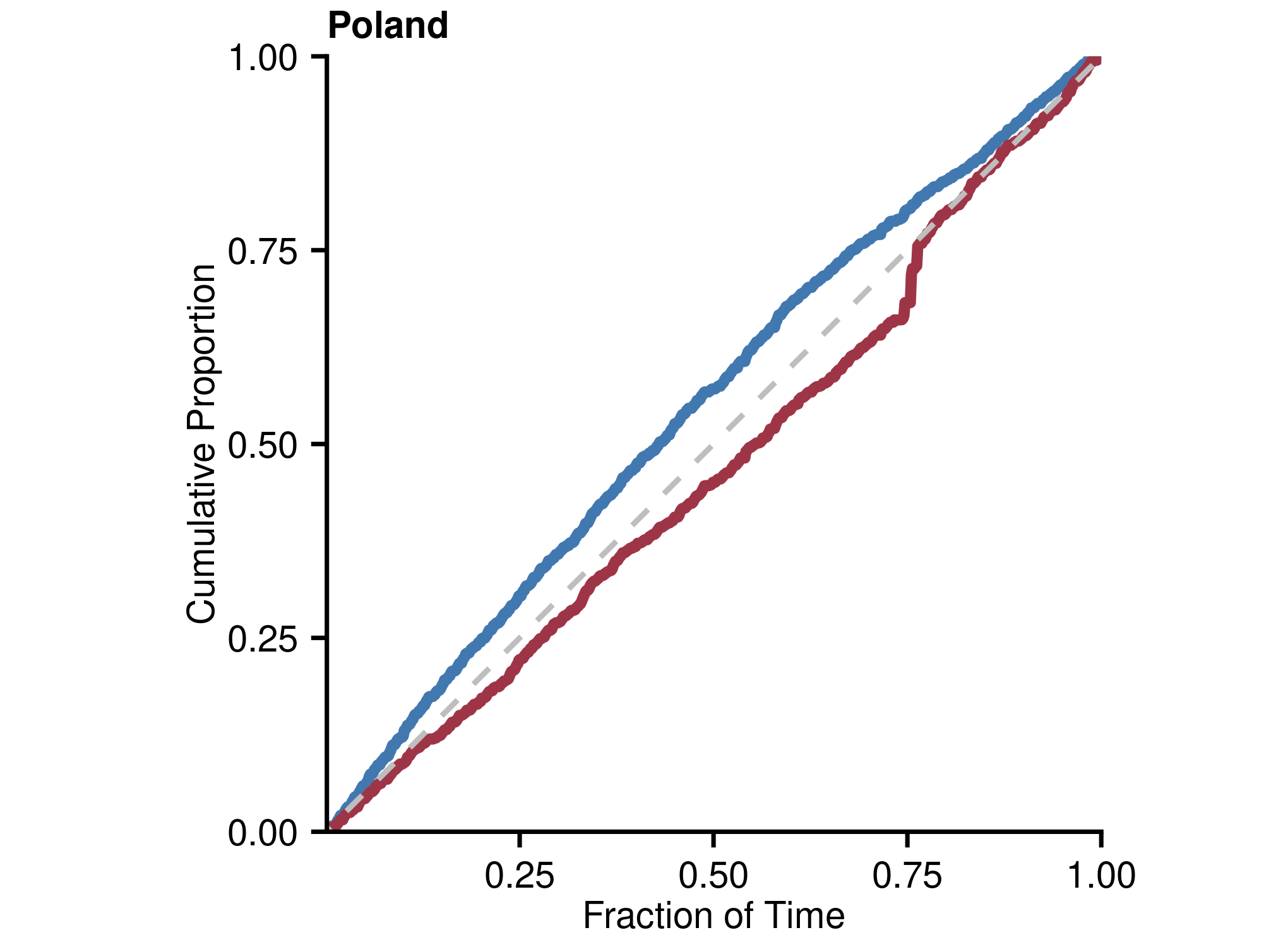} &
\includegraphics[width=0.31\textwidth]{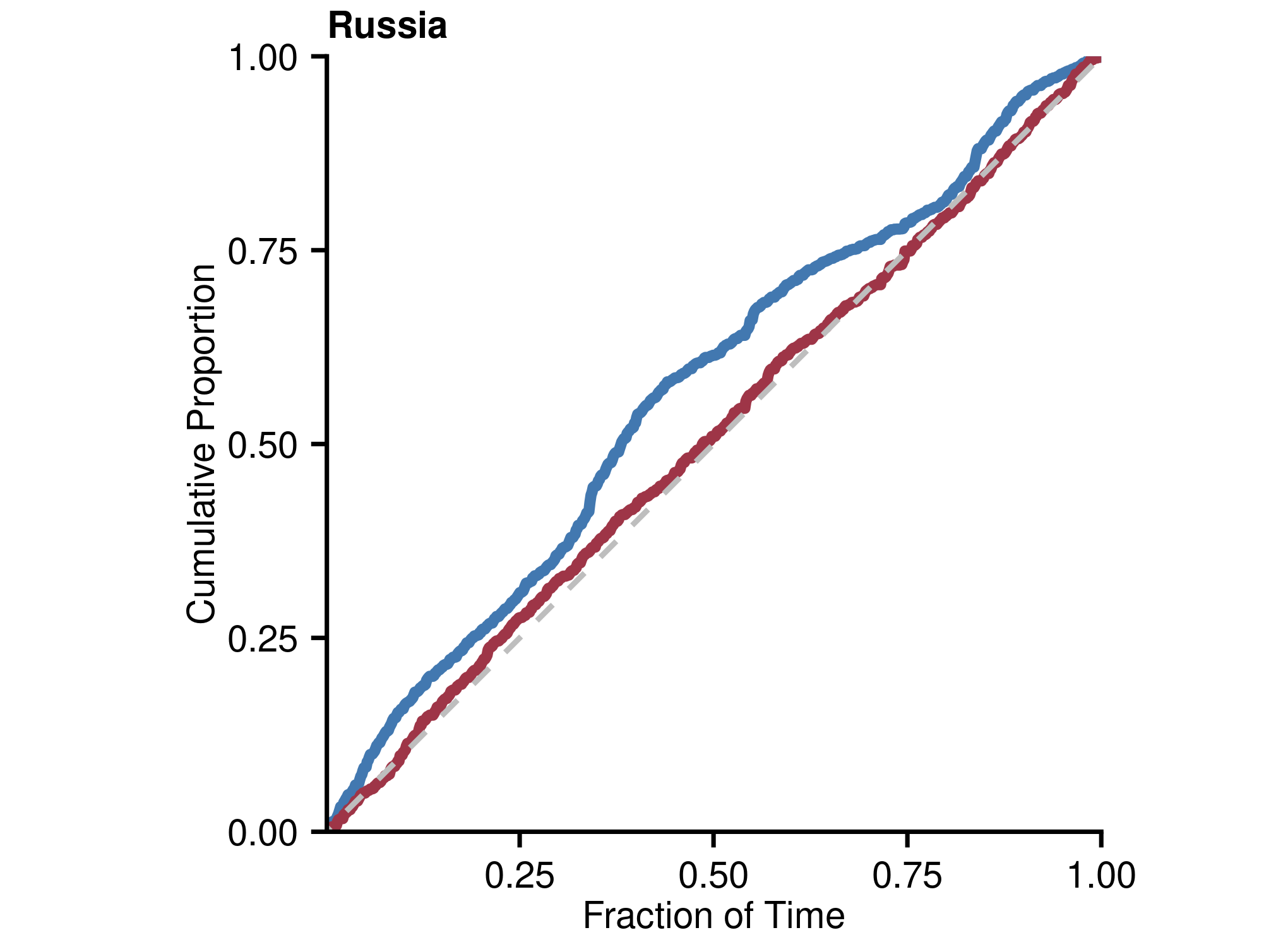} \\

\includegraphics[width=0.31\textwidth]{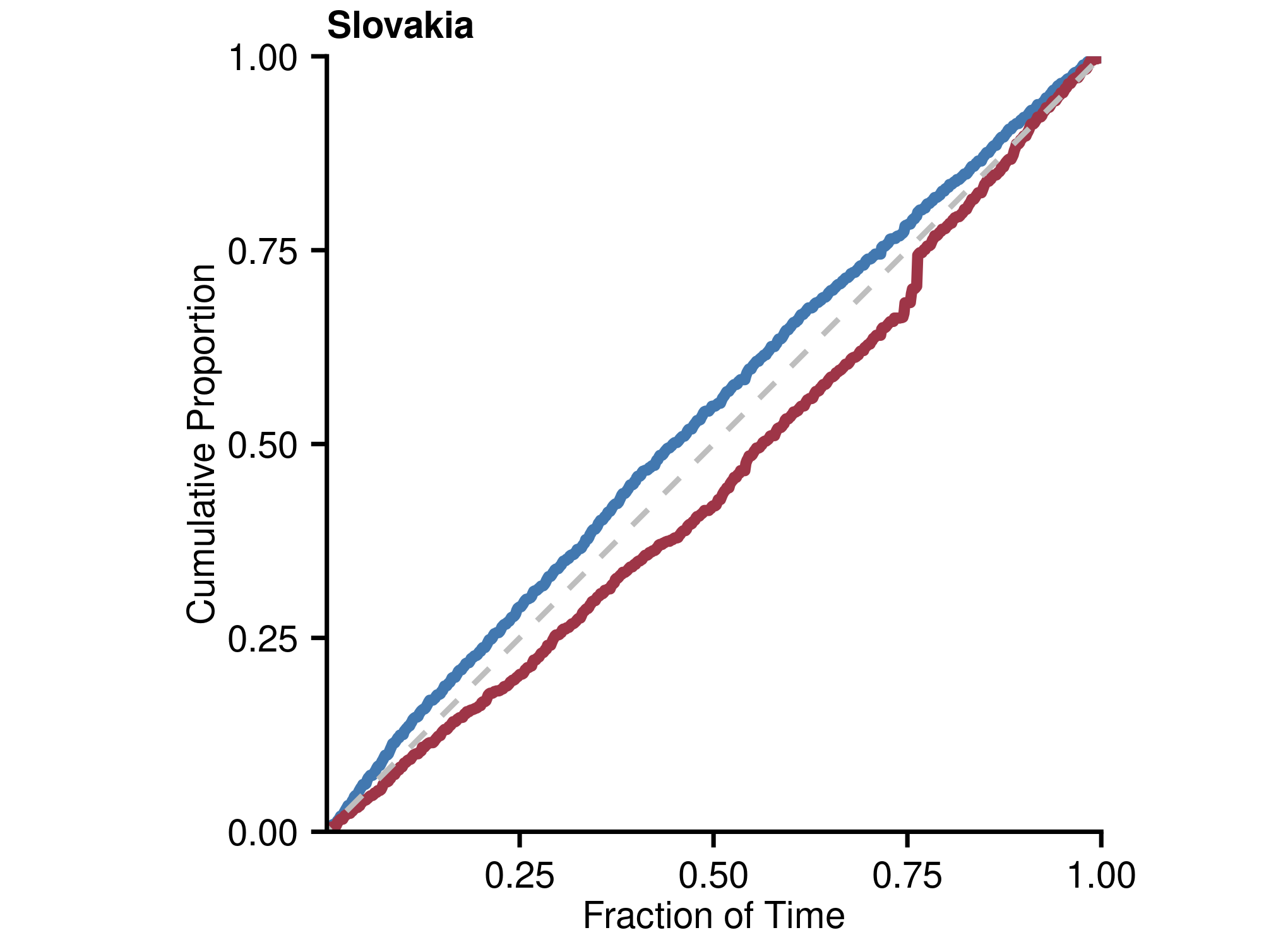} &
\includegraphics[width=0.31\textwidth]{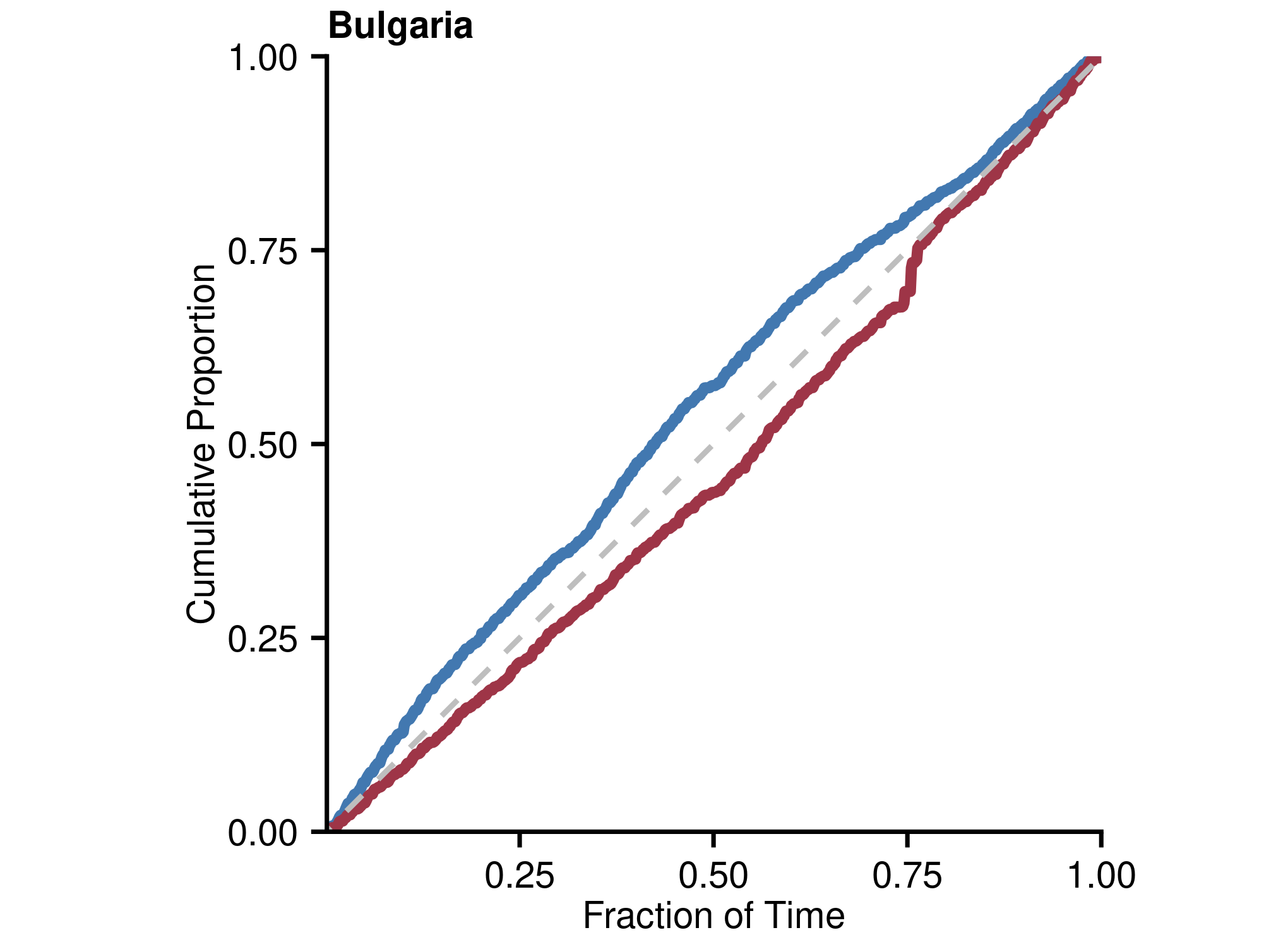} &
\includegraphics[width=0.31\textwidth]{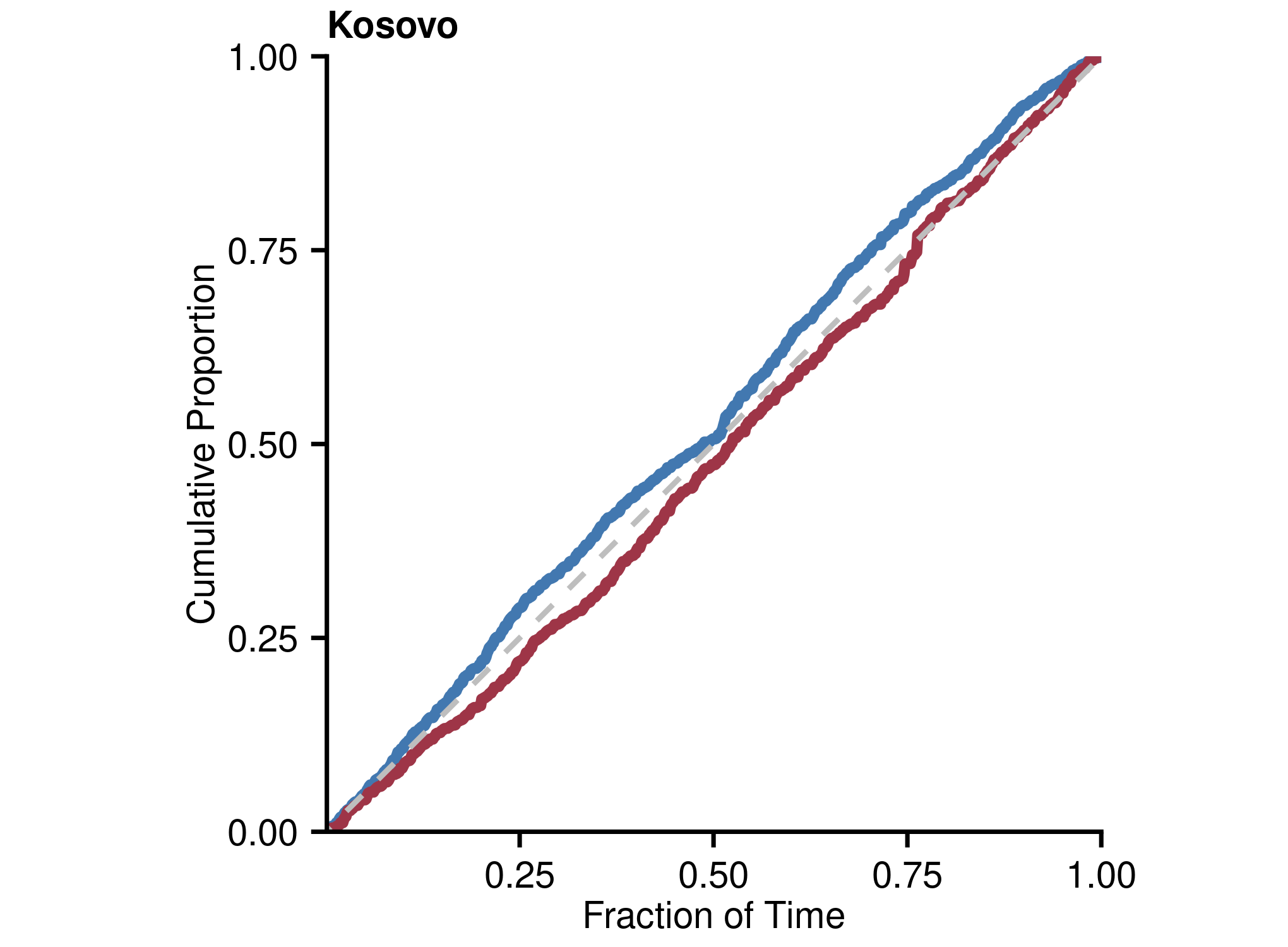} \\

\includegraphics[width=0.31\textwidth]{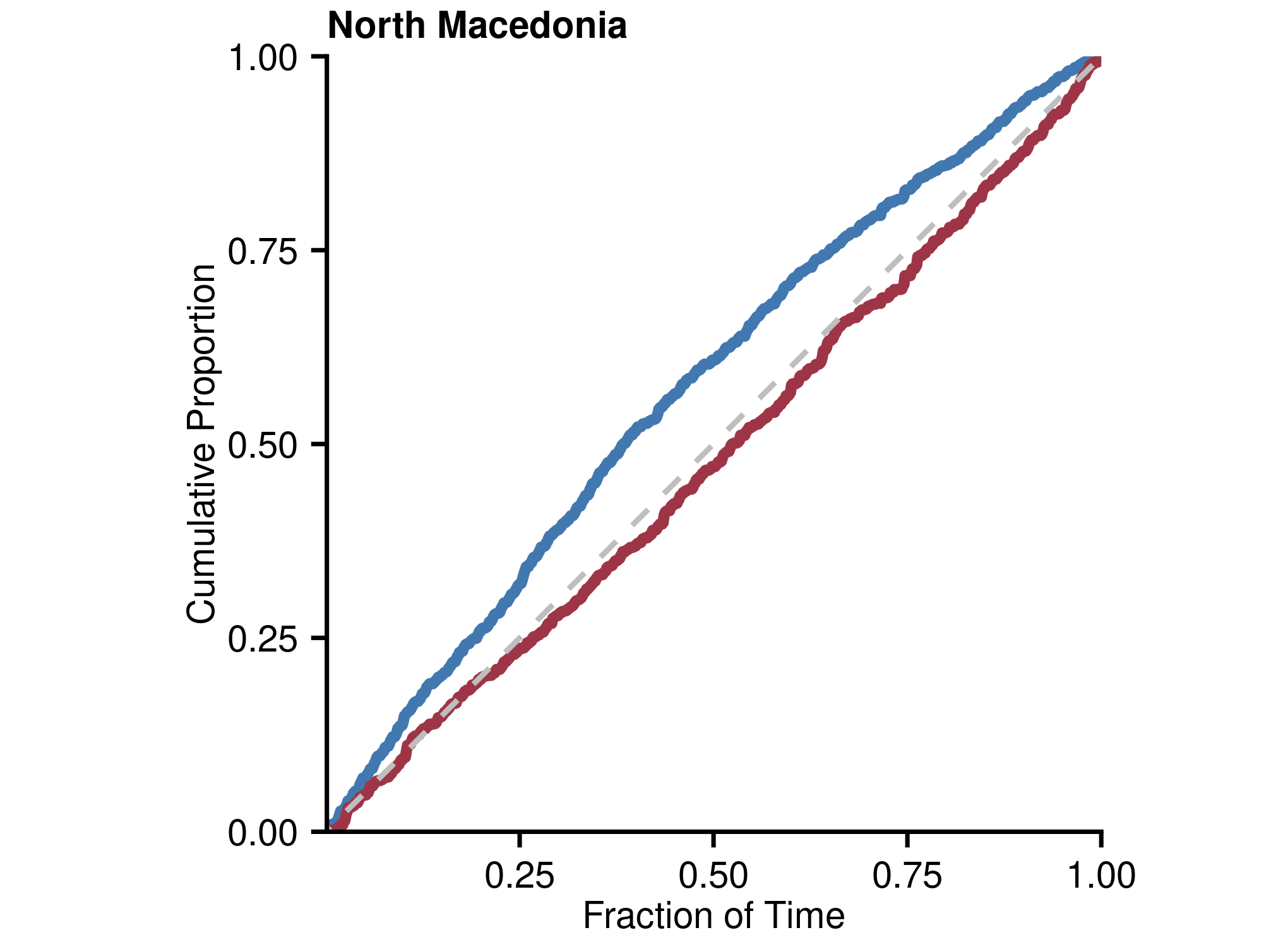} &
\includegraphics[width=0.31\textwidth]{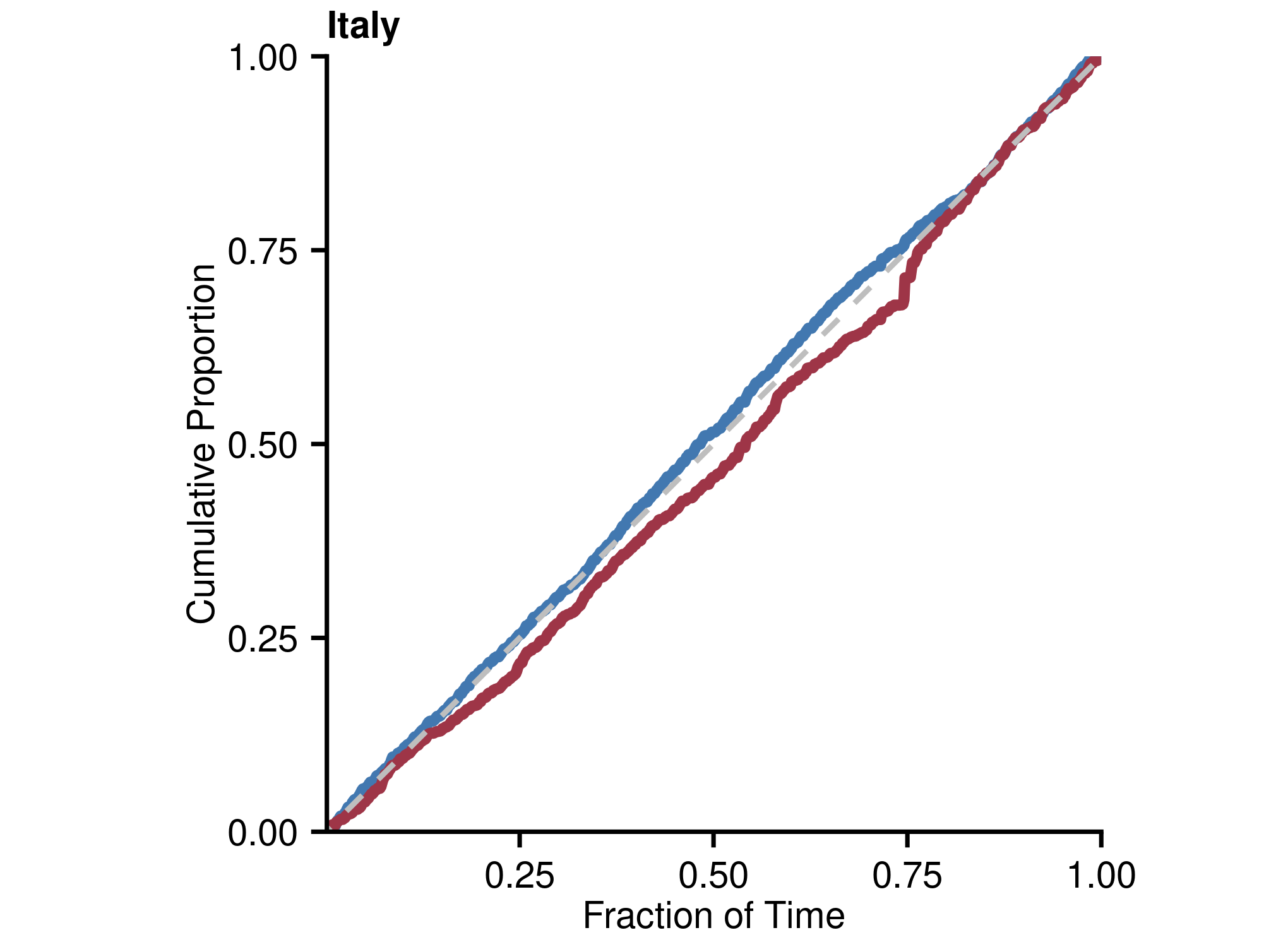} &
\includegraphics[width=0.31\textwidth]{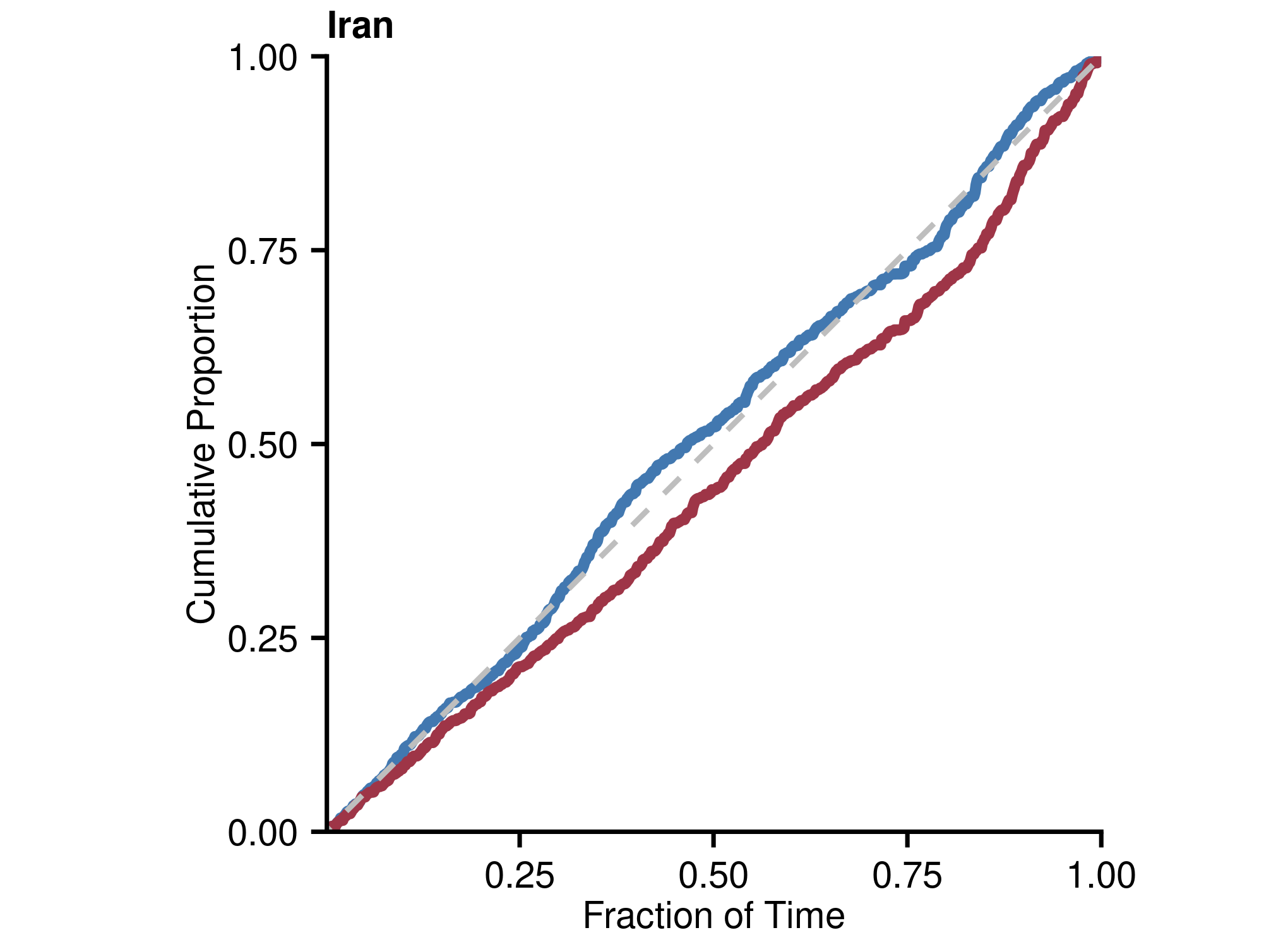} \\

\includegraphics[width=0.31\textwidth]{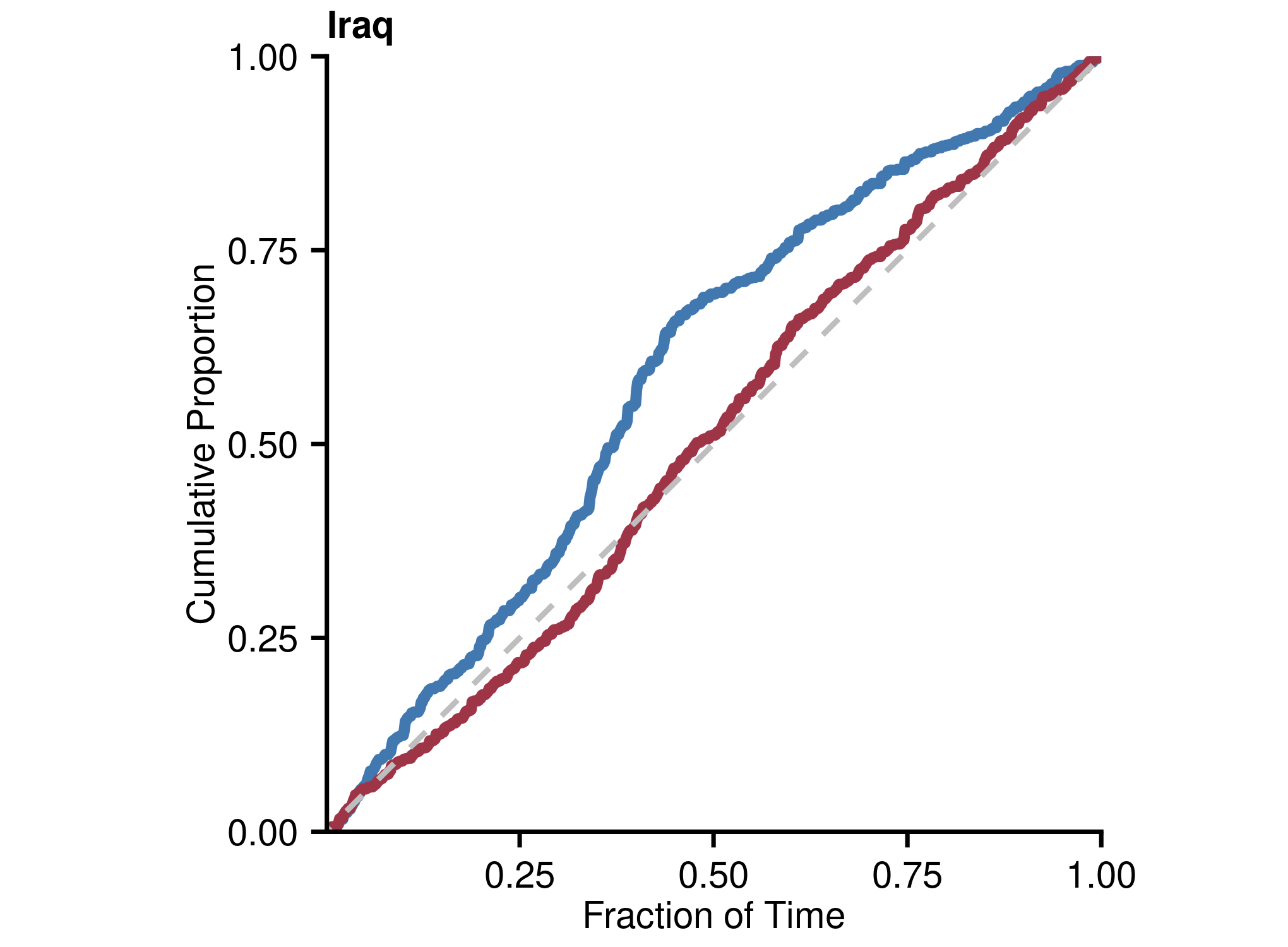} &
\includegraphics[width=0.31\textwidth]{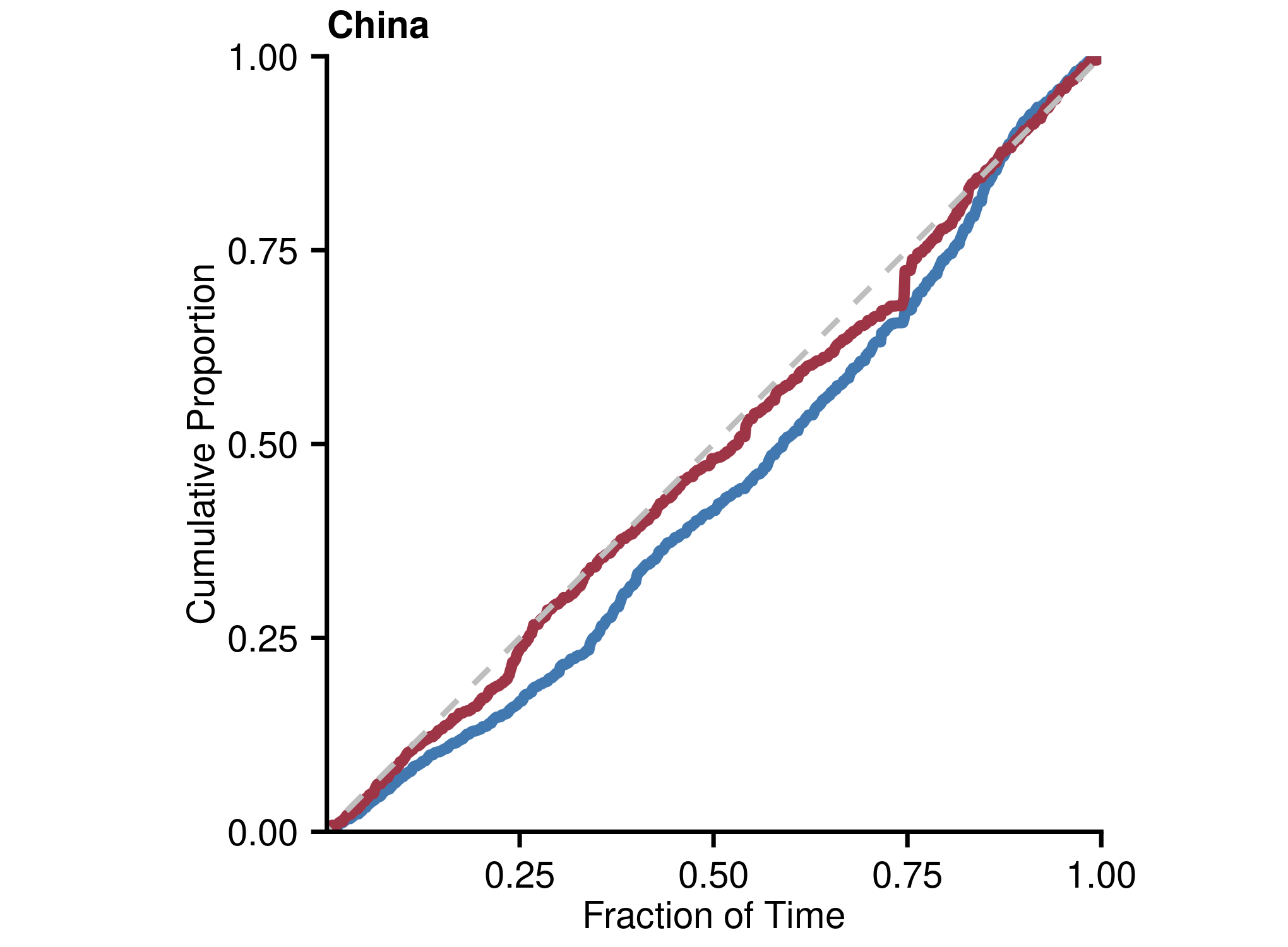} &
\includegraphics[width=0.31\textwidth]{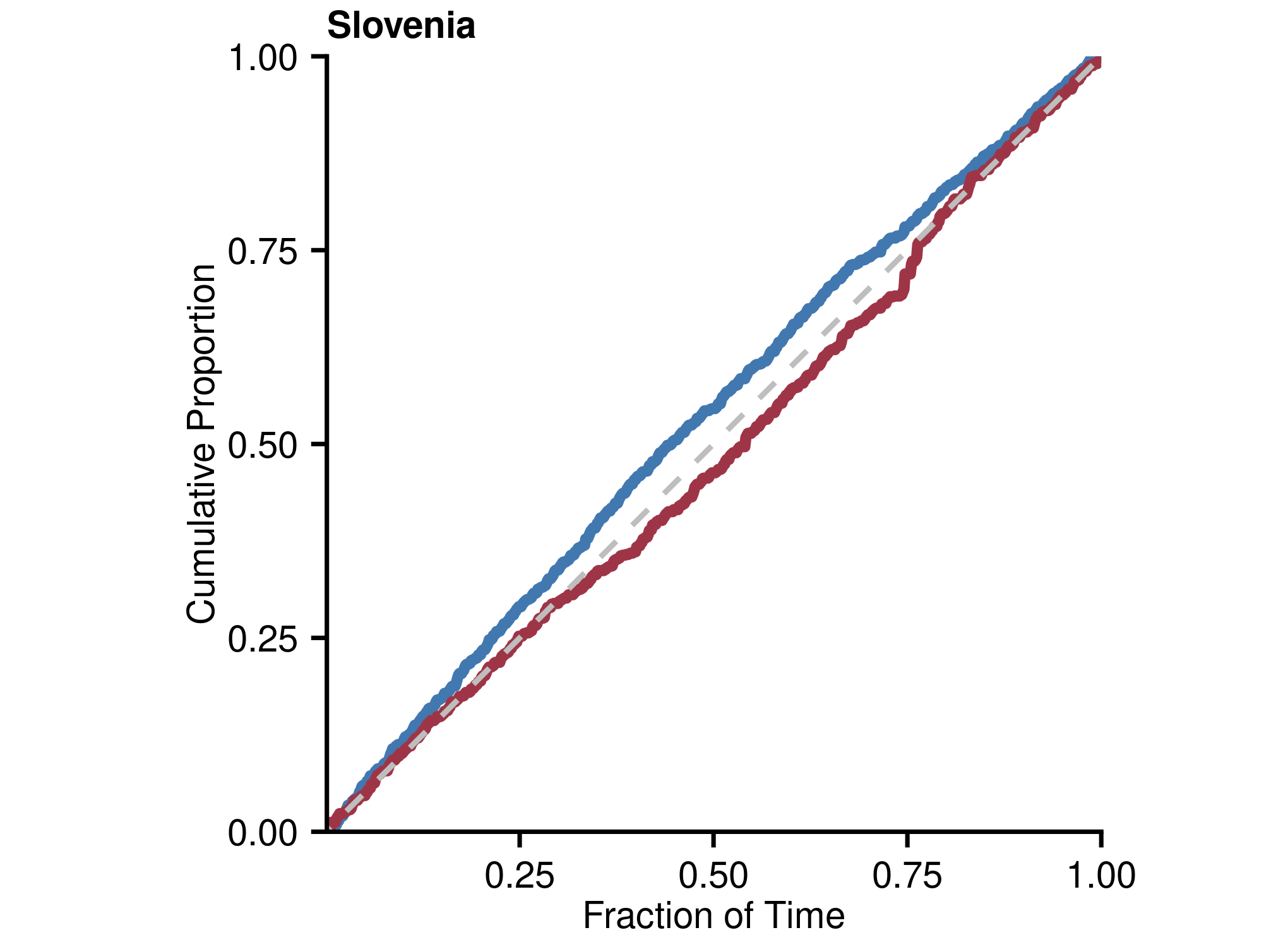} \\

\includegraphics[width=0.31\textwidth]{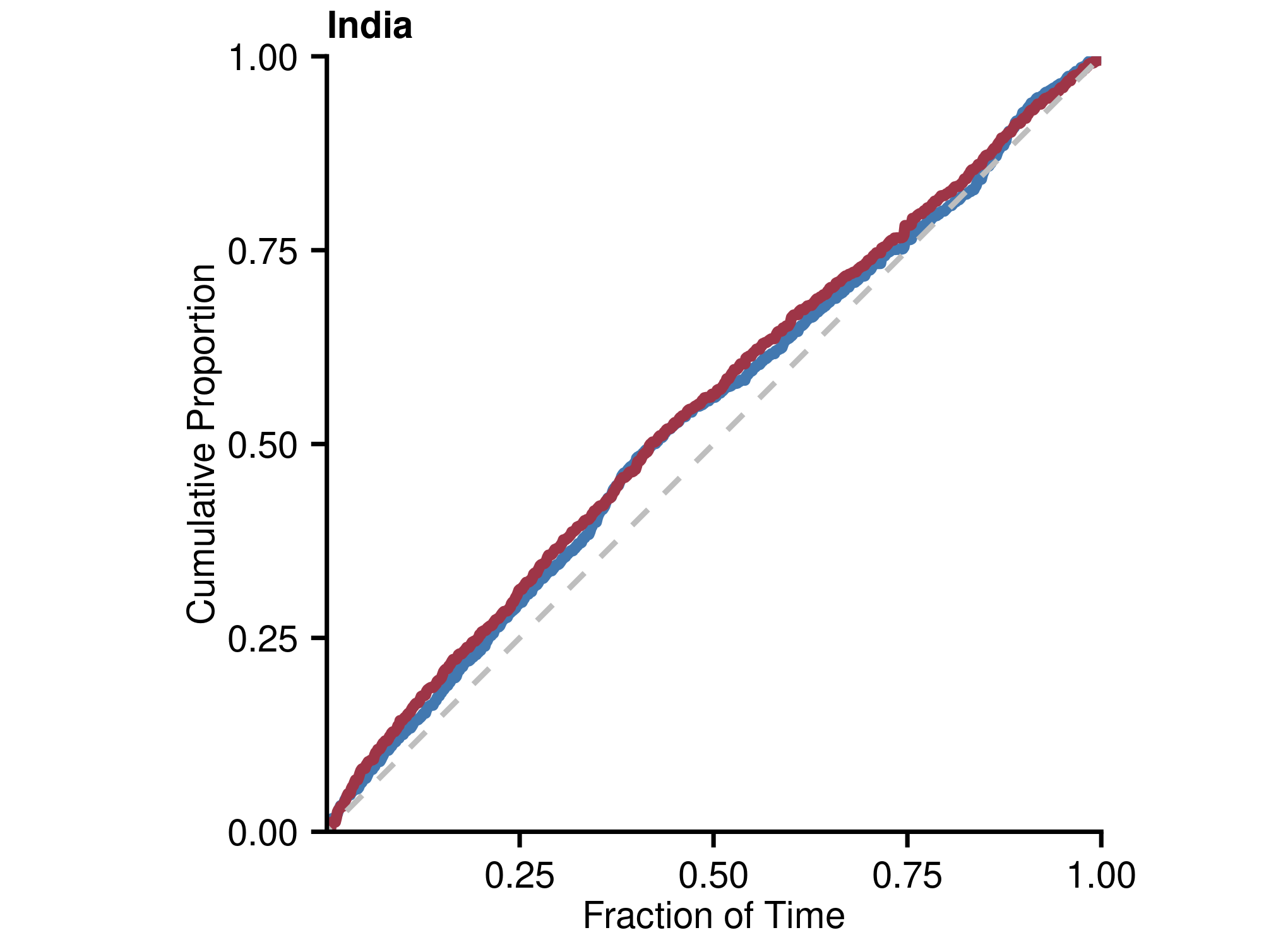} &
\includegraphics[width=0.31\textwidth]{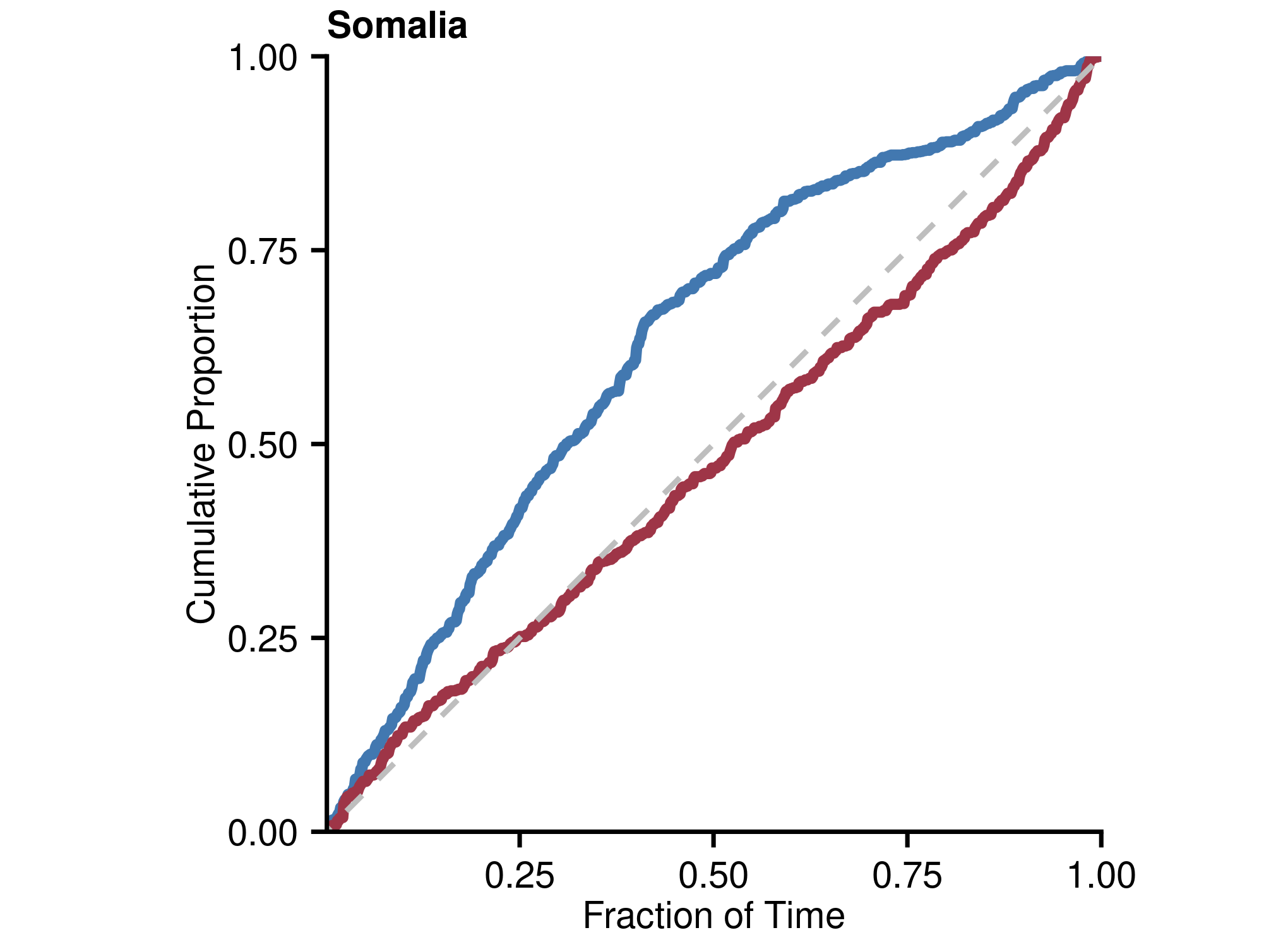} &
\includegraphics[width=0.31\textwidth]{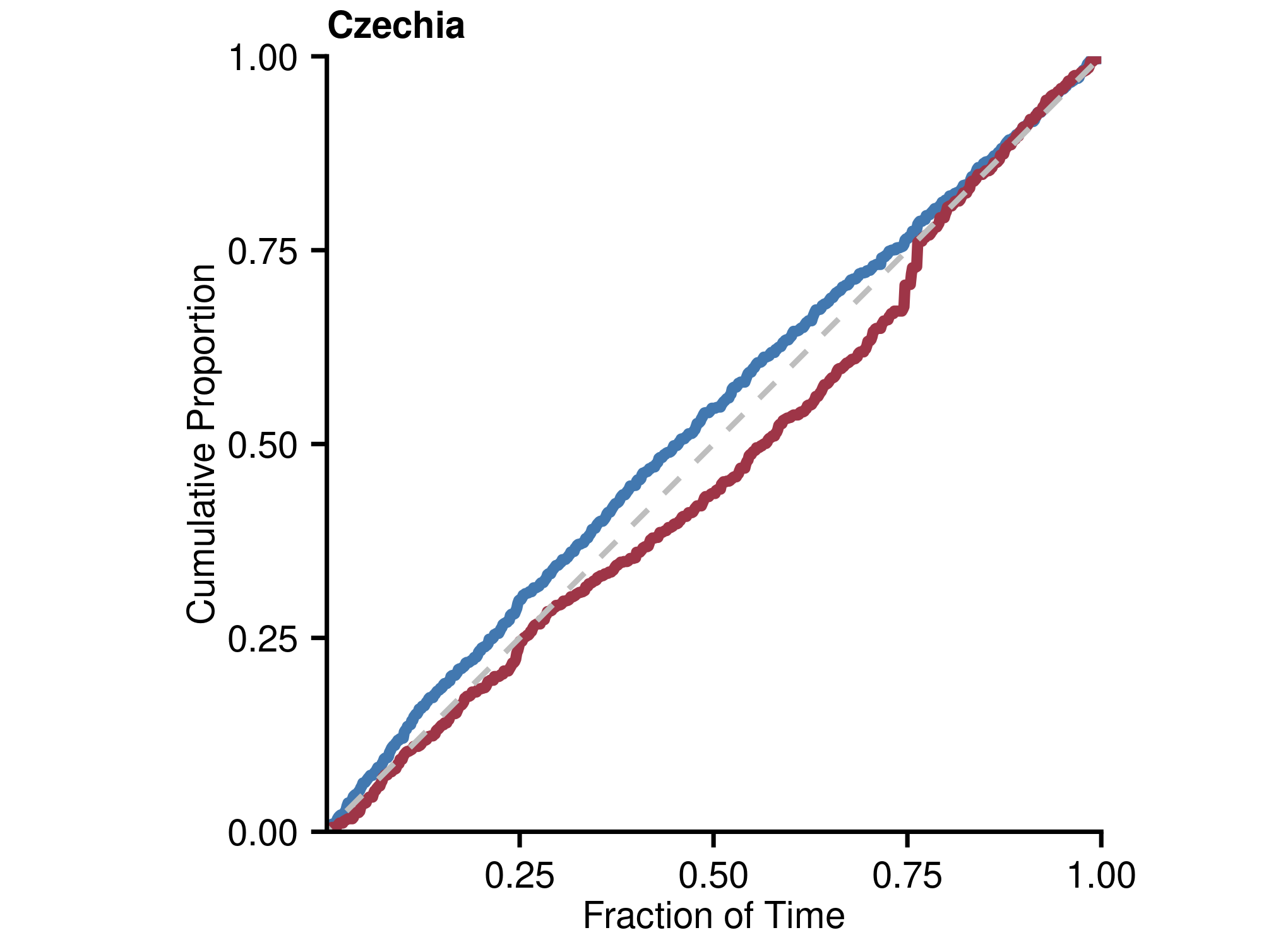} \\

\includegraphics[width=0.31\textwidth]{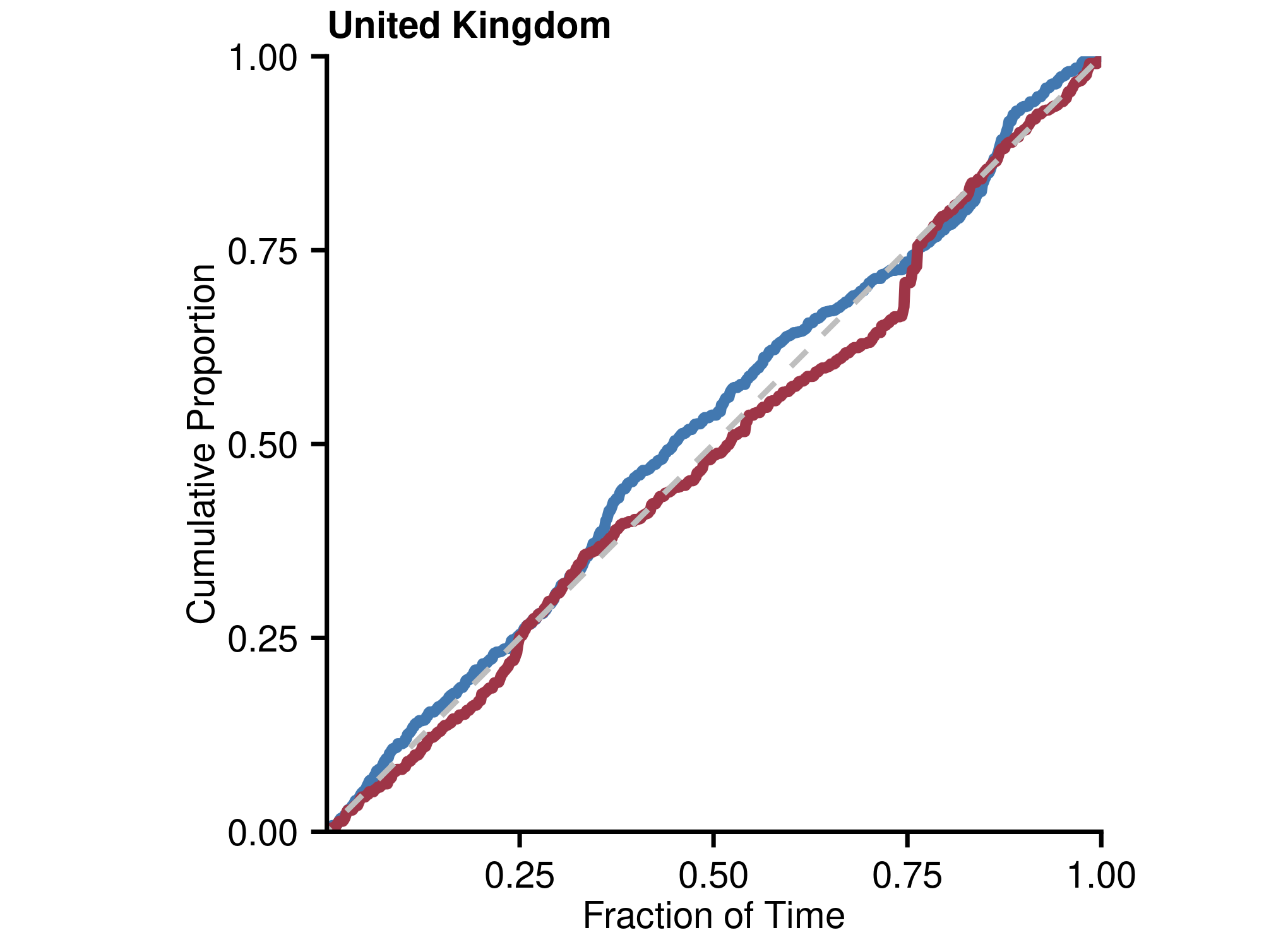} &
& \\
\end{longtable}
}

\subsubsection{By Age-Cohort}
%%% PLOTS Age-Cohorts
{
\renewcommand{\arraystretch}{1.0}
\setlength{\tabcolsep}{4pt}
\begin{longtable}{ccc}
\caption{\textbf{Flow Stability by Age Cohort:} Cumulative distribution of arrivals (blue) and exits (red) over the observed period. The x-axis represents the fraction of time elapsed; the y-axis shows the cumulative share of migration. Dashed lines indicate a constant migration rate.} \label{fig:flow_stability_age_cohorts} \\
\endfirsthead

\multicolumn{3}{r}%
{{\bfseries \tablename\ \thetable{} -- continued from previous page}} \\
\\
\endhead
\includegraphics[width=0.31\textwidth]{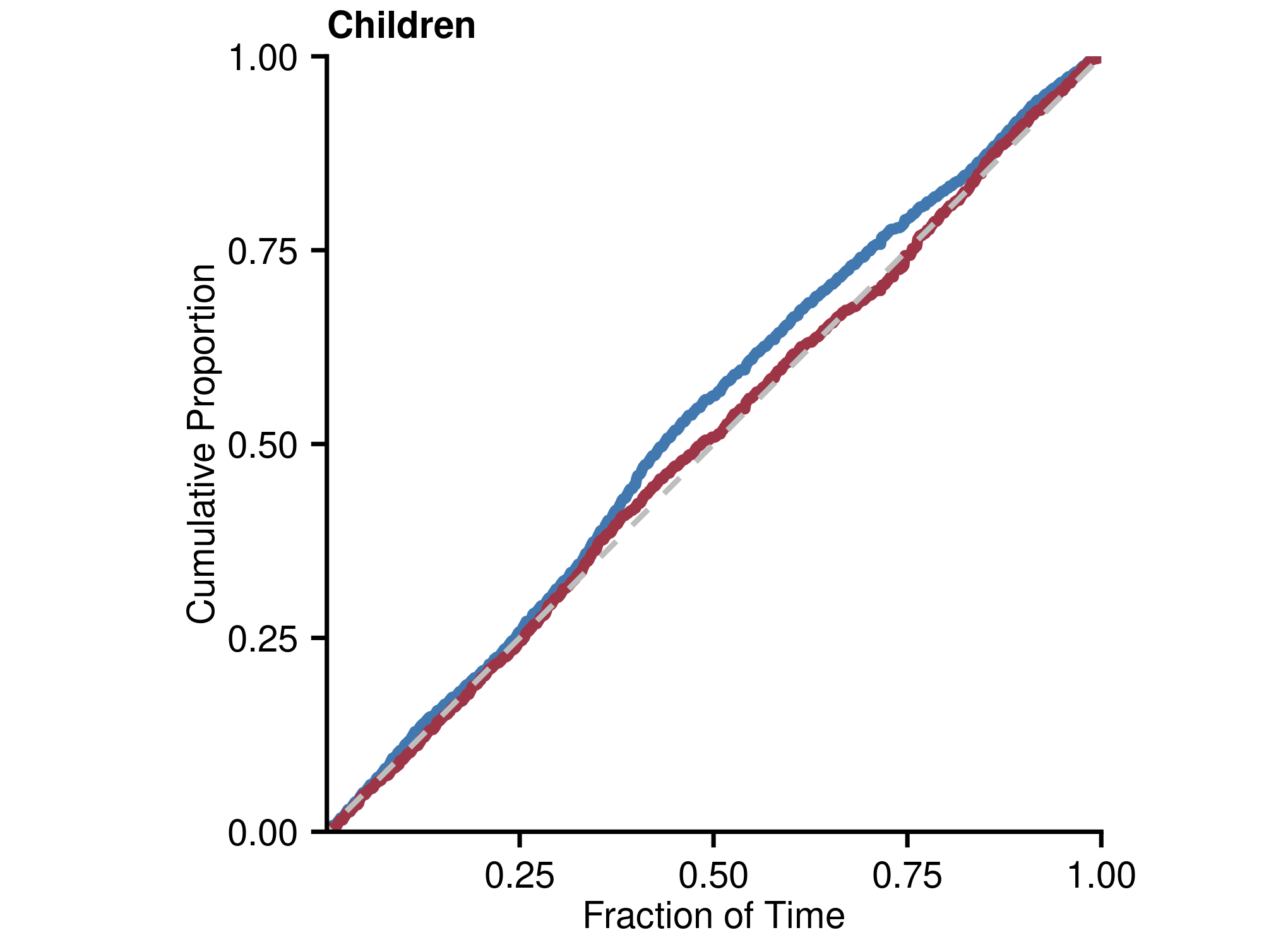} &
\includegraphics[width=0.31\textwidth]{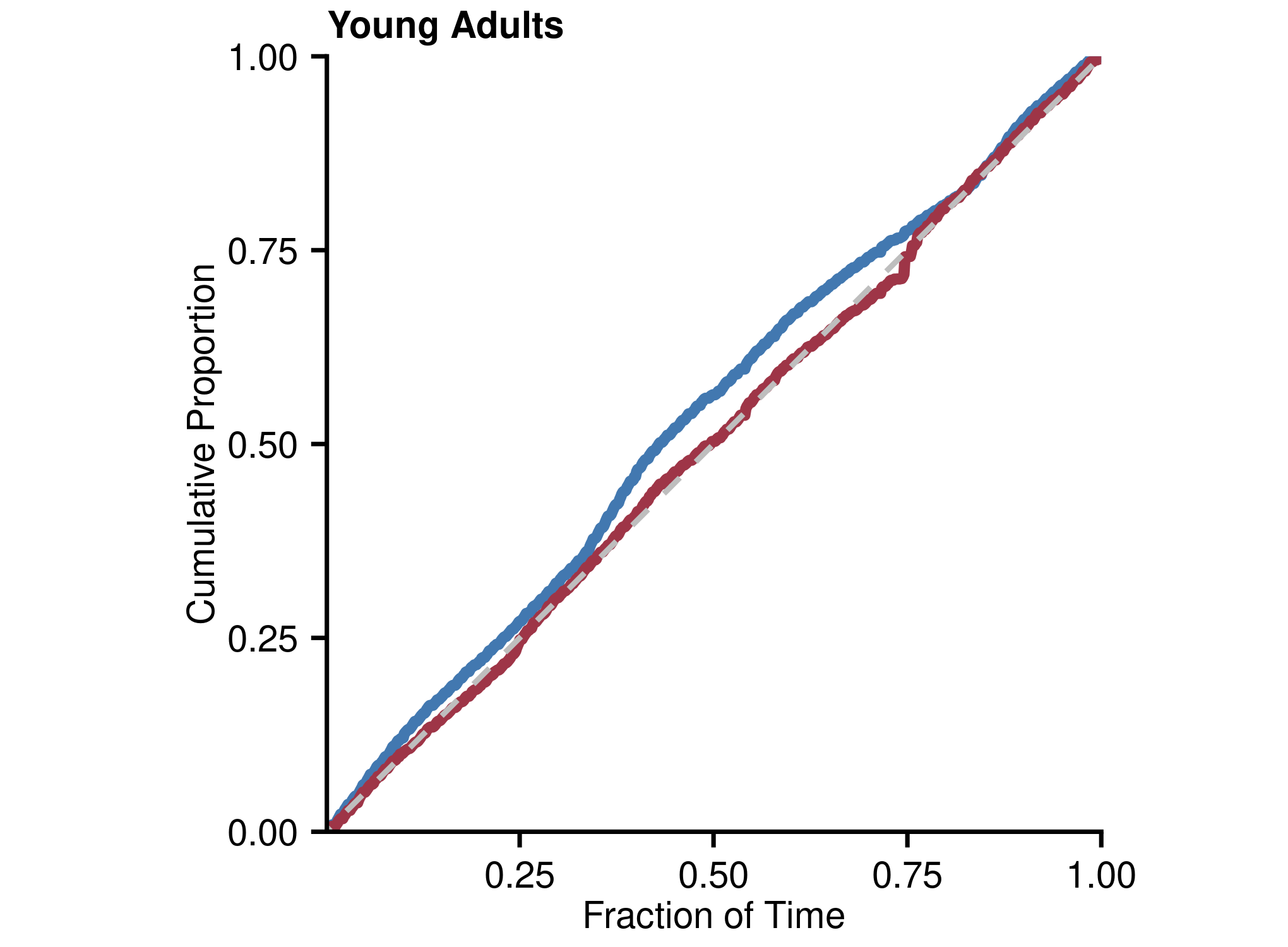} &
\includegraphics[width=0.31\textwidth]{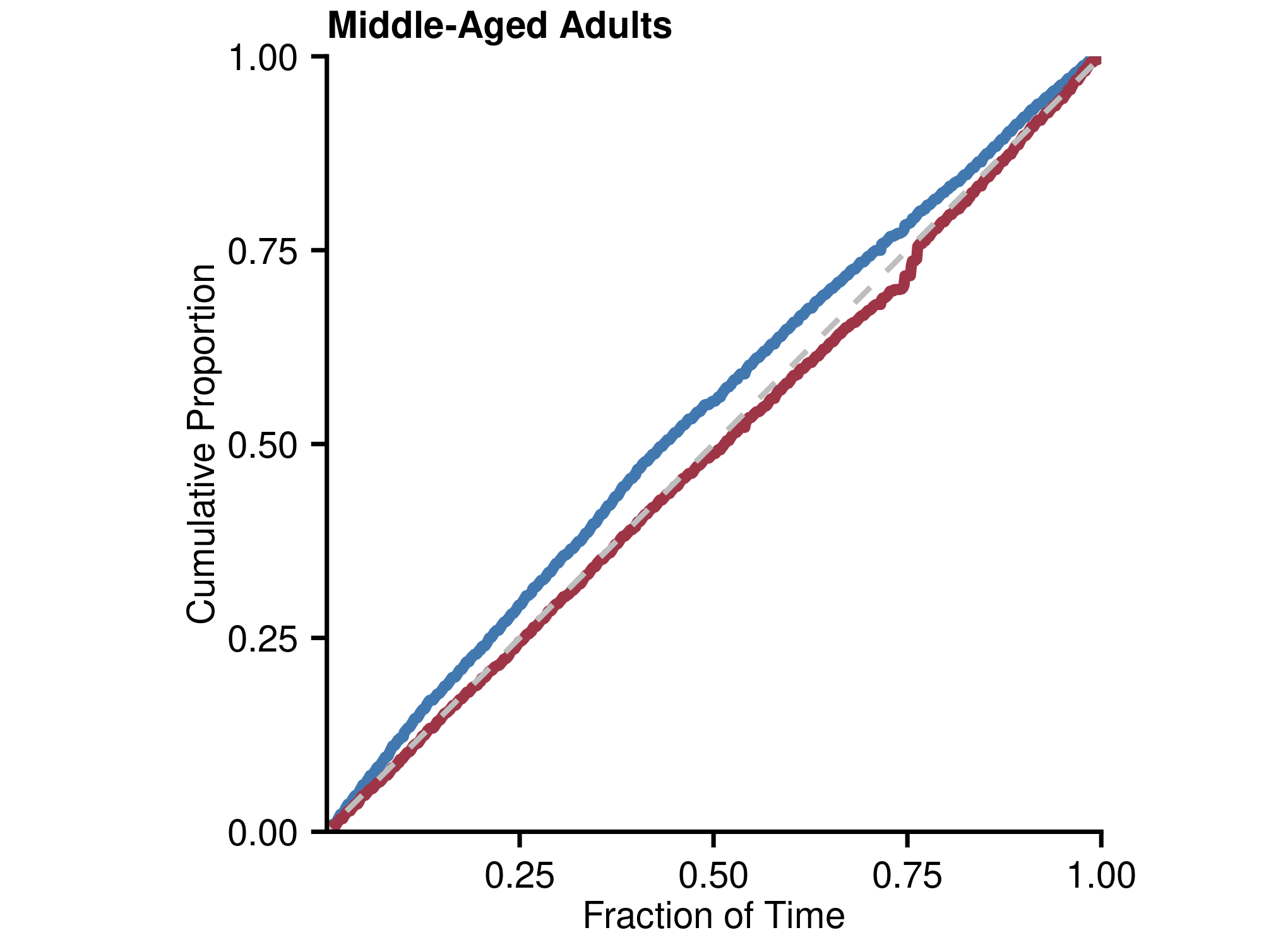} \\

\includegraphics[width=0.31\textwidth]{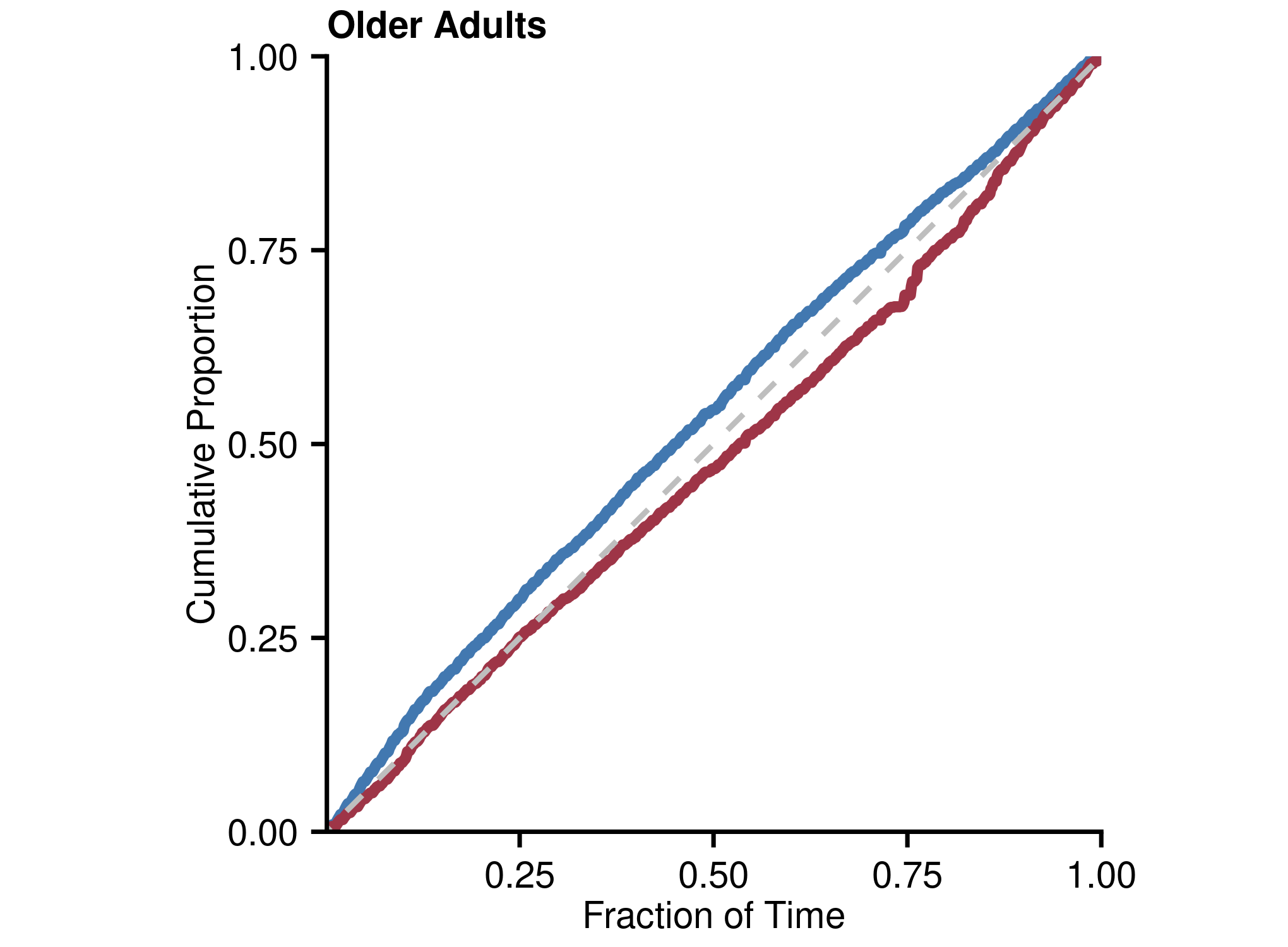} &
\includegraphics[width=0.31\textwidth]{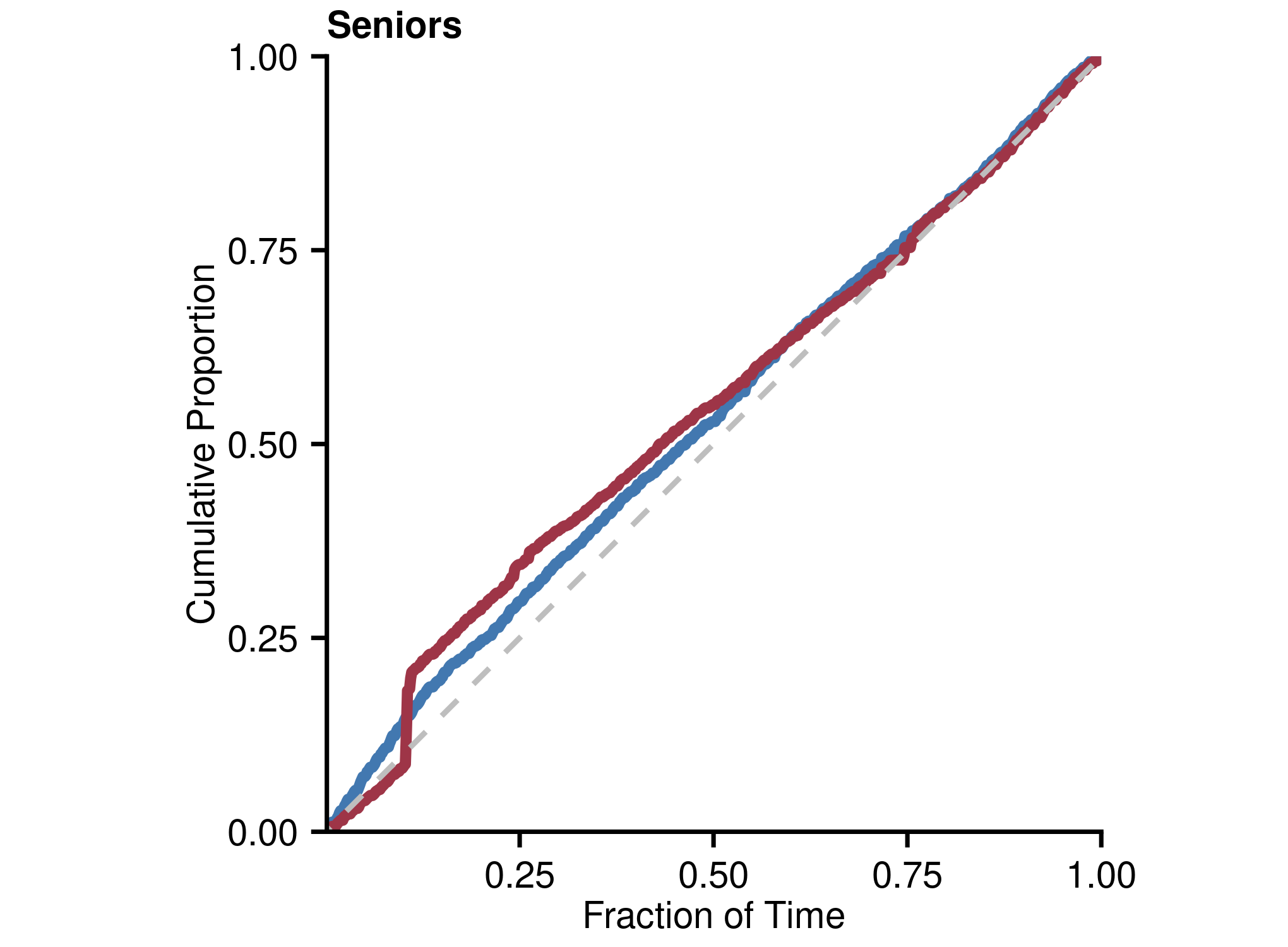} &
\\
\end{longtable}
}

\end{document}